\documentclass[12pt,english,aps,prd,preprintnumbers,superscriptaddress,floatfix,
showpacs,nofootinbib]{revtex4-2}
\usepackage{color}
\usepackage[dvipsnames]{xcolor}%
\usepackage{amsmath}
\usepackage{hyperref}
\usepackage[pdftex]{graphicx}
\usepackage{bm}
\usepackage[normalem]{ulem}

\usepackage{soul}
\usepackage[utf8]{inputenc}

\newcommand{\be}{\begin{eqnarray}}
\newcommand{\ee}{\end{eqnarray}}

\def\lsim{\mathrel{\rlap{\lower4pt\hbox{$\sim$}}
    \raise1pt\hbox{$<$}}}               
\def\gsim{\mathrel{\rlap{\lower4pt\hbox{$\sim$}}
    \raise1pt\hbox{$>$}}}  

\begin{document}
\setstcolor{red} 

\title{The BEST framework for the search for the QCD critical point and 
the chiral magnetic effect}


\author{Xin An}
\affiliation{Department of Physics and Astronomy, 
University of North Carolina, Chapel Hill, North Carolina 27599, USA}

\author{Marcus Bluhm}
\affiliation{SUBATECH UMR 6457 (IMT Atlantique, Université de Nantes, IN2P3/CNRS), 
4 rue Alfred Kastler, 44307 Nantes, France}

\author{Lipei Du}
\affiliation{Department of Physics, The Ohio State University, Columbus, OH 43210, USA}

\author{Gerald V. Dunne}
\affiliation{Department of Physics, University of Connecticut, Storrs CT 06269-3046, USA}

\author{Hannah Elfner}
\affiliation{GSI Helmholtzzentrum fuer Schwerionenforschung, Planckstr. 1, 64291 
Darmstadt, Germany}
\affiliation{Institute for Theoretical Physics, Goethe University, Max-von-Laue-Strasse 1,
60438 Frankfurt am Main, Germany}
\affiliation{Frankfurt Institute for Advanced Studies, Ruth-Moufang-Strasse 1, 
60438 Frankfurt am Main, Germany}
\affiliation{Helmholtz Research Academy Hesse for FAIR (HFHF), GSI Helmholtz Center, 
Campus Frankfurt, Max-von-Laue-Straße 12, 60438 Frankfurt am Main, Germany}

\author{Charles Gale}
\affiliation{Department of Physics, McGill University, 3600 University Street, 
Montreal, QC, Canada H3A 2T8}

\author{Joaquin Grefa}
\affiliation{Department of Physics, University of Houston, Houston, TX 77204, USA}

\author{Ulrich Heinz}
\affiliation{Department of Physics, The Ohio State University, Columbus, OH 43210, USA}

\author{Anping Huang}
\affiliation{
Physics Department and Center for Exploration of Energy and Matter,
Indiana University, 2401 N Milo B. Sampson Lane, Bloomington, IN 47408, USA}

\author{Jamie M. Karthein}
\affiliation {Department of Physics, University of Houston, Houston, TX 77204, USA}

\author{Dmitri E. Kharzeev}
\affiliation{Center for Nuclear Theory, Department of Physics and Astronomy, 
Stony Brook University, NY 11794, USA}
\affiliation{Physics Department, Brookhaven National Laboratory, Upton, NY 11973, 
USA}

\author{Volker Koch}
\affiliation{Nuclear Science Division, Lawrence Berkeley National Laboratory, 
1 Cyclotron Road, Berkeley, CA 94720, USA}

\author{Jinfeng Liao}
\affiliation{
Physics Department and Center for Exploration of Energy and Matter,
Indiana University, 2401 N Milo B. Sampson Lane, Bloomington, IN 47408, USA}

\author{Shiyong Li}
\affiliation{Department of Physics, University of Illinois, Chicago, IL 60607, USA}

\author{Mauricio Martinez}
\affiliation{Department of Physics, North Carolina State University, Raleigh, NC 27695, USA}

\author{Michael McNelis}
\affiliation{Department of Physics, The Ohio State University, Columbus, OH 43210, USA}

\author{Debora Mroczek}
\affiliation{Illinois Center for Advanced Studies of the Universe, 
Department of Physics, 
University of Illinois at Urbana-Champaign, Urbana, IL 61801, USA}

\author{Swagato Mukherjee}
\affiliation{Physics Department, Brookhaven National Laboratory, Upton, NY 11973, USA}

\author{Marlene Nahrgang}
\affiliation{SUBATECH UMR 6457 (IMT Atlantique, Université de Nantes, IN2P3/CNRS), 
4 rue Alfred Kastler, 44307 Nantes, France}

\author{Angel R.~Nava Acuna}
\affiliation{Department of Physics, University of Houston, Houston, 
TX 77204, USA}

\author{Jacquelyn Noronha-Hostler}
\affiliation{Illinois Center for Advanced Studies of the Universe, 
Department of Physics, 
University of Illinois at Urbana-Champaign, Urbana, IL 61801, USA}

\author{Dmytro Oliinychenko}
\affiliation{Institute for Nuclear Theory, University of Washington, Box 351550, 
Seattle, WA, 98195, USA}

\author{Paolo Parotto}
\affiliation{Department of Physics, University of Wuppertal, Gaussstr. 20, 
D-42119 Wuppertal, Germany}

\author{Israel Portillo}
\affiliation{Department of Physics, University of Houston, Houston, TX 77204, USA}

\author{Maneesha Sushama Pradeep}
\affiliation{Department of Physics, University of Illinois, Chicago, IL 60607, USA}

\author{Scott Pratt}
\affiliation{Department of Physics and Astronomy and National Superconducting 
Cyclotron Laboratory Michigan State University, East Lansing, MI 48824 USA}

\author{Krishna Rajagopal}
\affiliation{Center for Theoretical Physics, Massachusetts Institute of 
Technology, Cambridge, MA 02139, USA}

\author{Claudia Ratti}
\affiliation{Department of Physics, University of Houston, Houston, TX 77204, USA}

\author{Gregory Ridgway}
\affiliation{Center for Theoretical Physics, Massachusetts Institute of 
Technology, Cambridge, MA 02139, USA}

\author{Thomas Sch\"afer}
\affiliation{Department of Physics, North Carolina State University, Raleigh, NC 27695, USA}

\author{Bj\"orn Schenke}
\affiliation{Physics Department, Brookhaven National Laboratory, Upton, NY 11973, USA}

\author{Chun Shen}
\affiliation{Department of Physics and Astronomy, Wayne State University Detroit,  MI 48201, USA}
\affiliation{RIKEN BNL Research Center, Brookhaven National Laboratory Upton,  NY 11973, USA}

\author{Shuzhe Shi}
\affiliation{Department of Physics, McGill University, 3600 University Street, Montreal, QC, 
Canada H3A 2T8} 

\author{Mayank Singh}
\affiliation{Department of Physics, McGill University, 3600 University Street, Montreal, QC, 
Canada H3A 2T8} 
\affiliation{School of Physics and Astronomy, University of Minnesota, 
Minneapolis, MN 55455, USA}

\author{Vladimir Skokov}
\affiliation{Department of Physics, North Carolina State University, Raleigh, NC 27695, USA}
\affiliation{RIKEN BNL Research Center, Brookhaven National Laboratory Upton,  NY 11973, USA}

\author{Dam T. Son}
\affiliation{Kadanoff Center for Theoretical Physics, University of Chicago, 
Chicago, Illinois 60637, USA} 

\author{Agnieszka Sorensen}
\affiliation{Nuclear Science Division, Lawrence Berkeley National Laboratory, 
1 Cyclotron Road, Berkeley, CA 94720, USA}
\affiliation{Department of Physics and Astronomy, University of California, 
Los Angeles, CA 90095, USA}

\author{Mikhail Stephanov}
\affiliation{Department of Physics, University of Illinois, Chicago, IL 60607, USA}

\author{Raju Venugopalan}
\affiliation{Physics Department, Brookhaven National Laboratory, Upton, NY 11973, USA}

\author{Volodymyr Vovchenko}
\affiliation{Nuclear Science Division, Lawrence Berkeley National Laboratory, 
1 Cyclotron Road, Berkeley, CA 94720, USA}

\author{Ryan Weller}
\affiliation{Center for Theoretical Physics, Massachusetts Institute of 
Technology, Cambridge, MA 02139, USA}

\author{Ho-Ung Yee}
\affiliation{Department of Physics, University of Illinois, Chicago, IL 60607, USA}

\author{Yi Yin}
\affiliation{Quark Matter Research Center, Institute of Modern Physics, 
Chinese Academy of Sciences, Lanzhou 730000, China}
\affiliation{University of Chinese Academy of Sciences, Beijing 100049, China}

\date{\today}

\begin{abstract}
The Beam Energy Scan Theory (BEST) Collaboration was formed with the goal
of providing a theoretical framework for analyzing data from the Beam Energy 
Scan (BES) program at the relativistic heavy ion collider (RHIC) at Brookhaven
National Laboratory. The physics goal of the BES program is the search for a 
conjectured QCD critical point as well as for manifestations of the chiral
magnetic effect. We describe progress that has been made over the previous
five years. This includes studies of the equation of state and equilibrium 
susceptibilities, the development of suitable initial state models, progress
in constructing a hydrodynamic framework that includes fluctuations and 
anomalous transport effects, as well as the development of freezeout 
prescriptions and hadronic transport models. Finally, we address the challenge
of integrating these components into a complete analysis framework. This 
document describes the collective effort of the BEST Collaboration and its 
collaborators around the world.
\end{abstract}

\maketitle

\tableofcontents

\section{Introduction}
\label{sec:intro}

The properties of hot and dense strongly interacting matter have been an important
focus of research for many years. Experiments at RHIC and the LHC have 
revealed several interesting and unexpected properties of the Quark Gluon Plasma 
(QGP), most prominently its near perfect fluidity \cite{Adcox:2004mh,Back:2004je,
Adams:2005dq}, see \cite{Schafer:2009dj,Heinz_2013} for reviews. The QGP
created at LHC and top RHIC energies consists in nearly equal parts of 
matter and antimatter, implying, in particular, that the baryon number chemical 
potential $\mu_B$ is much smaller than the temperature $T$ \cite{Andronic:2011yq}. 
Lattice calculations~\cite{Aoki:2006we,Bazavov:2011nk} at vanishing $\mu_B$ show 
that QCD predicts a crossover transition from the QGP to a hadron gas with many 
thermodynamic properties changing dramatically but continuously within a narrow 
range around the transition temperature, which lies in the interval $154\,
\mathrm{MeV}\leq T\leq 158.6\, \mathrm{MeV}$ \cite{Aoki:2006br,Aoki:2009sc,
Bazavov:2011nk,Bazavov:2014pvz,Bazavov:2018mes,Borsanyi:2020fev}. 
\begin{figure}[htb!]
\centering
\vspace*{0.3cm}
\includegraphics[width=0.7\textwidth]{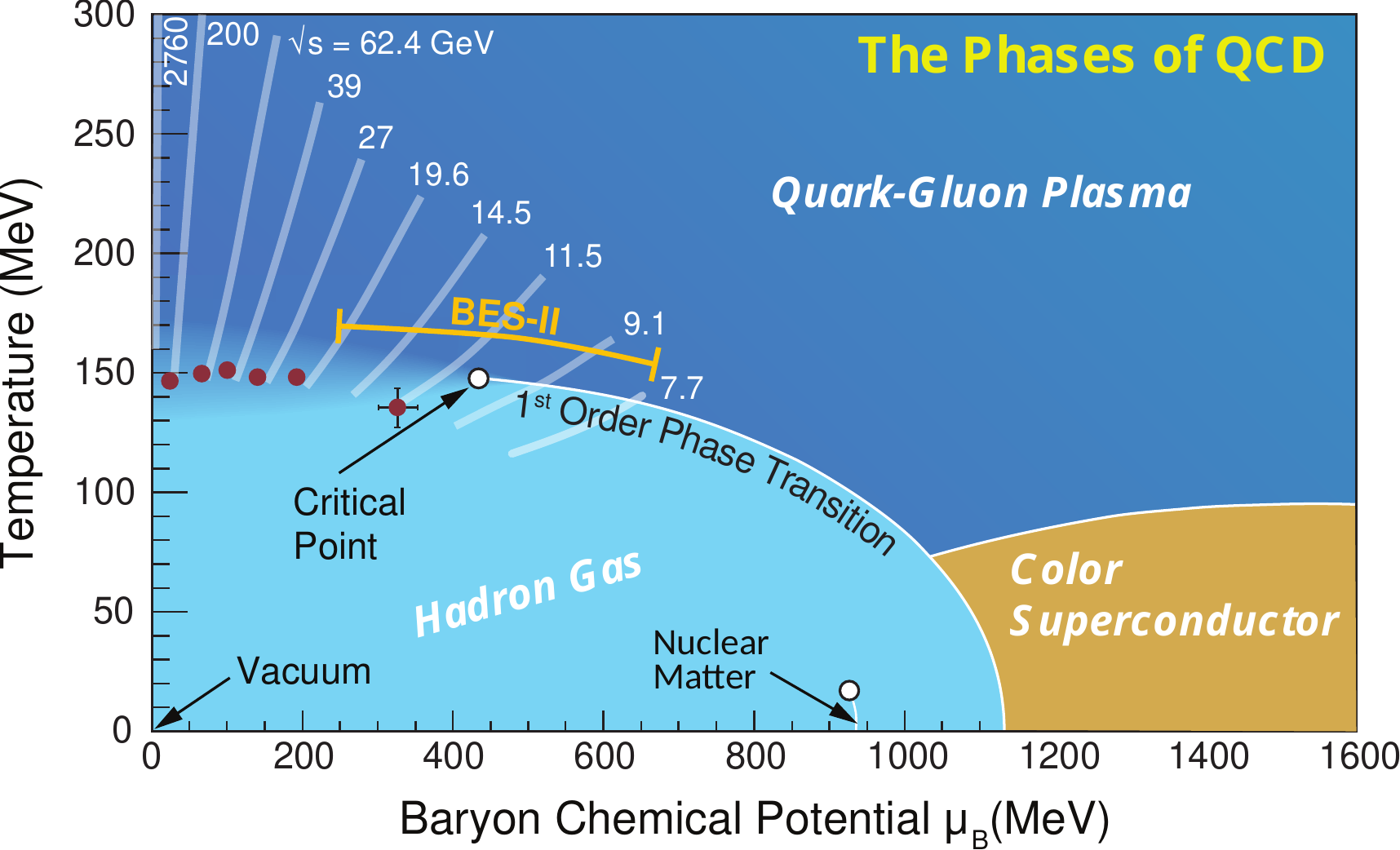}
\caption{A sketch illustrating the experimental and theoretical exploration of the
QCD phase diagram. Although the matter produced in collisions at the highest energies
and smallest baryon chemical potentials is known to change from QGP to a hadron gas
through a smooth crossover, lower energy collisions can access higher baryon chemical
potentials where a first order phase transition line is thought to exist. The reach
of the BES-II program  at RHIC is shown, as are approximate locations at which 
cooling droplets of QGP produced in collisions with given beam energy cross
the phase boundary. Note that the actual trajectory of excited matter in the 
phase diagram is more complicated, and specific examples of these trajectories
are shown in Fig.~\ref{fig:hydro:phase_diagram_BEST_EoS}.
Figure adapted from \cite{Geesaman:2015fha}. }
\label{fig:intro:BES}
\end{figure}
In contrast, a droplet of QGP at large baryon number chemical potential may
experience a sharp first order phase transition as it cools, with bubbles of 
QGP and hadronic matter coexisting at a well-defined co-existence temperature. 
If the first order regime exists, then the co-existence region must eventually 
end in a critical point. It is not yet known whether QCD has a first order 
co-existence region and an associated critical point
\cite{Stephanov:1998dy,Stephanov:1999zu,Fodor:2004nz,Allton:2005gk,Gavai:2008zr,
deForcrand:2008zi,Hatsuda:2006ps}, nor is it known where in the phase diagram it 
might lie. Many model calculations predict the existence of a critical point, but 
do not reliably constrain its location (see e.g. \cite{Stephanov:2004wx} for an 
overview). Model-independent lattice QCD calculations, on the other hand, become 
more difficult with increasing $\mu_B$ and, thus, do not yet provide definitive 
answers about the existence of a critical point. While lattice calculations have 
advanced significantly, both in terms of new techniques and advances in computing 
(see e.g. ~\cite{Fodor:2004nz,Gavai:2008zr,Datta:2012pj,Bazavov:2017dus,
Fodor:2018wul,Giordano:2020roi,Steinbrecher:2018vxm}), at present only experimental 
measurements can answer these questions definitively. 

In order to systematically survey the high baryon density region of the QCD phase 
diagram major experimental programs are under way (see e.g.  \cite{Bzdak:2019pkr} 
for an overview). In particular, the so called Beam Energy Scan (BES) at RHIC studies 
strongly interacting matter at different net-baryon densities by varying the collision
energy, as illustrated in Fig.~\ref{fig:intro:BES}. Besides a general survey of the QCD
matter, this energy or rather baryon-density scan aims at two potential discoveries that 
would have a significant impact on our understanding of the QCD phase diagram: 

\begin{itemize}
\item {\it The discovery of a QCD critical point}: If, as a function of the 
beam energy, the path of a heavy ion collision in the phase diagram changes
from passing through the first order co-existence line to traversing the 
crossover regime we expect to observe non-monotonic behavior in various observables. 
The most dramatic effects are predicted to occur in fluctuation observables, as 
discussed in \ref{sec:theory_fluct}. Additional evidence is provided by 
hydrodynamic effects on the lifetime and collective expansion of the fireball 
controlled by the softening of the equation of state near a critical point.

\item {\it The discovery of the onset of the chirally restored phase}:
In chirally restored quark gluon plasma the handedness of fermions
is conserved, but at the quantum level these conservation laws are modified
by triangle anomalies. In the presence of an external magnetic field, such 
as the one generated by the current of the colliding highly charged ions, 
these anomalies lead to novel transport effects, in particular the chiral 
magnetic effect (CME) \cite{Kharzeev:2004ey,Kharzeev:2007jp,Fukushima:2008xe} 
which predicts electric charge separation induced by an anomalous current.
\end{itemize}

Experimental results obtained during the first phase of the BES program have 
already provided interesting signals (see \cite{Bzdak:2019pkr} for a recent 
review) and, thus, have suggested that these discoveries may be possible. 
However, several improvements are needed in order to advance from a collection 
of tantalizing hints to a claim of discovery. The first issue is that the data 
collected in the exploratory phase of the the RHIC beam energy scan do not 
have sufficient statistics to claim any definitive signals for either a QCD 
critical point or for anomalous transport processes. This situation is being 
addressed during the second phase of the RHIC BES, BESII. Second, and this is 
at the heart of the effort we are reporting on here, to definitively claim or 
rule out the presence of a QCD critical point or anomalous transport requires 
a comprehensive framework for modeling the salient features of heavy ion collisions 
at BES energies which allows for a {\em quantitative} description of the data. 
A crucial aspect of this effort is the need to embed equilibrium quantities 
like the critical equation of state and anomalous conservation laws into a 
dynamical scheme. This framework correlates different observables, predicts 
the magnitude of the expected effects, includes  ``conventional'' backgrounds, 
and relates a possible discovery at a given beam energy, nuclear species and impact 
parameter to the existence of a phase boundary or a critical point at a location 
$(\mu_B,T)$ in the phase diagram. 

 This task requires advances on many theoretical frontiers, ranging from lattice 
QCD to hydrodynamics, magnetohydrodynamics, and kinetic theory, and finally to 
the tasks of model validation and data analysis. Specifically, the dynamical framework
which has been developed to successfully describe the evolution of a system created 
at top RHIC and LHC energies needs to be extended in several key aspects:

\begin{itemize}
\item {\it Initial condition:} At energies relevant for the BES the colliding nuclei 
are not sufficiently Lorentz contracted to be considered thin sheets in the longitudinal
direction. As a consequence, the transition to hydrodynamics does not happen at one 
given (proper) time but over a time interval of several fm/$c$. Therefore, as parts 
of the system already evolve hydrodynamically others are still in the pre-hydrodynamic
stage. In addition, at the lower collision energies the dominance of gluons in the initial state 
is no longer given, and quark degrees of freedom together with their conserved charges 
need to be taken into account. 

\item {\it Hydrodynamic evolution:} Viscous hydrodynamics, which has been successfully 
applied to the systems at the highest energies, needs to be amended to account for the 
propagation of the non-vanishing conserved currents of QCD: baryon number, strangeness 
and electric charge, together with their respective dissipative (diffusive) corrections. 
In addition, the description of anomalous transport requires the inclusion of anomalous 
currents together with their dissipative terms. Finally, in order to evolve (critical) 
fluctuations and correlations, one has to develop a framework that incorporates
higher moments of the hydrodynamic variables, and that takes into account 
out-of-equilibrium effects such as critical slowing down near a critical point, 
or domain formation near a first order transition. Two approaches are currently
being pursued, stochastic hydrodynamics, as well as deterministic evolution equation 
for second- and higher-order correlation functions.
    
\item {\it Equation of state:} The hydrodynamic evolution of systems created at the BES
requires an equation of state (EOS) at finite and possibly large net baryon number chemical
potential with a potential phase transition and critical point, the location of which is 
still unknown.
Additional dependencies on net strangeness and electric charge densities are essential to 
reproduce the hadronic chemistry at different collision energies.
Therefore, one has to develop a model equation of state which includes 
a phase transition and which at the same time has a solid footing in QCD. To this end it 
is important to calculate in Lattice QCD higher order coefficients for the Taylor expansion
of the pressure in terms of the baryon number chemical potential. These will then serve 
QCD constraints for any equation of state with a phase transition and critical point. 
    
\item {\it Particlization:} The transition from hydrodynamic fields into particles, often 
referred to as particlization, which typically is implemented at the phase boundary, needs to 
ensure that fluctuations and correlations are preserved and do not receive additional 
(spurious) contributions \cite{Steinheimer:2017dpb}.
    
\item {\it Hadronic phase:} The relative time the system spends in the hadronic phase
increases with decreasing collision energy. Therefore, the kinetic out-of-equilibrium evolution of the 
hadronic phase requires special attention. In addition, one needs to allow for (mean 
field) interactions in order to match a possible phase transition and critical point in 
the hadronic phase and evolve the system in the presence of these interactions.
    
\item {\it Data analysis:} A Bayesian global analysis, similar to that already successfully
applied to the highest energy collisions \cite{Bernhard:2019bmu, JETSCAPE:2020shq, JETSCAPE:2020mzn, Nijs:2020ors, Nijs:2020roc}, is required to constrain and extract the physical parameters of the model,
such as transport coefficients and the location of critical point etc. Since at lower energies 
we encounter additional relevant dynamical variables, such as diffusion coefficients, critical point, 
mean fields etc, the presently available Bayesian analysis frameworks need to be extended 
considerably. 
\end{itemize}

It is the purpose of this paper to report on the progress made by the Beam Energy Scan
Theory (BEST) Collaboration towards developing a dynamical framework which takes into 
account these essential new aspects. We start with an overview of the recent, pertinent,  
results from lattice QCD. Next we briefly review the relevant theoretical concepts with 
regards to critical fluctuations and anomalous transport. Before we turn to the various 
new developments concerning the initial state and hydrodynamics we discuss the
modeling of the equation of state with a critical point. After discussing several new 
methods for particlization and the kinetic treatment of the hadronic phase we finally 
present the Bayesian data analysis framework which will be applied in order to constrain
the physical model parameters with experimental data.

\section{Lattice QCD results}
\subsection{Phase diagram}
\label{sec:PD}

\begin{figure}[ht!]
\centering
\includegraphics[width=0.49\linewidth,height=0.4\textwidth]{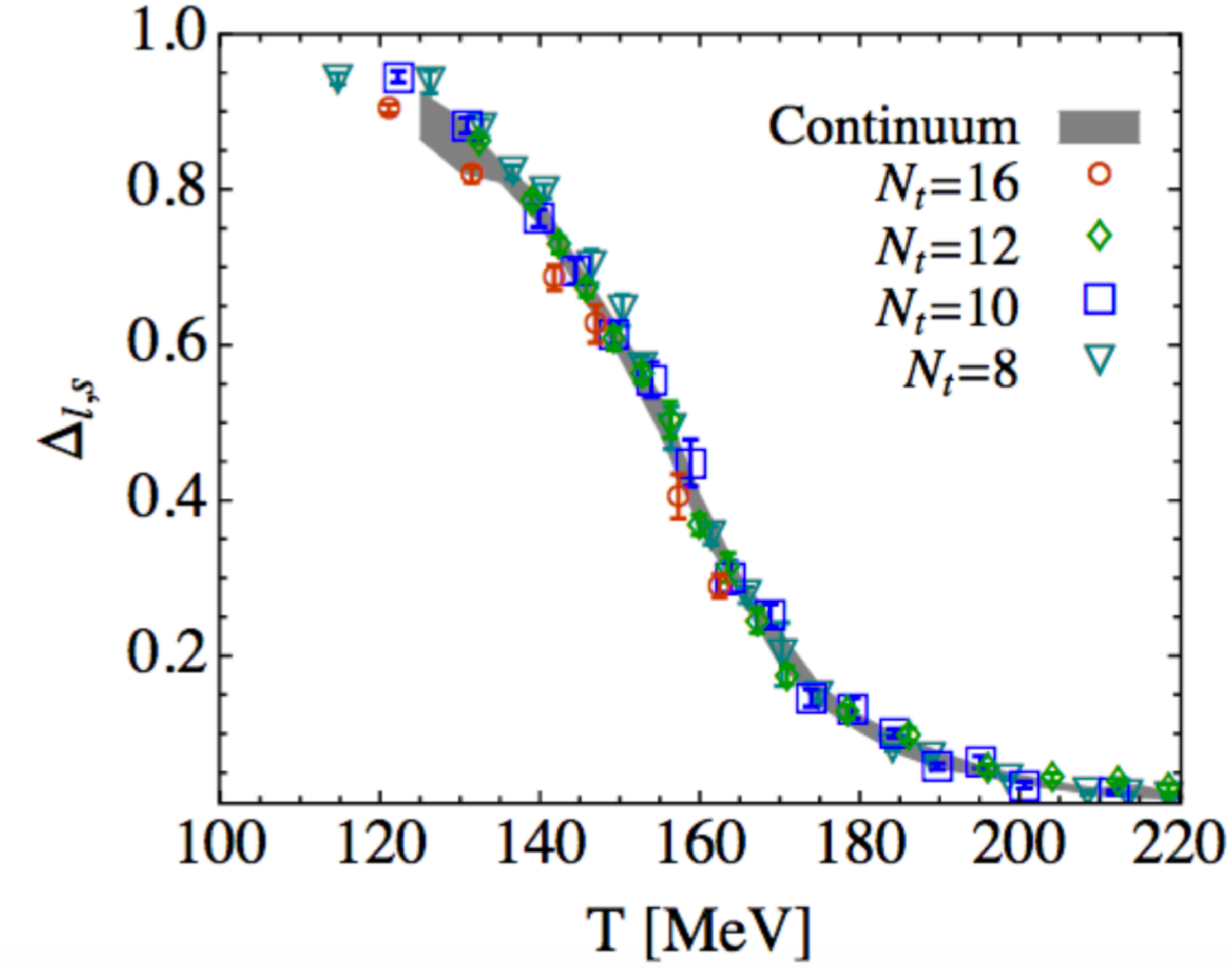}
\includegraphics[width=0.49\textwidth,height=0.4\textwidth]{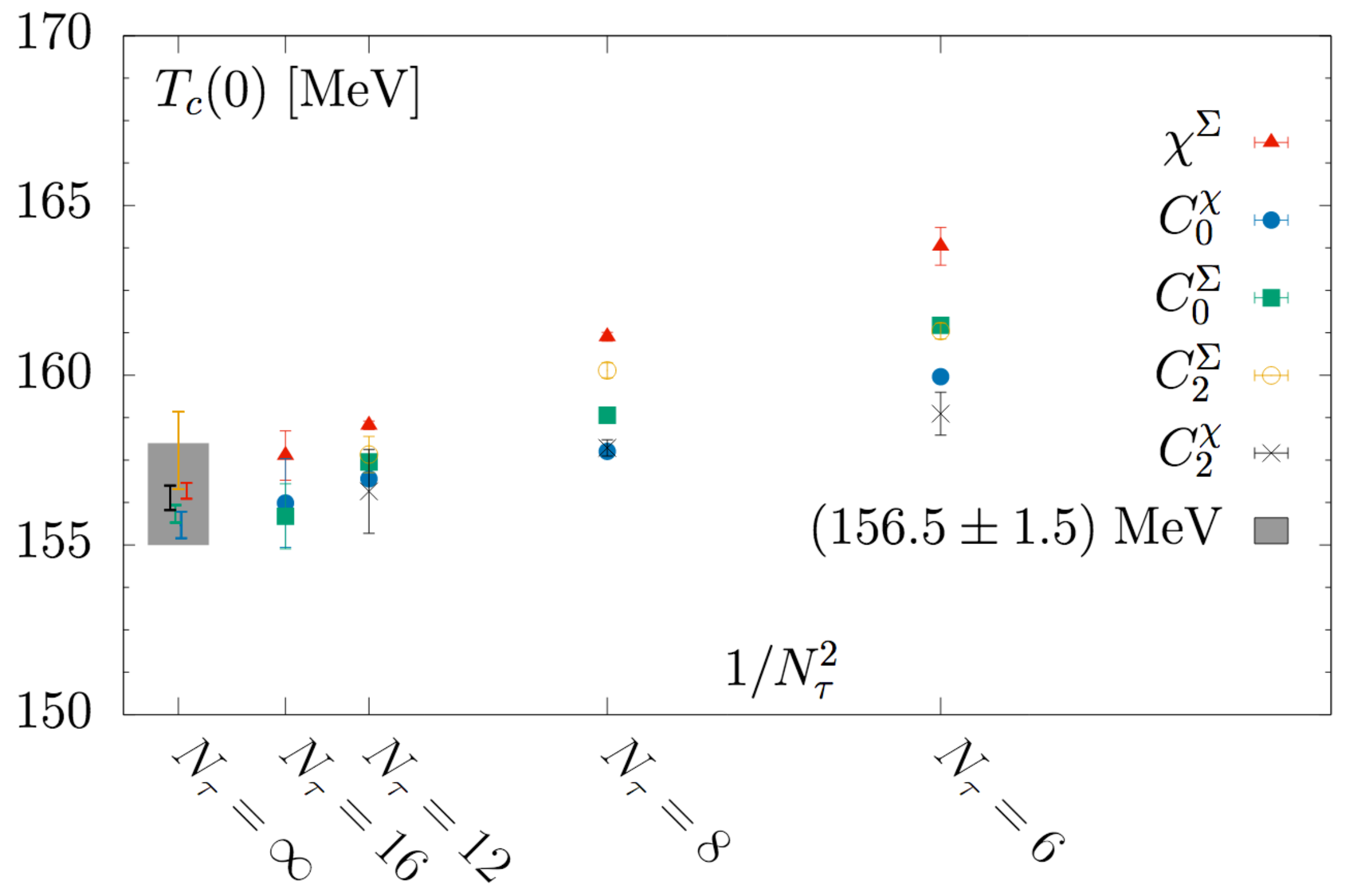}
\caption{
Left: from Ref. \cite{Borsanyi:2010bp}. Continuum extrapolated result for the subtracted 
chiral condensate as a function of the temperature. Right: from Ref.~\cite{Bazavov:2018mes}. 
Continuum extrapolations of pseudo-critical temperatures $T_c(0)$ using five different 
chiral observables defined in Ref.~\cite{Bazavov:2018mes}.}
\label{chiralc}
\end{figure}

 Numerical simulations have demonstrated that at zero baryon chemical potential the 
QCD phase transition between hadronic matter at low temperature and a QGP at high 
temperature is a smooth crossover \cite{Aoki:2006we}. The QCD Lagrangian is symmetric 
under chiral transformations of the fermion fields in the case of massless quarks. 
However, chiral symmetry is spontaneously broken by the QCD vacuum, and the chiral 
condensate
\begin{equation}
    \langle\bar{\psi}\psi\rangle_f=\frac{T}{V}\frac{\partial\ln Z}{\partial m_f}
\end{equation}
has a non-zero expectation value at low temperatures. Here, $Z$ is the QCD partition
function, $V$ is the volume, and $m_f$ is the mass of a quark with flavor 
$f=\{ {\it up,\,down},\ldots\}$. As quarks deconfine and the transition to the 
QGP takes place, chiral symmetry is restored. This is evident from the fact that 
the chiral condensate features a rapid decrease in the vicinity of the transition 
temperature, and approaches zero at high temperatures, see the left panel of 
Fig.~\ref{chiralc}. 

Because of the crossover nature of the transition, 
a definition of the transition temperature is ambiguous. A common choice is to locate 
the peak of the chiral susceptibility $\chi_l = \partial\langle\bar{\psi}\psi\rangle_l /
\partial m_l$ as a function of the temperature. By extrapolating this observable to 
finite chemical potential it is possible to follow the location of the transition 
temperature with increasing $\mu_B$:
\begin{equation}
    \frac{T_c(\mu_B)}{T_c(0)}=1 
    - \kappa_2\left(\frac{\mu_B}{T_c(\mu_B)}\right)^2
    + \kappa_4\left(\frac{\mu_B}{T_c(\mu_B)}\right)^4.
\end{equation}
The right panel of Fig.~\ref{chiralc} shows the pseudo-critical temperature at 
$\mu_B=0$, extrapolated to the continuum using five different chiral observables to
define its location. The state of the art results for the transition temperature at
$\mu_B=0$ ($T_0=156.5\pm1.5$ MeV \cite{Bazavov:2018mes} and $T_0=158.0\pm0.6$ MeV
\cite{Borsanyi:2020fev}), the curvature of the phase diagram ($\kappa_2=0.012(4)$
\cite{Bazavov:2018mes} and $\kappa_2=0.0153(18)$ \cite{Borsanyi:2020fev}) and the
fourth-order correction ($\kappa_4=0.000(4)$ \cite{Bazavov:2018mes} and
$\kappa_4=0.00032(67)$ \cite{Borsanyi:2020fev}) have all been obtained within BEST by
members of the HotQCD and WB collaborations, respectively (for previous results see 
Ref.~\cite{Bhattacharya:2014ara,Bonati:2014rfa,Cea:2015cya,Bellwied:2015rza,Bonati:2015bha,Bonati:2018wdn,Bonati:2018nut}, for a determination of the QCD transition temperature 
in the chiral limit see Ref. \cite{Ding:2019prx}). It is worth pointing out that the
curvature is very small and that the fourth-order correction is compatible with zero;
besides, the results of the two collaborations, obtained with different lattice actions,
agree with each other within uncertainties.

\begin{figure}[ht!]
\vspace*{0.2cm}
\centering
\includegraphics[width=0.49\linewidth,height=0.4\textwidth]{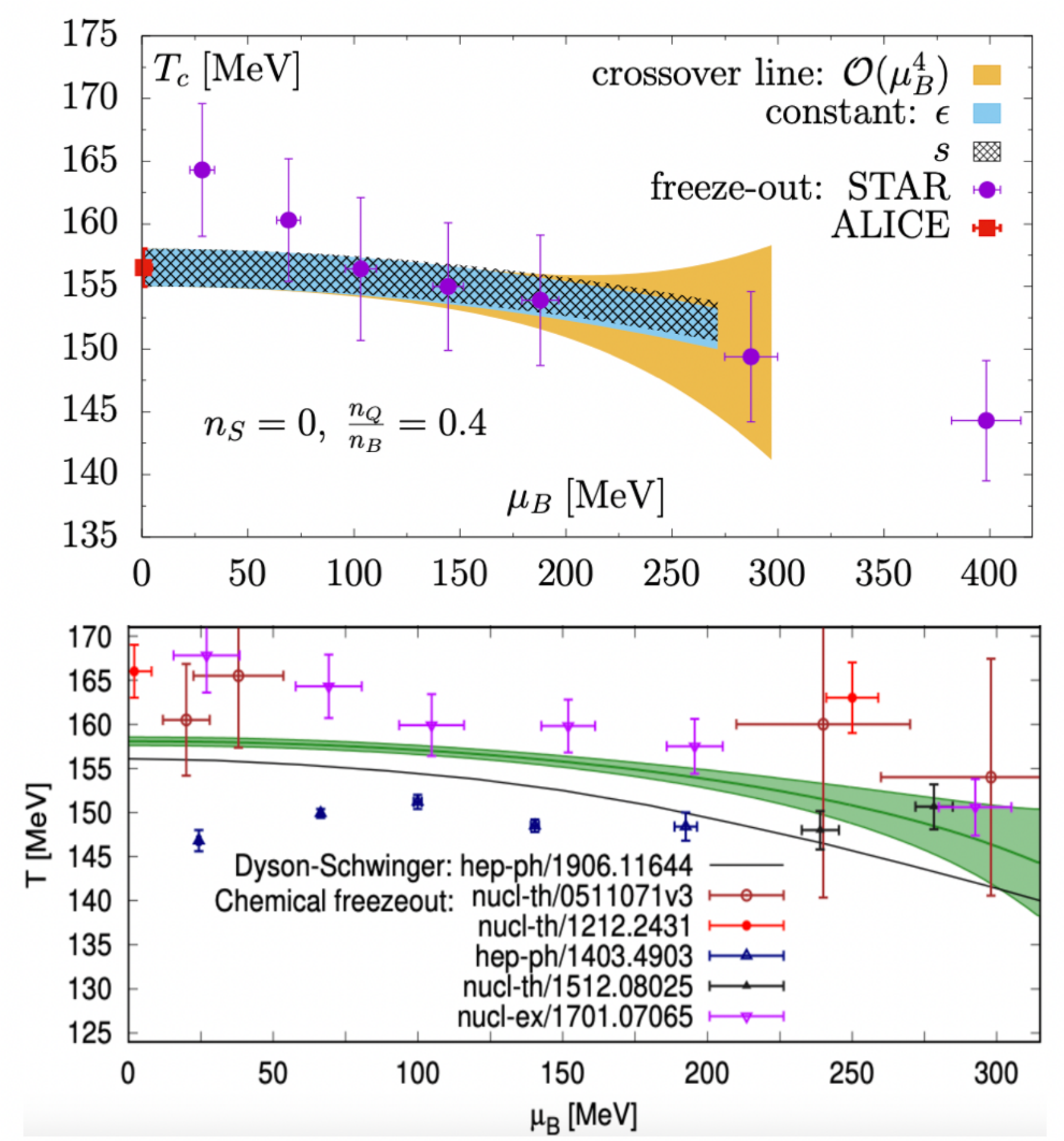}
\includegraphics[width=0.49\textwidth,height=0.4\textwidth]{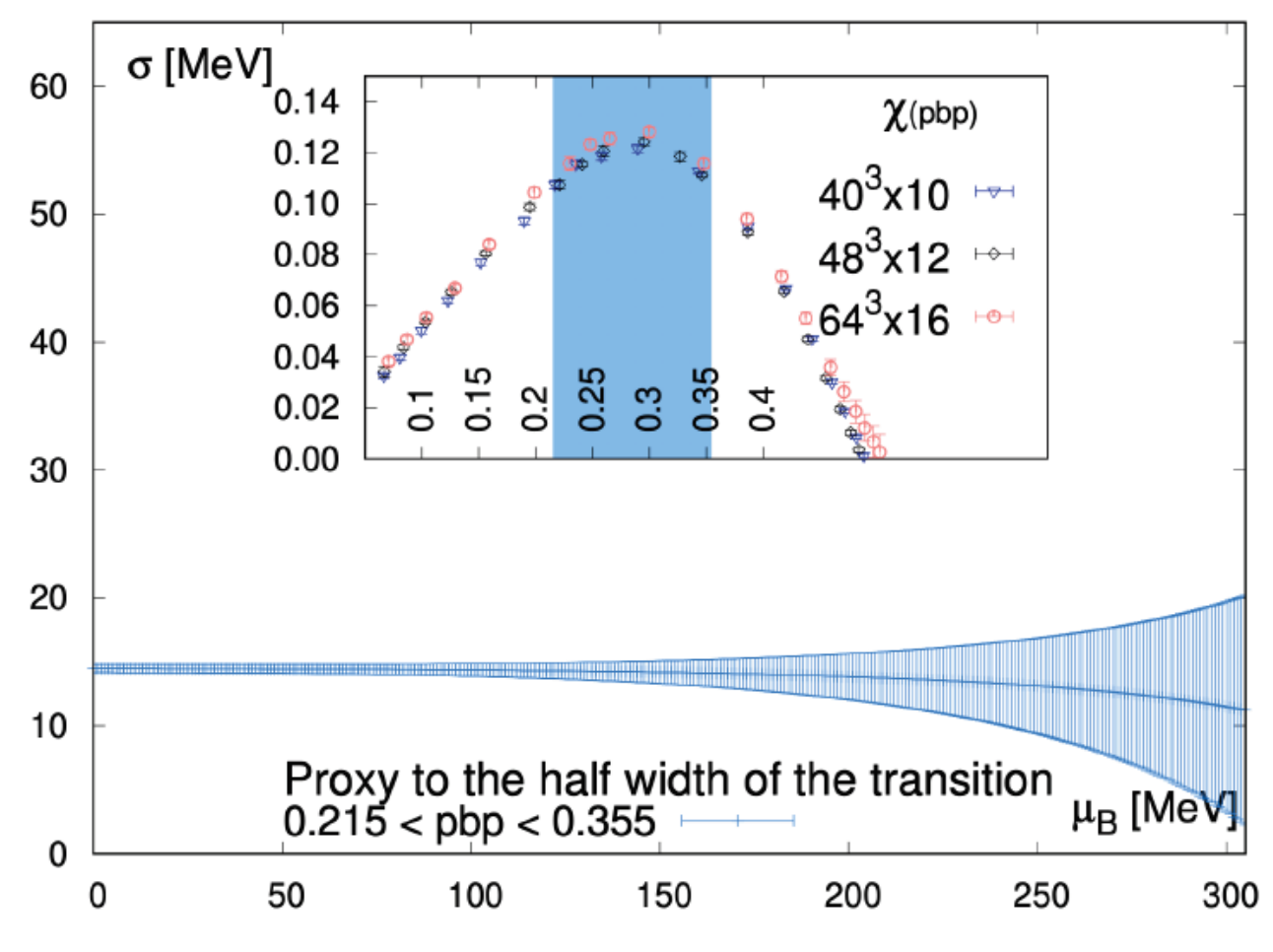}
\caption{Left: QCD transition line from Refs. \cite{Bazavov:2018mes} (top) and 
\cite{Borsanyi:2020fev} (bottom). Both panels show comparisons between the lattice 
QCD bands and freeze-out points from the literature. Right (from 
Ref.~\cite{Borsanyi:2020fev}): Width $\sigma$ of the peak of the chiral susceptibility 
as a function of the chemical potential. The inset shows the susceptibility $\chi$
as a function of $\langle\bar\psi\psi\rangle$.}
\label{phase}
\end{figure}

The left panel of Fig.~\ref{phase} shows the transition line obtained in 
Refs.~\cite{Bazavov:2018mes} (top) and \cite{Borsanyi:2020fev} (bottom). 
By calculating the second-order baryon number fluctuation along the transition line
\cite{Steinbrecher:2018vxm} or by looking at the height and width of the peak of the 
chiral susceptibility \cite{Borsanyi:2020fev} it was concluded that no sign of 
criticality is observed in lattice QCD simulations at $\mu_B<300$ MeV. This is 
evident from the right panel of Fig.~\ref{phase}, which shows the width of the chiral
susceptibility peak as a function of the chemical potential: while a decrease
is expected in the vicinity of the critical point, the curve is compatible with a
constant.

\subsection{Equation of state at finite density}

\begin{figure}[ht!]
\centering
\includegraphics[width=0.55\linewidth]{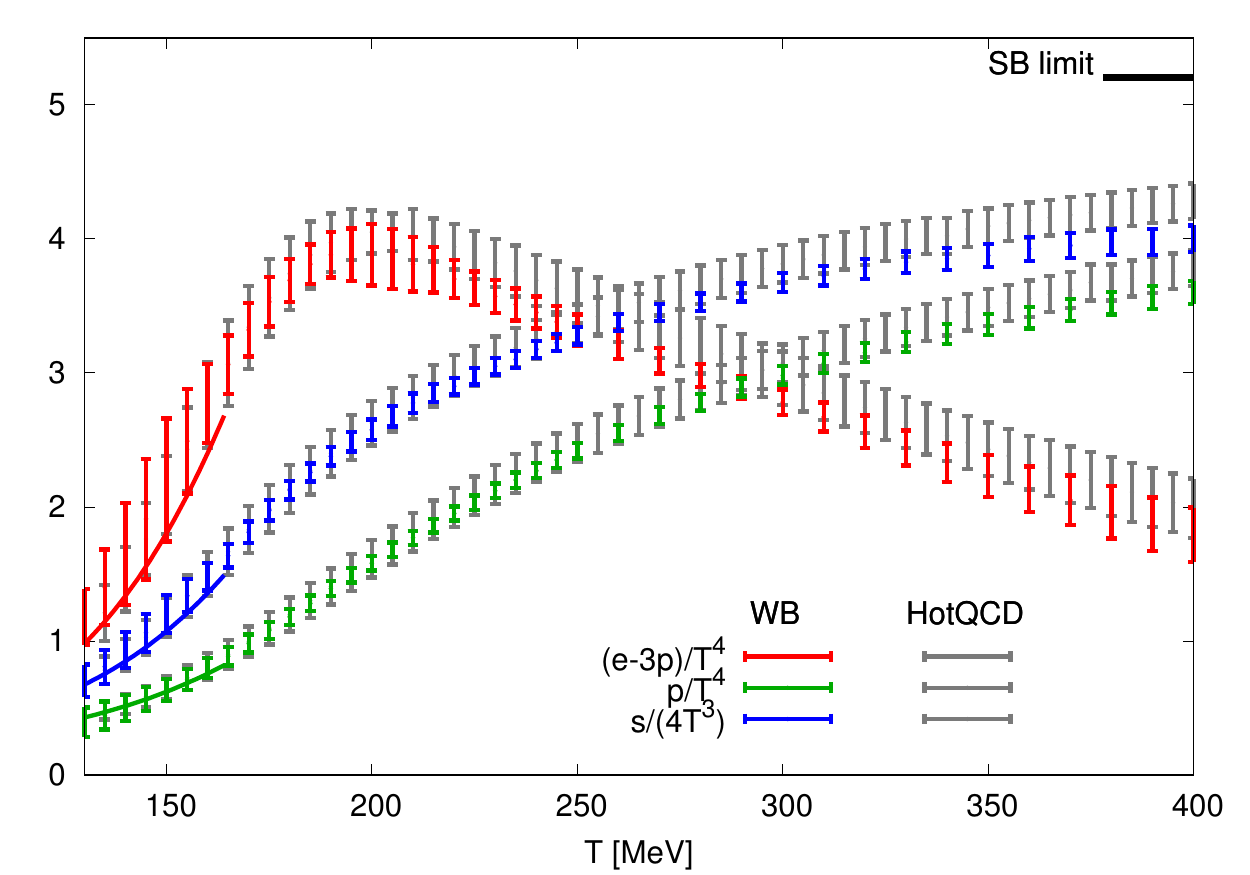}
\caption{From Ref. \cite{Ratti:2018ksb}. Continuum extrapolated results for trace anomaly, 
entropy density and pressure as functions of the temperature at $\mu_B=0$. The gray points 
are the results from the HotQCD Collaboration \cite{Bazavov:2014pvz}, while the colored ones 
are from WB \cite{Borsanyi:2013bia}. The solid lines at low temperature show the 
result of a hadron resonance gas model, and the line at high temperature indicates
the Stefan-Boltzmann limit.}
\label{EoS0}
\end{figure}

The equation of state of QCD at $\mu_B=0$ has been known from first principles for a
number of years. The WB Collaboration published continuum extrapolated results for 
pressure, energy density, entropy density, speed of sound and interaction measure
in Refs.~\cite{Borsanyi:2010cj,Borsanyi:2013bia}. These results were confirmed by 
the HotQCD Collaboration in Ref.~\cite{Bazavov:2014pvz}. A comparison between these
results is shown in Fig.~\ref{EoS0}.

Lattice QCD simulations at finite chemical potential are hindered by the well-known 
sign problem, which limits the range of available results for the thermodynamics of
strongly interacting matter. The equation of state of QCD at finite density is obtained
either as a Taylor series in powers of $\mu_B/T$ around $\mu_B=0$, or through simulations
at imaginary chemical potential and their analytical continuation to real $\mu_B$ 
\cite{[{See, for example, }][{, and references therein.}]Muroya:2003qs}. The Taylor 
expansion of the pressure can be written as
\begin{equation}
\frac{p(T,\mu_B)}{T^4}=\frac{p(T,0)}{T^4} 
+\sum_{n=1}^\infty  \left.\frac{1}{n!}
  \frac{\mathrm{\partial^{n}}(p/T^4)}
       {\partial(\frac{\mu_B}{T})^{n}}\right|_{\mu_B=0}
       \left(\frac{\mu_B}{T}\right)^{n}
       =\sum_{n=0}^{\infty}c_{n}(T)\left(\frac{\mu_B}{T}\right)^{n},
\end{equation}
where the $c_n$ coefficients are defined as
\begin{eqnarray}
c_n=\frac{1}{n!}\frac{\partial^n(p/T^4)}{\partial(\mu_B/T)^n}
\end{eqnarray}
and they are related to the susceptibilities of conserved charges as
\begin{equation}
    c_n^i=\frac{1}{n!}\chi_n^i,~~~\mathrm{where}~~~
    \chi_n^i=\frac{\partial^n(p/T^4)}{\partial(\mu_i/T)^n}
    \quad \mathrm{and} \quad i = B, Q, S.
\end{equation}
At temperatures of $T\sim100-160$ MeV it is possible to make direct comparisons between
a hadron resonance gas model and Lattice QCD, which show a good agreement for most
observables, when using enhanced particle lists which include either barely seen
\cite{Alba:2017mqu,Alba:2017hhe,Alba:2020jir} or predicted but not yet observed 
\cite{Bazavov:2014xya} resonant states as input. 

One has to keep in mind that there are three conserved charges in QCD: baryon number $B$,
electric charge $Q$ and strangeness $S$. When extrapolating to finite baryonic chemical
potential, a choice needs to be made also for $\mu_S$ and $\mu_Q$. Two common choices in
the literature are either $\mu_S=\mu_Q=0$, or $\mu_S$ and $\mu_Q$ as functions of $\mu_B$
and $T$ such that the following conditions are satisfied: $\langle n_S \rangle=0$,
$\langle n_Q \rangle \sim 0.4\langle n_B \rangle$, where $n_i$ is the density of conserved
charge $i$. The latter reflects the initial conditions in a heavy-ion collision, namely
the proton to neutron ratio in heavy nuclei such as Au and Pb and the absence of
net-strangeness in the colliding nuclei. After the early results for $c_2,~c_4$ 
and $c_6$ \cite{Allton:2005gk}, the first continuum extrapolated  $c_2$ results were
published in Ref.~\cite{Borsanyi:2012cr}; results for $c_4$ were shown in 
Ref.~\cite{Hegde:2014sta}, but only for a finite lattice spacing. The BEST Collaboration
obtained the state of the art lattice QCD Equation of State at finite density in three
distinct cases:

\begin{enumerate}
\item{Continuum extrapolated results for the Taylor expansion coefficients of the
pressure up to $\mathcal{O}((\mu_B/T)^6)$ in the case of $\langle n_s\rangle=0$ and
$\langle n_Q\rangle=0.4\langle n_B \rangle$ \cite{Gunther:2016vcp,Gunther:2017sxn}.}
    
\item{Continuum extrapolated results for the Taylor expansion coefficients of the
pressure up to $\mathcal{O}((\mu_B/T)^4)$ \cite{Bellwied:2015lba}, a continuum
estimate of the sixth-order coefficient $\chi_6^B$ \cite{Bazavov:2017dus} and 
results for $\chi_8^B$ at $N_t=8$ \cite{Bazavov:2020bjn} and $N_t=12$ \cite{Borsanyi:2018grb} at 
$\mu_S=\mu_Q=0$.}
    
\item{Reconstructed equation of state, up to fourth-order in chemical potential 
\cite{Noronha-Hostler:2019ayj}, or including some sixth-order terms \cite{Monnai:2019hkn} at finite 
$\mu_B,~\mu_S$ and $\mu_Q$.}
\end{enumerate}

\begin{figure}[htb!]
\vspace*{0.2cm}
\centering
\includegraphics[width=1.00\linewidth]{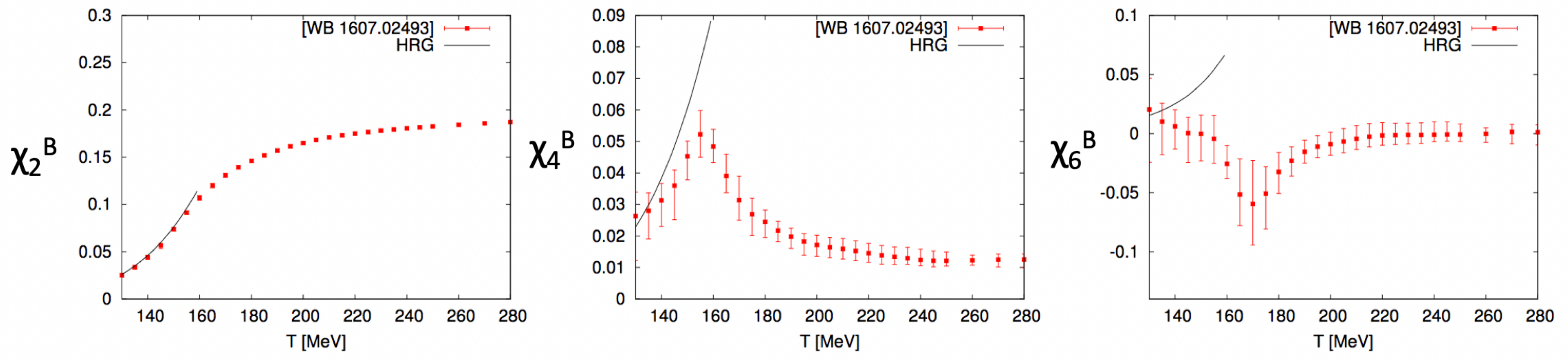}
\hspace*{0.05\hsize}\includegraphics[width=0.91\linewidth]{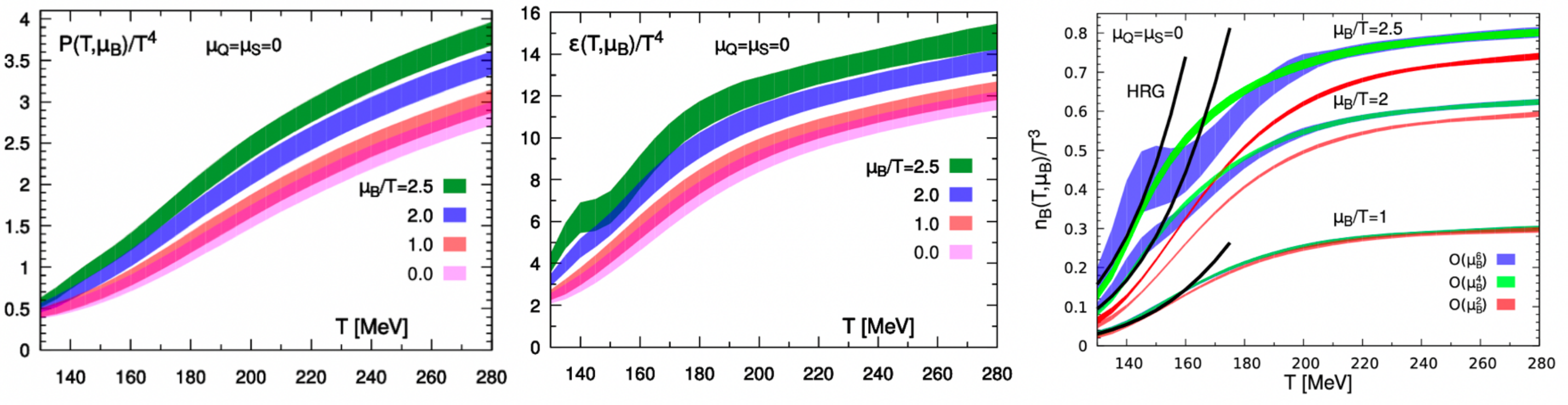}
\caption{Top: from Ref. \cite{Gunther:2016vcp}. Continuum extrapolated second-, fourth- and 
sixth-order Taylor expansion coefficients in the strangeness neutral case. 
Bottom: from Ref. \cite{Bazavov:2017dus}. Pressure (left), energy density (center) and baryonic 
density (right) as functions of the temperature, for several values of $\mu_B/T$.}
\label{EoS_chis}
\end{figure}

Figure \ref{EoS_chis} shows the Taylor coefficients for case 1 (upper panels, from 
Ref.~\cite{Gunther:2016vcp}) and some thermodynamic quantities for case 2 (lower panels,
from Ref.~\cite{Bazavov:2017dus}), respectively. More recently, a novel extrapolation
method has been proposed in Ref.~\cite{Borsanyi:2021sxv}, which considerably extends 
the range in $\mu_B$ and eliminates the wiggles around the transition temperature,
typical of the Taylor expansion method (see the blue bands in the bottom right panel of Fig.~\ref{EoS_chis}).

\subsection{Correlations and fluctuations of conserved charges}

Fluctuations of conserved charges are one of the most promising measurements from 
the Beam Energy Scan program, as they are sensitive to the presence of a critical point
\cite{Gavai:2008zr,Stephanov:1999zu,Cheng:2007jq} and allow a comparison between first
principle results and experiments \cite{Karsch:2012wm,Bazavov:2012vg,Borsanyi:2013hza}.
Results for fluctuations at small chemical potentials and their comparison to
data have been obtained in the past \cite{Borsanyi:2011sw,Bazavov:2012jq,DElia:2016jqh,Gavai:2005yk}.
Within the BEST Collaboration, several new results for equilibrium fluctuations of conserved charges
have been obtained on the lattice, and comparisons to the Hadron Resonance Gas model, 
to perturbation theory \cite{Haque:2014rua,Mogliacci:2013mca} and to experimental 
results have been discussed.

\begin{figure}[ht!]
\centering
\includegraphics[width=0.49\linewidth,height=0.30\linewidth]{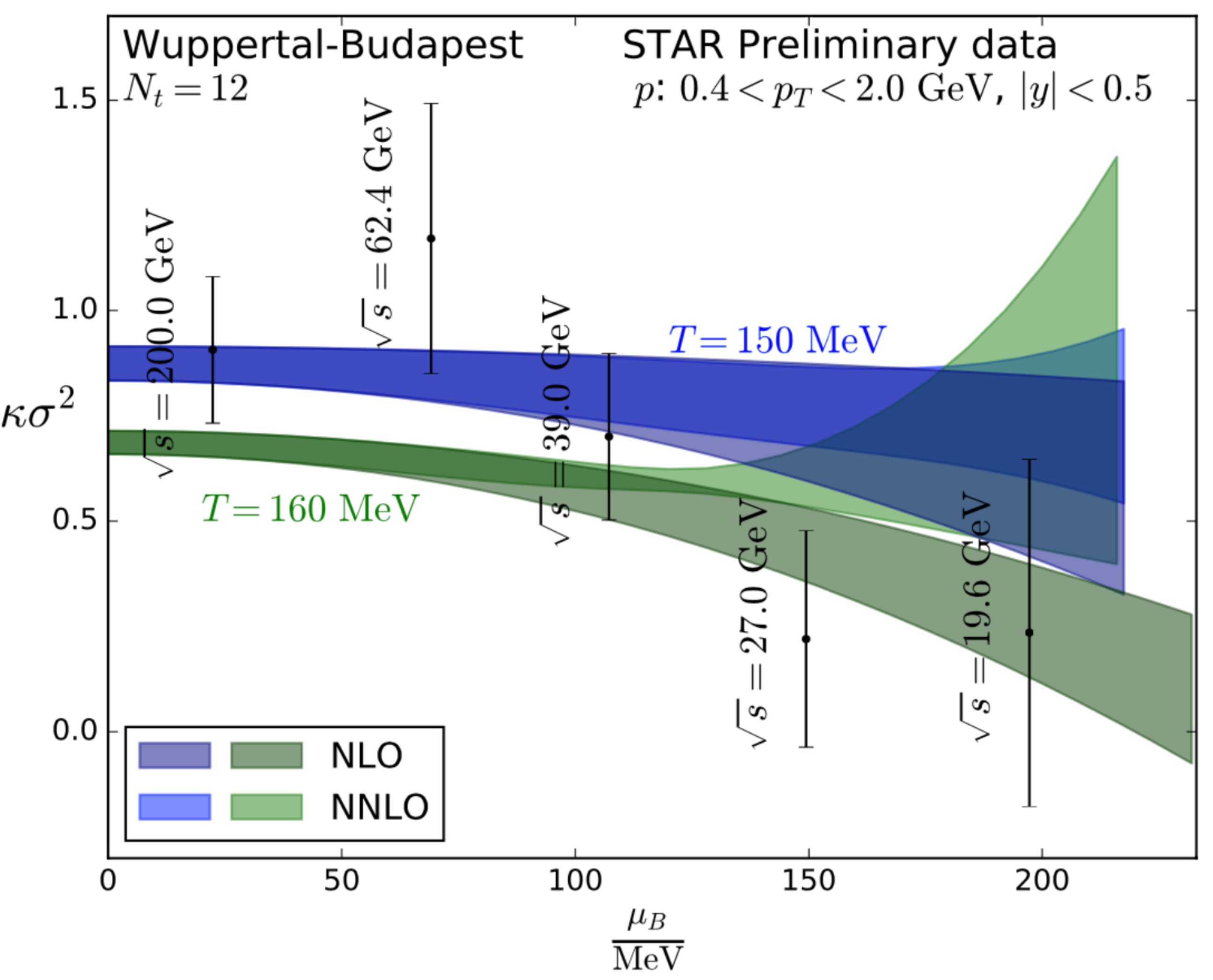}
\includegraphics[width=0.49\linewidth,height=0.32\linewidth]{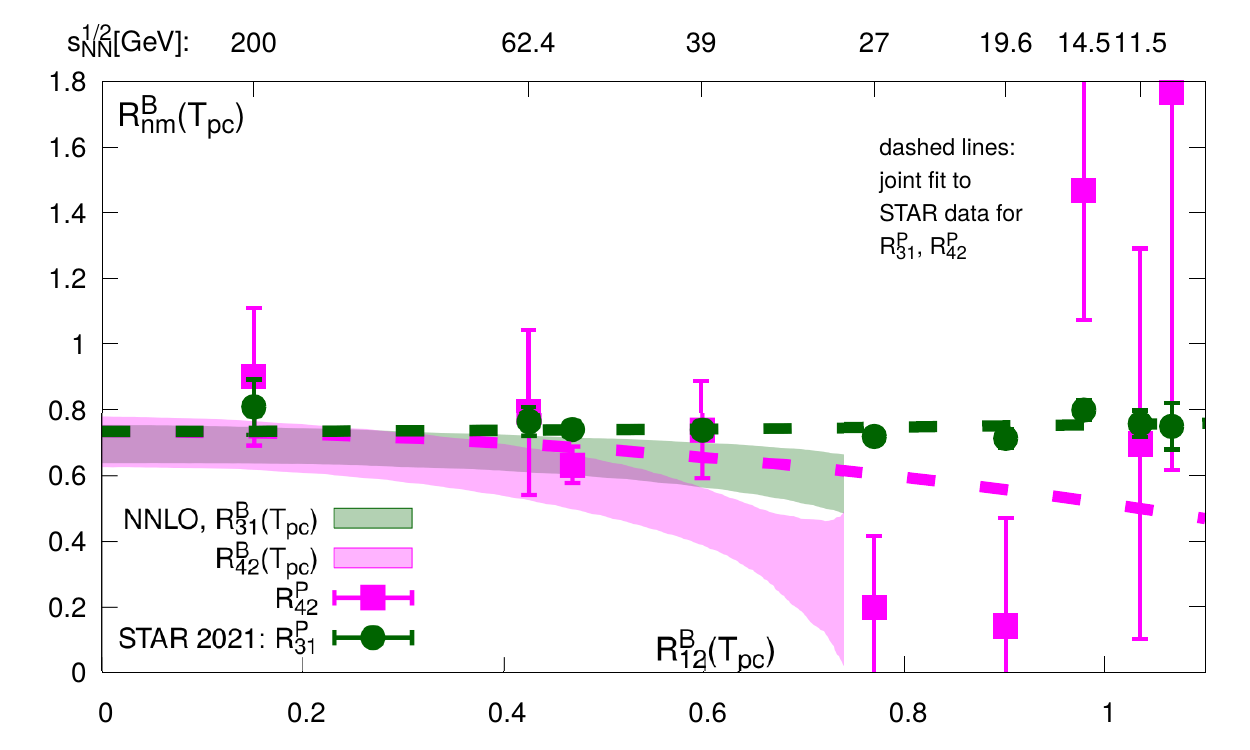}
\caption{
Left: from Ref.~\cite{Borsanyi:2018grb}. Ratio of fourth- to second-order 
baryon number fluctuations, extrapolated to finite chemical potential $\mu_B$, 
for different values of the temperature at $\mu_B=0$. The points are experimental 
results from the STAR Collaboration \cite{STAR:2020tga}. 
Right: from Ref.~\cite{Bazavov:2020bjn}. Ratio of third- to first- (green) 
and of fourth- to second-order (magenta) baryon number fluctuations as functions of 
the ratio of first- to second-order baryon number fluctuations (or collision 
energy, as indicated on top of the figure), compared to STAR results
\cite{STAR:2020tga}.}
\label{fig:kappak}
\end{figure}

The high-temperature behavior of fluctuations and correlations between different 
flavors was explored in Refs. \cite{Bellwied:2015lba,Ding:2015fca}. In 
Ref.~\cite{Borsanyi:2018grb}, several higher order diagonal and off-diagonal 
correlators between baryon number, electric charge and strangeness were explored. 
The higher order coefficients were used to expand the lower order ones to finite 
chemical potential and compare them to experiment. Results for the ratio of the  
fourth- to second-order baryon number fluctuations, $\chi_4^B/\chi_2^B$ from 
Ref.~\cite{Borsanyi:2018grb} are shown in the left panel of Fig.~\ref{fig:kappak}, 
in comparison to experimental results from the STAR Collaboration \cite{STAR:2020tga}. 
HADES has similar measurements at $\sqrt{s_\mathrm{NN}} = 2.4$ GeV \cite{HADES:2020wpc}.
For a determination of the curvature of the chemical freeze-out line, see 
Ref.~\cite{Bazavov:2015zja}.

The right panel of Fig.~\ref{fig:kappak} shows similar results from 
Ref.~\cite{Bazavov:2020bjn}. In particular, it was pointed out in that manuscript 
that the observed decrease in the experimental values with decreasing collision 
energy can be reproduced from first principles. One should however keep in mind 
that lattice QCD results correspond to the thermodynamic equilibrium fluctuations 
of the baryon number in the grand-canonical ensemble limit, while the experimental data show fluctuations of net-protons.
The relation between the two, in thermal equilibrium , has been explained in Ref.~\cite{Kitazawa:2012at}.
A quantitative analysis of the difference between protons and baryons at RHIC-BES was presented in \cite{Vovchenko:2021kxx}

More recently, triggered by forthcoming experimental measurements from the STAR collaboration, 
results for baryon number fluctuations up to sixth-order at small values of $\mu_B$ have 
been obtained in Ref.~\cite{Bazavov:2020bjn}. The same motivation was behind new results 
for $B$, $Q$, $S$ correlators at finite chemical potential, and the definition of their 
proxies to be compared to experimental results in Ref.~\cite{Bellwied:2019pxh}. For recent 
reviews of the state of the art of first principle simulations in comparison to experimental 
results see Refs.~\cite{Ding:2015ona,Ratti:2018ksb}.

\section{Critical fluctuations and anomalous transport}
\label{sec:theory}

\subsection{Fluctuation dynamics near the critical point}
\label{sec:theory_fluct}

\subsubsection{Introduction}

\begin{figure*}[!hbt]
\begin{center} \vspace{-0.1in}
\includegraphics[width=0.45\textwidth]{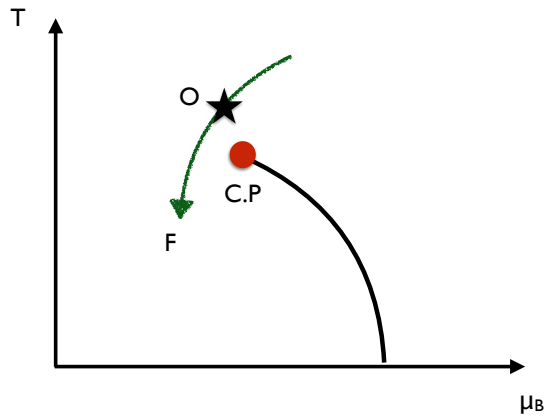} 
\caption{
Schematic trajectory of a fireball created in a heavy ion collision passing close 
to the conjectured QCD critical point (red circle). The point ``O'' denotes the 
point where, because of critical slowing down, the critical fluctuations fall out 
of equilibrium. The point ``F'' illustrates the location where the freeze-out 
happens.} 
\label{fig:CP}
\end{center}
\vspace{-0.1in}
\end{figure*} 

To turn high precision experimental data anticipated from BESII into definitive
information about the existence of a  QCD critical point a quantitative framework 
for modeling the salient features of these low energy collisions is indispensable. 
To this end, viscous hydrodynamics, which successfully describes the evolution of 
the fireballs created at top RHIC and LHC energies needs to be extended to be 
suitable for the conditions at lower energies. For example, as discussed in detail 
in Sec.~\ref{sec:hydro} the conserved currents of QCD need to be propagated 
explicitly. In addition, an equation of state (EOS) with a critical point in
the universality class of a possible QCD critical endpoint is needed. The 
QCD critical point is expected to be in the 3d Ising class \cite{Halasz_1998,
Berges:1998rc}. As we shall see in Sec.~\ref{sec:eos} such an EOS has been 
constructed in Ref.~\cite{Parotto:2018pwx}. 

However, as the fireball approaches the critical point, hydrodynamics is not 
sufficient to capture all the relevant dynamics.  In particular, the evolution of 
the long wavelength fluctuations (LWF) of the order parameter field close to the 
critical point  is beyond a hydrodynamic description. Due to critical slowing down, 
LWF inescapably fall out of equilibrium as the system approaches the critical point, 
see Fig.~\ref{fig:CP} for an illustration. An analysis of the resulting
out-of-equilibrium effect~\cite{Mukherjee:2015swa} (see Ref.~\cite{Yin:2018ejt} for 
more references) indicates  that in many phenomenological relevant situations the
real-time fluctuations differ from equilibrium expectations not just quantitatively, 
but even qualitatively. In addition, since the order parameter fluctuations themselves
contribute to the stress-energy tensor $T^{\mu\nu}$, their out-of-equilibrium dynamics  
back-react on the bulk evolution of the fireball. Therefore, a quantitative framework
that describes the intertwined dynamics among the fluctuations near the phase boundary 
and bulk evolution is crucial, see \cite{Bluhm:2020mpc} for a recent review. We now discuss the basic features of such a
framework, and the present status of their implementation will be presented in 
Sec.~\ref{sec:hydro_flcut}. 

Before turning to the quantitative framework, let us first explain qualitative 
features of critical dynamics. A key concept is Kibble-Zurek (KZ) dynamics (see
Ref.~\cite{Zurek:1996sj} for a review). Since the evolution of the critical 
fluctuations becomes effectively frozen at the point where the time remaining 
to reach the critical point is shorter than the relaxation time (the point "O" 
in Fig.~\ref{fig:CP}), one can use the frozen correlation length, known as the 
KZ length,  $l_{{\rm KZ}}$, and the aforementioned timescale at which critical
fluctuations become frozen, $\tau_{{\rm KZ}}$, to characterize the qualitative
features of out-of-equilibrium evolution near the critical point. The KZ timescale
$\tau_{{\rm KZ}}$ also determines the time interval during which out-of-equilibrium
effects are important. According to the benchmark estimate presented in
Ref.~\cite{Akamatsu:2018vjr}, $\tau_{{\rm KZ}}$ for a heavy-ion collision is 
around $6$~fm. Out-of-equilibrium scaling leads to a potentially unique signature 
of critical behavior, and Kibble-Zurek scaling for non-Gaussian cumulants has 
been studied in model calculations \cite{Mukherjee:2016kyu}.

The study of out-of-equilibrium fluctuations has already attracted much attention, 
prior to the works discussed here. The limitation of the growth of the critical 
correlation length due to finite time effects was originally studied in
Ref.~\cite{Berdnikov:1999ph}, and the dynamic universality class (model H) of a 
critical point in the QCD phase diagram was identified in \cite{Son:2004iv}.
A number of authors investigated the theory of fluctuations in relativistic fluid
dynamics \cite{Kovtun:2003vj,Kovtun:2011np,PeraltaRamos:2011es,Kapusta:2011gt}.
out-of-equilibrium effects on non-Gaussian cumulants were investigated in
Ref.~\cite{Mukherjee:2015swa} based on a set of cumulant equations. 
In parallel, the model of chiral fluid dynamics (CFD) was developed
\cite{Nahrgang:2011mg, Herold:2013bi,Herold:2014zoa, Herold:2016uvv} (see Ref.~\cite{Nahrgang:2016ayr} for an overview), and extended to a QCD-assisted transport approach \cite{Bluhm:2018qkf} by using an effective potential beyond mean field and the sigma spectral function from functional renormalization group calculations \cite{Pawlowski:2017gxj}.
In the CFD framework, the chiral condensate is identified as a dynamical variable 
while the slow modes relevant for the QCD critical point are related to conserved 
baryon densities (see Refs.~\cite{Son:2004iv,Fujii:2004jt,Fukushima:2010bq}).
We note that the dynamics of the chiral condensate has interesting phenomenological consequences \cite{Bluhm:2020rha} and its quantitative impact on critical fluctuations in full nonequilibrium calculations remains to be evaluated.

\subsubsection{Theoretical developments}

The appropriate quantitative framework for describing those out-of-equilibrium LWF 
modes is \textit{fluctuating hydrodynamics} supplemented with the salient feature 
of a critical point. Fluctuating hydrodynamics describes the evolution of (average)
hydrodynamic variables and their fluctuations. In the traditional stochastic approach,
described by Landau and Lifshitz~\cite{landau1980statistical}, the effects of
fluctuations are accounted for by adding stochastic noise terms to the conservation 
equations. The magnitude of the noise, encoded in noise correlation functions, 
is fixed by the fluctuation-dissipation theorem. This approach has been extended 
to relativistic hydrodynamics in Ref.~\cite{Kapusta:2011gt}. Even though numerical
simulations based on the stochastic approach are computationally demanding, 
we shall discuss encouraging new progress along this direction in Sec.~\ref{sec:hydro}. 

In contrast to the more familiar stochastic approach, the same dynamics may also be
captured in a deterministic approach \cite{Akamatsu:2016llw,Akamatsu:2017rdu, Stephanov:2017ghc,Akamatsu:2018vjr,Martinez:2018wia,An:2019csj,An:2020vri,Pratt:2018ebf,Pratt:2019pnd,Rajagopal:2019xwg,Du:2020bxp}. In this approach, wavenumber-dependent correlation 
functions of hydrodynamic variables are treated as additional slow variables in 
addition to the hydrodynamic ones. The resulting equations of motion are deterministic 
and describe the coupled evolution of the correlation functions and the conventional 
hydrodynamic variables. This approach successfully describes several non-trivial 
out-of-equilibrium effects. For example, the authors of Ref.~\cite{Martinez:2018wia} 
studied the impact of hydrodynamic fluctuations on correlation functions in a fluid 
with a conserved charge (such as baryon charge) undergoing a scaling (Bjorken) 
expansion. In Ref.~\cite{An:2019csj}, the deterministic approach is extended for 
a general fluid background. Simulations using the deterministic approach are 
less computationally demanding than those based on stochastic hydrodynamics, because
the equations of motion are similar in structure to those of ordinary
fluid dynamics. First applications for 3-dimensionally expanding systems, albeit with residual symmetry constraints, were reported in \cite{Rajagopal:2019xwg,Du:2020bxp}. However, the "deterministic approach" becomes more and more complex
if one wants to go beyond two-point functions, as would be required in order to study
non-Gaussian fluctuations.  

The Hydro+ formalism, which we will discuss in more detail, follows the deterministic
approach. Hydro+ was developed to describe the intertwined dynamics of critical
fluctuations and bulk evolution~\cite{Stephanov:2017ghc}. The key new ingredient in 
Hydro+ is the Wigner transform of the equal-time (in the LRF) two-point function of the fluctuation 
of the order parameter field $M(t,\bm{x})$:
\begin{eqnarray}
\label{phi-def}
\phi_{\bm{Q}}(t,{\bm x})&\equiv&
 \int d^3{\bm{y}}\, \langle\delta M\left(t,\bm{x}-\bm{y}/2\right)\,\delta M\left(t,\bm{x}+\bm{y}/2\right)\rangle\, e^{-i\bm{y}\cdot\bm{Q}}\, .
\end{eqnarray}
Here, $\phi_{\bm{Q}}(t,{\bm x})$ describes the magnitude of the critical fluctuation at 
wavelength $1/Q$,  and depends on time and spatial coordinate $\bm{x}$. The quantity 
$\phi_{\bm{Q}}(t,{\bm x})$ is treated as a dynamical variable in Hydro+ and obeys a 
relaxation rate equation:
\begin{eqnarray}
\label{phi-eqn}
\left( u^{\tau}\partial_{\tau}+u^{i}\partial_{i}\right)\,\phi_{\bm{Q}}(t,x)
= - \Gamma_{\bm{Q}}
\left( \phi_{\bm{Q}}(t,x)- \bar{\phi}_{\bm Q}(t,x)  \right)\, ,
\end{eqnarray}
where $\bar{\phi}_{\bm{Q}}$ is the equilibrium value of $\phi_{\bm{Q}}$. The stress-energy tensor
$T^{\mu\nu}$ and baryon number current $J^{\mu}$ are still conserved, and their
conservation equations, $\partial_{\mu}T^{\mu\nu}=0$ and $\partial_{\mu}J^{\mu}=0$,
together with Eq.~\eqref{phi-eqn} are the equations of motion for Hydro+. However, 
the transport coefficients and EOS are generalized in Hydro+. In particular, the 
constitutive relation for $T^{\mu\nu}$  is given by:
\begin{eqnarray}
\label{Tmunu}
T^{\mu\nu}=
\varepsilon u^{\mu}u^{\nu}+ p_{(+)}\,\left(g^{\mu\nu}+u^{\mu}u^{\nu}\right)
+\textrm{viscous terms}\, ,
\end{eqnarray}
with a similar expression for $J^{\mu}$, see Ref.~\cite{Stephanov:2017ghc}.
Note that the generalized pressure $p_{(+)}$ depends not only on the hydrodynamic 
variables $\varepsilon$ and $n_B$, the energy and baryon number densities, but 
also on the additional Hydro+ variable $\phi_{\bm{Q}}(t,{\bm x})$. $p_{(+)}$ is 
related to the generalized entropy density $s_{+}$  by generalized thermodynamic 
relations \cite{Stephanov:2017ghc}. Since hydrodynamic (collective) flow is induced 
by the gradient of the generalized pressure, in  Hydro+ the bulk evolution is 
intrinsically coupled with that of $\phi_{\bm{Q}}(t,{\bm x})$. Therefore,  Hydro+ 
couples LWF with hydrodynamics self-consistently.

Before closing this discussion, we would like to mention that the application of 
the deterministic approach is not limited to studying critical fluctuations. 
For example, the evolution of the fluctuations of conserved charges was investigated 
in Refs.~\cite{Ling:2013ksb,Pratt:2018ebf,Pratt:2019pnd} in order to constrain the charge diffusive
constant of the quark-gluon plasma from balance function measured experimentally at 
top RHIC energy. In Ref.~\cite{An:2019csj}, a general description of fluctuating 
hydrodynamics based on the deterministic approach has been formulated. This framework 
matches the  Hydro+ description of fluctuations near the QCD critical point and 
non-trivially extends inside and outside the critical region (see also Du et al. \cite{Du:2021zqz} for a related analysis of critical baryon diffusion effects).  Finally, we note 
that with suitable generalizations, the formalism of Hydro+ can also be used to study
hydrodynamics with chiral anomaly which couples non-conserved axial charge densities to 
hydrodynamic modes. 

In spite of the significant progress made with regards to the evolution of LWF there 
is still need for further development:

\begin{enumerate}
\item The formalism of fluctuating hydrodynamics discussed in this section  only 
applies to the crossover side of the phase boundary. How to extend this to the first 
order transition region requires further investigation. The authors of
Refs.~\cite{Steinheimer:2012gc,Steinheimer:2013gla,Pratt:2017lce} have investigated
the role of the spinodal instability based on hydrodynamics with an EOS that 
contains a first order transition as well as a finite range term to model the 
interface tension. However, it remains to be investigated how these results are 
affected by critical and non-critical fluctuations.

\item Most of the studies based on the deterministic approach are limited to
two-point functions of fluctuations. The extension of the existing formalism 
to higher-point functions, perhaps following the method of
Ref.~\cite{Mukherjee:2015swa}, is desirable, see Ref.~\cite{Pratt:2019fbj} 
and Ref.~\cite{An:2020vri} for recent developments along this direction. 

\item The additional, deterministic variables as propagated in Hydro+ are 
already averaged quantities and as such cannot directly be included into
standard event generators of heavy-ion collisions. Additional modeling of how 
to couple the initial state fluctuations is necessary. Final state fluctuations,
diffusion and dependence on kinematic cut as occurs in the hadronic phase need 
to be modeled and coupled consistently. A model for particlizing, or freezing 
out, a hydrodynamic fluid with fluctuations as described by Hydro+ is 
described in Sect.~\ref{sec:part-crit}.

\end{enumerate}

We shall report the first simulations of Hydro+ in Sec.~\ref{sec:hydro}, where we also
report progress using the stochastic approach. 

\subsection{Chiral Magnetic Effect and related phenomena}
\label{sec:theory_cme}

As already discussed in Sec.~\ref{sec:PD}, the spontaneous breaking of chiral symmetry
by the formation of a chiral condensate in the vacuum is a fundamental feature of QCD.
An equally important prediction of QCD is that the chiral condensate will eventually
disappear at high temperature, and that chiral symmetry is restored. Chiral restoration
above a critical temperature $T_c \approx 155\rm MeV$ has been established by lattice
QCD calculations~\cite{Bazavov:2011nk,Borsanyi:2010bp}. The system created in heavy 
ion collisions at RHIC and LHC is expected to reach the chiral transition temperature
and it is important to devise a measurement that directly probes chiral symmetry 
restoration. 

A promising approach is to look for the so-called Chiral Magnetic Effect
(CME)~\cite{Kharzeev:2004ey,Kharzeev:2007jp,Fukushima:2008xe} which predicts the
generation of an electric current by an external magnetic field under the presence 
of chirality imbalance:  
\begin{eqnarray} \label{eq:cme}
\mathbf{J}_e =\sum_f \frac{Q_f^2}{2\pi^2}   \mu_5 \mathbf{B}
\end{eqnarray}
where the sum is over all light flavors with electric charge $Q_f$, and $\mu_5 $ is 
the axial chemical potential that quantifies the chirality imbalance i.e. the 
difference in densities between right-handed and left-handed quarks. The CME is an 
important example of anomalous chiral transport processes. 

The CME requires a chirality imbalance (i.e. $\mu_5 \neq 0$). In the initial
state of a heavy ion collision such an imbalance can arise from topological 
transitions in the gluon sector, such as instantons and sphalerons. These 
objects are a key feature of non-perturbative dynamics in QCD, but they are 
hard to observe directly, because gluons do not carry conserved quantum numbers 
such as baryon number or electric charge. The QCD axial anomaly implies that  
every topological transition in the gluon sector induces a change in the  
chirality by $2N_f$ units. Consequently, an experimental observation
of chirality imbalance via the CME would also be a direct probe of the elusive gluon 
topological transitions. 

 The CME also requires that the chirality imbalance, once created by topological 
transitions, is not destroyed by explicit or spontaneous chiral symmetry 
breaking in QCD. Explicit symmetry breaking is encoded in current quark masses,
while spontaneous symmetry breaking is related to the quark condensate and
the so-called constituent mass $m_Q\sim 300$ MeV. Both effects correspond to 
operators that violate chirality by two units. While the effect of a small 
current quark mass on the CME is negligible, the effect of a constituent mass
is not. For this reason, the observation of the CME can provide important 
evidence for chiral symmetry restoration.  

 In addition to the CME, there are other anomalous chiral transport phenomena 
such as the so-called Chiral Vortical Effect 
(CVE)~\cite{Kharzeev:2007tn,Son:2009tf,Landsteiner:2011cp,Hou:2012xg},  the Chiral 
Electric Separation Effect (CESE)~\cite{Huang:2013iia,Jiang:2014ura} as well as the
Chiral Magnetic Wave (CMW)~\cite{Kharzeev:2010gd,Burnier:2011bf}. These ideas
have attracted strong interdisciplinary interest, particularly in condensed matter 
physics (for a review see \cite{Armitage:2017cjs}). For a more detailed discussion
and an extensive bibliography, see the recent  reviews~\cite{Kharzeev:2020jxw,
Bzdak:2019pkr,Kharzeev:2015znc,Li:2020dwr,Liu:2020ymh,Zhao:2019hta,Wang:2018ygc,
Fukushima:2018grm}. 
 
While the scientific significance of a possible CME discovery is high, the experimental 
search has encountered considerable challenges since the program was initiated in
2004~\cite{Kharzeev:2004ey,Voloshin:2004vk}. The key issues were identified at the 
start of the BEST Collaboration around 2015. The past several years have seen significant
progress in addressing these issues, as well as new opportunities for experimental
signatures, as we discuss next.

For the CME to occur in heavy ion collisions requires a net axial charge $N_5$ in 
a given event, as well as a strong magnetic field. Let us first discuss the axial 
charge generation. In a typical collision the fireball acquires considerable initial 
axial charge $N_5$ from random topological fluctuations of the strong initial 
color fields. This has been demonstrated by recent classical-statistical simulations 
performed in the so-called glasma framework~\cite{Mueller:2016ven,Mace:2016svc,
Mace:2016shq,Mace:2017wcl,Lappi:2017skr}, which provides a quantitative tool for 
constraining the axial charge initial conditions that would be necessary for evaluating  
CME signals in these collisions.  

Axial charge is not conserved due to the quantum anomaly
and the nonzero quark masses. That is, starting with a certain non-vanishing initial 
$N_5$, it will subsequently relax toward a vanishing equilibrium value. The rate for 
such relaxation is controlled by the random gluonic topological fluctuations at finite 
temperature and also receives contribution from finite quark masses. Here the key issue 
is whether the initial axial charge can survive long enough to induce a measurable 
CME signal.  Realistic estimates including both gluonic and mass contributions to 
axial charge relaxation~\cite{Guo:2016nnq,Hou:2017szz,Lin:2018nxj,Liang:2020sgr} 
suggest that the QGP maintains its finite chirality for a considerable time. For 
the CME modeling, it is important to account for such non-equilibrium dynamics 
of axial charges. One approach is chiral kinetic theory, which has recently been 
developed~\cite{Son:2012wh,Stephanov:2012ki,Chen:2014cla,Chen:2015gta,Stephanov:2015roa,
Hidaka:2016yjf,Hidaka:2017auj,Huang:2018wdl,Gao:2018wmr,Gao:2019znl,Sheng:2018jwf,
Weickgenannt:2019dks,Wang:2019moi,Li:2019qkf,Liu:2018xip,Mueller:2017arw,Mueller:2017lzw,
Shi:2020htn}, and applied to the phenomenology of heavy ion collisions in 
Refs.~\cite{Kharzeev:2016sut,Huang:2017tsq,Ebihara:2017suq,Sun:2018idn}. 
Alternatively, one can adopt the stochastic hydrodynamic description for axial 
charge dynamics~\cite{Lin:2018nxj}, which can be naturally integrated into a 
hydro-based modeling framework for both the bulk evolution and the CME transport.

The other key element is the magnetic field $\bf{B}$. Heavy ion collisions create an
environment with an extreme magnetic field -- at least at very early times -- which 
arises from the fast-moving, highly-charged nuclei. A simple estimate gives
$|e\mathbf{B}|\sim \frac{\alpha_{EM} Z\gamma b}{R_A^2} \sim m_\pi^2$ at the center 
point between the two colliding nuclei upon initial impact.  Given a magnetic field 
of this strength and a chiral QGP, we expect the CME to occur. However, for a 
quantitative analysis of possible CME signals, two crucial factors need to be
understood: the azimuthal orientation as well as time duration of the magnetic
field. A randomly oriented magnetic field which is not correlated to any other 
observable such as elliptic flow prevents the CME, even if present, to be observed 
in heavy ion experiments. A magnetic field which, although very strong initially, 
decays too fast would lead to an  undetectably small signal~\cite{Shi:2017cpu}. 

As first shown in  \cite{Bzdak:2011yy, Bloczynski:2012en}, strong fluctuations of the initial protons 
in the colliding nuclei lead to significant fluctuations in the azimuthal orientation 
of the $\mathbf{B}$ field relative to the bulk matter geometry.  Fortunately one can 
use simulations to quantify the azimuthal correlations between magnetic field and 
various geometric orientations (e.g. reaction plane, elliptic and triangular participant
planes) in the collision.  Such magnetic field fluctuations turn out to be useful 
features for experimental analysis,  by comparing relevant charge-dependent correlations
measured with respect to reaction plane as well as elliptic and triangular event planes, 
see the discussions in \cite{Bzdak:2019pkr,Wang:2018ygc,Li:2020dwr}.

 The strong initial magnetic field rapidly decays over a short period of time due to 
the rapid motion of spectator protons along the beam direction. Understanding the
dynamical evolution of the residual magnetic field in the mid-rapidity region is a 
very challenging problem. Many studies based on different levels of approximation
have been made~\cite{McLerran:2013hla,Tuchin:2015oka,Inghirami:2016iru,Inghirami:2019mkc,
Gursoy:2018yai,Roy:2017yvg,Pu:2016ayh,Muller:2018ibh,Guo:2019joy,Guo:2019mgh}.
Generically we expect an electrically conducting QGP to increase the lifetime of 
the $\mathbf{B}$ field, but quantitative determinations are difficult. Simulations 
were performed based on a magneto-hydrodynamic (MHD) framework~\cite{Inghirami:2016iru,
Inghirami:2019mkc,Hernandez:2017mch,Denicol:2018rbw,Denicol:2019iyh,Shokri:2018qcu,
Siddique:2019gqh}. However the QGP may not have a sufficiently large electric 
conductivity to be in an ideal MHD regime. Another, perhaps more realistic approach
aims to solve the in-medium Maxwell's equations in an expanding and conducting 
fluid while neglecting the feedback of the $\mathbf{B}$ field on the medium bulk 
evolution~\cite{Gursoy:2018yai}. The BEST Collaboration effort has focused on 
developing a robust simulation  framework for $\bf{B}$ field evolution along this 
latter approach, with significant progress achieved recently. 
See  further discussions in Sec.~\ref{sec:hydro_anomal}.  Additionally, there are  
interesting studies of other effects induced by a strong magnetic field which could
be used to constrain the in-medium $\mathbf{B}$ field in heavy ion 
collisions~\cite{Inghirami:2019mkc,Gursoy:2018yai,Guo:2015nsa,Muller:2018ibh,
Guo:2019joy,Guo:2019mgh,Xu:2020sui,Aaboud:2018eph,Adam:2018tdm,Zha:2018tlq,Zha:2018ywo,
Klein:2018fmp}.

On the experimental side, the CME-induced transport is expected to result in a 
dipole-like charge separation  along $\mathbf{B}$ field direction~\cite{Kharzeev:2004ey}, 
which could be measured as a charge asymmetry in two-particle azimuthal 
correlations~\cite{Voloshin:2004vk}. Extensive searches have been carried out over 
the past decade to look for this correlation by the STAR Collaboration at the 
Relativistic Heavy Ion Collider (RHIC), as well as by ALICE and CMS Collaborations at the 
Large Hadron Collider (LHC)~\cite{Voloshin:2004vk,Xu:2017qfs,Zhao:2017nfq,
Voloshin:2018qsm,Magdy:2017yje,Magdy:2018lwk,Tang:2019pbl}.  Encouraging hints 
of the CME have been found, in particular in the regime studied by the RHIC Beam 
Energy Scan program. However, the interpretation of these data remains inconclusive  
due to significant background contamination. For a more in-depth  discussions see,
e.g. \cite{Bzdak:2019pkr,Kharzeev:2015znc,Li:2020dwr,Zhao:2019hta,Bzdak:2012ia}. A 
new opportunity of potential discovery for the CME is provided by a decisive isobar
collision experiment, carried out in the 2018 run at RHIC~\cite{Kharzeev:2020jxw,
Kharzeev:2019zgg,Skokov:2016yrj,Voloshin:2010ut,Adam:2019fbq}, whose data are still being analyzed. 

Critical to the success of the experimental program is a precise and realistic 
characterization of the CME signals as well as backgrounds in these collisions. To 
achieve this requires a  framework that addresses the main theoretical challenges
discusses above: (1) dynamical CME transport in the relativistically expanding 
viscous QGP fluid; (2) initial conditions and subsequent relaxation for the axial 
charge; (3) co-evolution of the dynamical magnetic field with the medium; 
(4) proper implementation of major background correlations.
A framework that addresses most of these effects, dubbed EBE-AVFD (Event-By-Event Anomalous-Viscous Fluid Dynamics)~\cite{Shi:2019wzi, Shi:2017cpu,Jiang:2016wve}, has been developed
by the BEST Collaboration and will be discussed in detail in 
Sec.~\ref{sec:hydro_anomal}. 

\section{EoS with 3D-Ising model critical point}
\label{sec:eos}

\begin{figure}[htb]
\centering
\includegraphics[width=0.99\linewidth]{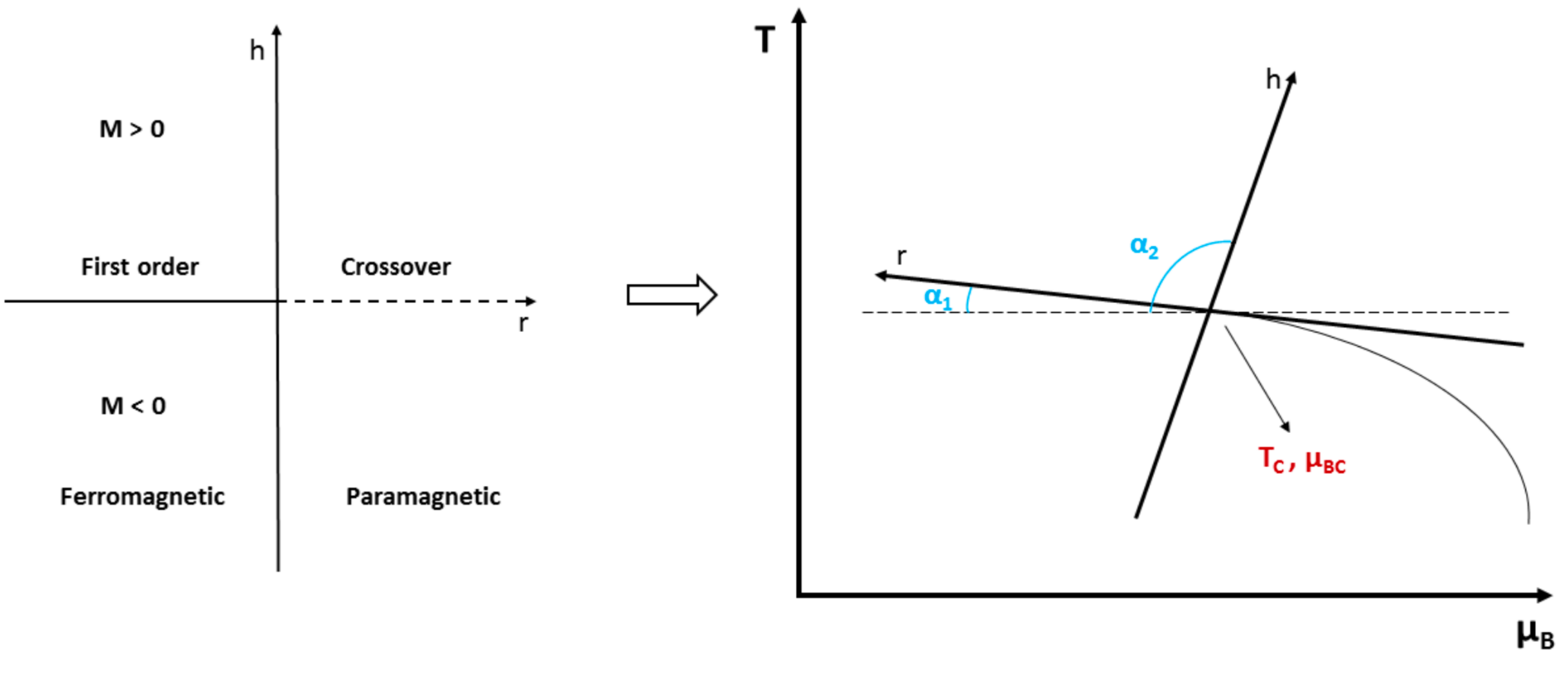}
\caption{\label{Ising_Axes} 
From Ref.~\cite{Parotto:2018pwx}. Non-universal mapping 
from Ising variables $(r,h)$ to QCD coordinates $(T,\mu_B)$.}
\end{figure}

Starting from the results discussed above, a family of Equations of State was created, 
each one containing a critical point in the 3D Ising model universality class, and 
constrained to reproduce the lattice QCD results up to $\mathcal{O}(\mu_B/T)^4$ 
\cite{Parotto:2018pwx}. Earlier, an equation of state containing a 3D Ising model 
critical point was obtained in Refs.~\cite{Nonaka:2004pg,Kampfer:2005nt}, but the critical
effects were built on top of a quasi-particle, MIT bag or Hadron Resonance Gas model 
equation of state, rather than systematically matching them to lattice QCD results.  In our
work the mapping between the Ising model phase diagram (in terms of reduced temperature 
$r$ and magnetic field $h$) and the QCD one (in terms of $T$ and $\mu_B$) is performed 
in terms of six parameters: the location of the critical point ($T_C,~\mu_{BC}$), the 
angles $\alpha_1$ and $\alpha_2$ that the $r$ and $h$ axes form with the $T=0$ QCD one, 
and the $(w,~\rho)$ parameters that indicate a global and a relative scaling of the axes. 
Such a mapping is shown in Fig.~\ref{Ising_Axes}.

Two of these parameters are fixed imposing that the critical point lies on the phase
transition line obtained in lattice QCD simulations (see details on the QCD transition 
line in Section \ref{sec:PD}). The other four parameters can be freely varied by the user,
who can download the code from the BEST Collaboration repository \cite{code:2018}. The goal 
is then that a systematic comparison between the predictions of hydrodynamic codes that
use this EoS as an input and the experimental data, will help to constrain these parameters,
including the location of the critical point.

\begin{figure}[htb]
\centering
\includegraphics[width=0.49\linewidth]{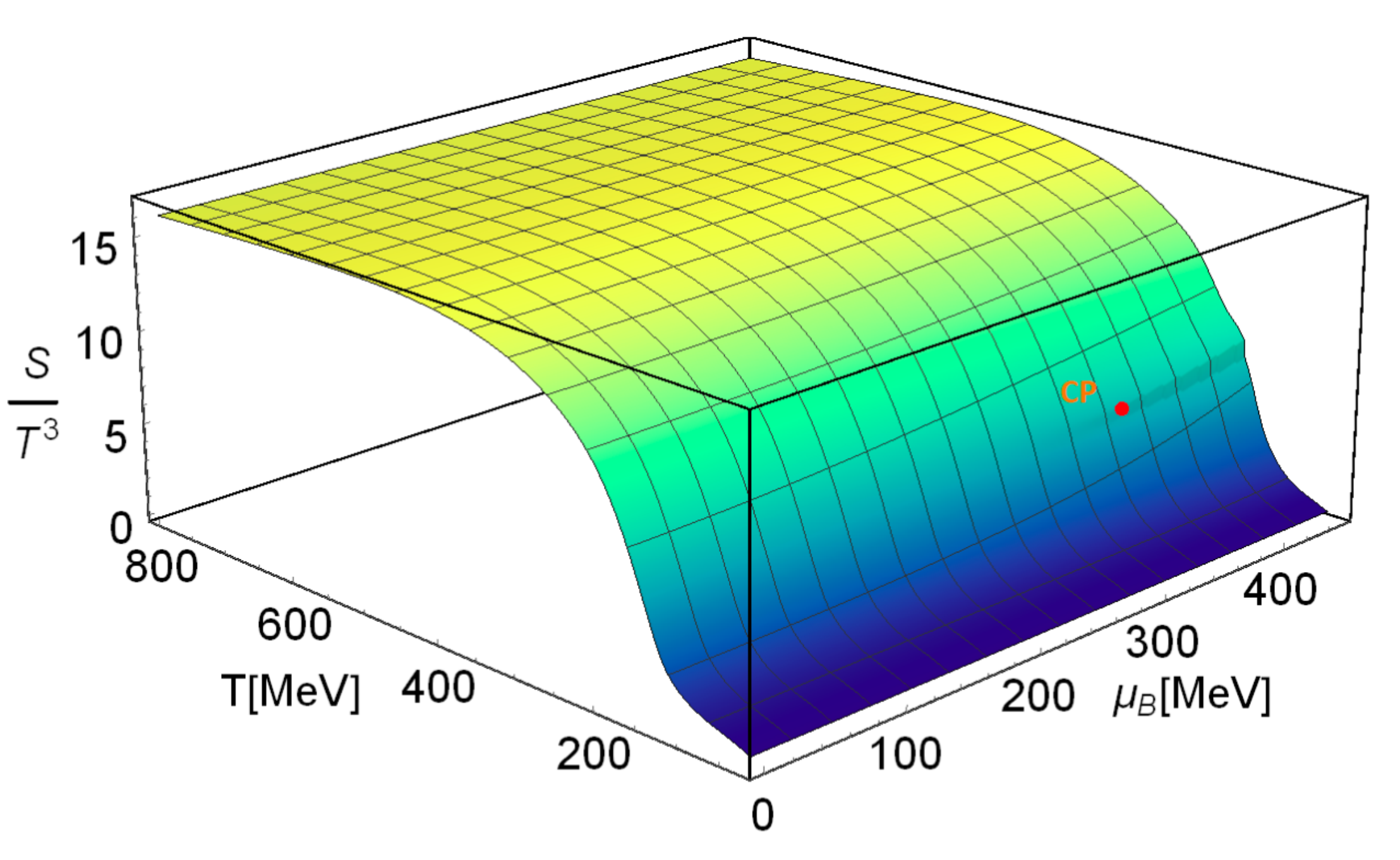}
\includegraphics[width=0.49\linewidth]{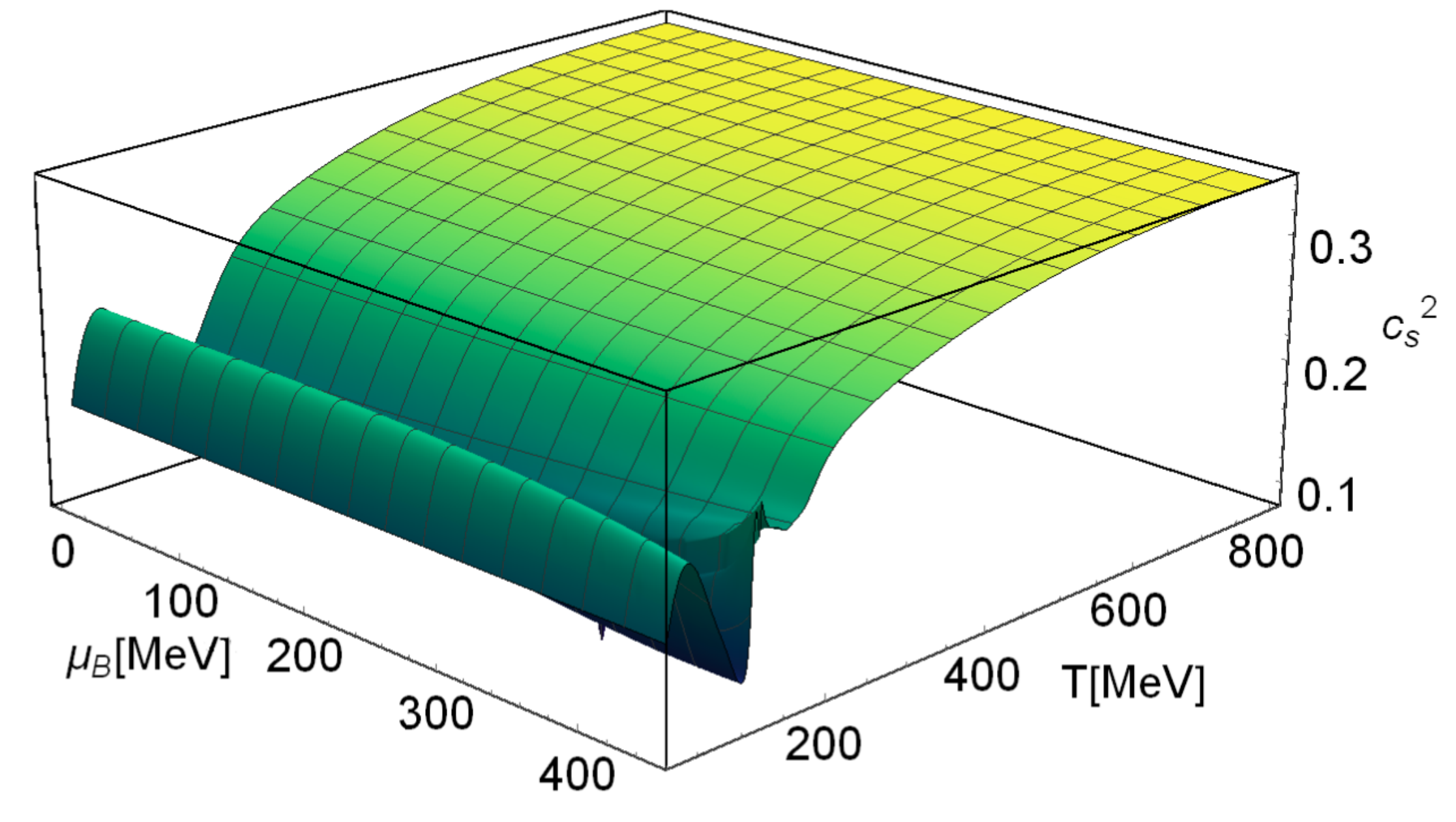}
\caption{\label{EoS_Ising}
From Ref.~\cite{Parotto:2018pwx}. Entropy density (left) and speed of sound 
$c_s^2$ (right) as functions of temperature and chemical potential, for the parameter 
choice described in Ref. \cite{Parotto:2018pwx}. In particular, the critical point 
for this choice is located at $\mu_{Bc}=350$ MeV and $T_c=143.2$ MeV.}
\end{figure}

The assumption is that the lattice QCD Taylor expansion coefficients can be written as 
the sum of the Ising contribution and a non-critical one, that can be obtained as 
the difference between lattice and Ising:
\begin{eqnarray}
T^4c_n^{LAT}(T)=T^4c_n^{Non-Ising}(T)+T_c^4c_n^{Ising}(T).
\end{eqnarray}
The full pressure is then reconstructed as
\begin{eqnarray}
P(T,\mu_B)=T^4\sum_nc_{2n}^{Non-Ising}(T)\left(\frac{\mu_B}{T}\right)^{2n}
 + P_{crit}^{QCD}(T,\mu_B).
\end{eqnarray}
Figure \ref{EoS_Ising} shows the entropy density and the speed of sound for the 
parameter choice used in Ref.~\cite{Parotto:2018pwx}.
This equation of state has been recently extended to the phenomenologically relevant case of strangeness neutrality and fixed electric charge/baryon number ratio in Ref. \cite{Karthein:2021nxe}.
Recently, the BEST EoS has been used to study the behavior of the critical fourth-order
cumulant of baryon number on conjectured freezeout trajectories in the QCD phase diagram
\cite{Mroczek:2020rpm}. It was found that subleading and non-singular terms have a
significant effect on the behavior of the fourth order cumulant (kurtosis). The original
prediction based on Ising universality is that, as the baryon chemical potential
increases along the freezeout curve, the kurtosis first exhibits a  dip, followed
by a peak \cite{Stephanov:2011pb}. However, when subleading terms are taken into account,
the dip is not a robust feature of the kurtosis along the freezeout line, and only
the enhancement is a generic feature of the equation of state.

\section{Initial Conditions}
\label{sec:init_cond}
\newcommand{\bps}[1]{{\color{red} \bf BPS: #1}}

\subsection{Challenges of the Beam Energy Scan}

At the highest RHIC energies and at the LHC, the approaching nuclei are highly Lorentz
contracted. At top RHIC energy the nuclei pass through one another in less than 0.15
fm/$c$, and at the LHC the time is even shorter. Particles are produced over a range 
of rapidities, roughly defined by the beam rapidities, $\pm 5.4$ at RHIC and $\pm 8$ 
at the LHC. This large rapidity range implies that comoving observers within $\sim$ 
one unit of rapidity around central rapidity see essentially the same physics, strongly 
Lorentz contracted, highly excited, target and projectile nuclei receding at a 
velocity close to the speed of light. This observation motivated Bjorken to propose 
a hydrodynamic model \cite{Bjorken:1982qr} of relativistic heavy ion collisions based 
on longitudinal boost invariance. The Bjorken model allows us to reduce the 3+1 
dimensional evolution to a 2+1 dimensional problem. This approximation appears to hold 
at the 5\% level for the highest RHIC energies, when considering mid-rapidity measurements.
The variation of the baryon density with rapidity can also be ignored at these energies. 
For a given beam energy the initial state for hydrodynamics can be characterized by
$\lesssim 6$ parameters describing the magnitude and shape of the transverse energy 
density profile, the baryon density, the anisotropy of the initial stress-energy tensor, 
and the initial transverse flow \cite{Novak:2013bqa,Moreland:2014oya}. 

At BES energies none of these simplifications are warranted. Nuclei require up to $4$ fm/$c$
to pass through one another, and a significant fraction of the transverse collective flow 
has developed before the incoming nuclei have finished depositing energy. The deposition 
of energy and baryon number vary over a much smaller rapidity range, invalidating any
assumptions of boost invariance. Describing the initial state for hydrodynamics is much 
more difficult as one must quantify the variations of energy density, baryon density, 
and initial transverse flow with rapidity. Further, one must account for the fact that 
energy and baryon density are deposited over a significant amount of time \cite{Shen:2020mgh}. Without a doubt, 
modeling this phase of the collision is one of the most daunting challenges faced by the 
BEST Collaboration. In the next section, the status of the Pre-BEST 3D collision models is
reviewed. The following two sections then present two schemes developed by the BEST 
Collaboration that address the challenges described above.

\subsection{Pre-BEST status of 3D initial condition models}

We present a brief summary of available models that have been or can be used to provide initial
conditions for 3+1 dimensional hydrodynamic simulations. First 3+1D hydrodynamic simulations
were performed with smooth initial conditions, and the typical approach to include a
longitudinal structure to a transverse optical Glauber model geometry was to apply an 
envelope function consisting of a plateau around space-time rapidity zero and two
half-Gaussians in the forward and backward directions \cite{Hirano:2001eu}. The parameters 
of the model, i.e., the plateau and Gaussian widths, could then be tuned to fit experimental
data. The same method could also be used when fluctuations in the transverse geometry are
included (see e.g. \cite{Schenke:2010rr}). Another approach, that similarly factorizes the
transverse from the longitudinal dependence, was followed in \cite{Ke:2016jrd}, extending 
the Trento model \cite{Moreland:2014oya} to three dimensions.

Early simulations of 3+1D hydrodynamics with initial state fluctuations in all three
dimensions were performed using UrQMD \cite{Bass:1998ca,Bleicher:1999xi} or NEXUS
\cite{Drescher:2000ec} to provide the initial conditions \cite{Steinheimer:2007iy,
Takahashi:2009na, Andrade:2009em, Karpenko:2015xea}. When using UrQMD, for example, all
produced point-like particles are assigned a 3D spatial Gaussian with a tuneable width to
generate smeared out energy, baryon, and momentum densities as input for the hydrodynamic
equations \cite{Steinheimer:2007iy}.

Also AMPT \cite{Zhang:1999bd}, which is based on HIJING \cite{Wang:1991hta}, has been used 
to generate fluctuating initial conditions for 3+1D hydrodynamics. Here, one has mini-jets
and soft partons (from melted strings) with varying formation times. Typically, after running
AMPT's parton cascade, one can determine a proper time surface on which most partons have
formed, use it as the initial time for hydrodynamics, and neglect late time interactions in
the cascade, that occur mainly at forward rapidities \cite{Pang:2012he}. Each parton is then
treated similarly to the UrQMD case above and 3D Gaussians are assigned to form an energy
momentum tensor in every hydro grid cell.

Aside from these models, which are based on generators that initially produce hadrons,
several other options are available, including some that are based on the AdS/CFT
correspondence \cite{Maldacena:1997re}. These can provide initial conditions for 
energy and momentum \cite{Chesler:2015fpa}, as well as baryon densities 
\cite{Casalderrey-Solana:2016xfq}, but so far they typically neglect geometric fluctuations.
Another possibility that has been explored is to extend the color glass condensate based
models to three dimensions. This has been done, for example, by employing JIMWLK evolution
\cite{JalilianMarian:1997jx,JalilianMarian:1997gr,JalilianMarian:1997dw,Iancu:2001ad,
Iancu:2000hn,Ferreiro:2001qy,Weigert:2000gi} to determine the Bjorken $x$ dependence 
of the gluon distributions in the incoming nuclei, and from that deduce the rapidity
dependence \cite{Schenke:2016ksl} of the initial energy momentum tensor, or by fully
extending the Yang-Mills computations, done in 2D in the IP-Glasma model 
\cite{Schenke:2012wb,Schenke:2012hg}, to three dimensions \cite{McDonald:2020oyf,
Schlichting:2020wrv}. While these models provide an initial energy momentum tensor, 
baryon stopping in a saturation framework was separately addressed in
\cite{MehtarTani:2009dv,Kolbe:2020hem}.

 Except for the implementations discussed in \cite{Akamatsu:2018olk} and 
 \cite{Du:2018mpf}, the models discussed above do not address the issue of the 
relatively long overlap time of the two colliding nuclei at low beam energies, which
makes an initialization on a constant eigentime surface problematic. Indeed, up to
now, even initial conditions based on UrQMD are based on particles propagated to a
constant eigentime surface, at which hydrodynamics is initialized 
\cite{Karpenko:2015xea}. Within BEST, a fully dynamical initial state model, based on
string deceleration, which provides three dimensional source terms for energy, momentum,
and baryon currents, was developed and is implemented dynamically into the 3+1 dimensional
\textsc{music} code \cite{Schenke:2010nt,Schenke:2010rr,Paquet:2015lta}. Its advantages 
over existing models that have been coupled to
hydrodynamics are that energy (and charge) deposition is linked to the dynamical
deceleration of the string ends, which leads to a realistic space time picture, and
that the model only requires a limited number of parameters, so that it can be 
incorporated into a Bayesian analysis framework. We will discuss this model in 
Section \ref{sec:dynIn}. In the following, we first describe another new development,
a minimal extension of the conventional Glauber model, that describes the longitudinal
structure of the initial state based on energy and longitudinal momentum conservation 
arguments.

\subsection{Simple collision geometry based 3D initial condition}

The conventional Glauber model assumes the colliding nuclei to be infinitely Lorentz
contracted along the beam direction. The produced energy/entropy densities in the 
transverse plane depend on the nuclear thickness functions $T_A$ and $T_B$. The authors 
of Ref.~\cite{Shen:2020jwv,Ryu:2021lnx} proposed a minimal extension of the Glauber model to 3D 
which respects the constraints imposed by energy and momentum conservation locally at 
every transverse position in the collision. At any point in the transverse plane $(x, y)$, 
conservation of energy $E(x,y)$ and longitudinal momentum $P_z(x,y)$ imply that 
\begin{eqnarray}
    E(x, y) = [T_A(x,y) + T_B(x,y)]m_N \cosh(y_\mathrm{beam}) 
    = \int d \eta_s\, \tau_0\, T^{\tau t} (\tau_0, x,y,\eta_s)\,,
    \label{eq:IS:simple3D:Econservation}
\end{eqnarray}
and
\begin{eqnarray}
    P_z(x,y) = [T_A(x,y) - T_B(x,y)]m_N \sinh(y_\mathrm{beam}) 
    = \int d \eta_s\, \tau_0\, T^{\tau z} (\tau_0, x,y,\eta_s)\,,
    \label{eq:IS:simple3D:Pzconservation}
\end{eqnarray}
respectively. Here, $T^{\mu\nu}(\tau,x,y,\eta_s)$ are the components of the stress
tensor at transverse position $(x,y)$, proper time $\tau$, and spatial rapidity 
$\eta_s$. These relations ensure that the space-momentum correlations in the initial state 
are continuously passed to the hydrodynamic phase. Assuming Bjorken flow, the local energy-momentum tensor of 
the fluid at the hydrodynamic starting time $\tau_0$ matches with the Glauber model collision
geometry, and especially the global angular momentum is smoothly mapped from the colliding nuclei 
to the fluid fields.

 Ref.~\cite{Shen:2020jwv} shows that at sufficiently high collision energies a flux-tube-like parameterization of the longitudinal
distribution of energy density $T^{\tau\tau}(\tau_0,x,y,\eta_s)=e(\tau_0,x,y,\eta_s)$,
combined with local energy-momentum conservation, results in a transverse energy density
scaling $e(x, y) \propto \sqrt{T_A(x, y) T_B(x, y)}$, which is preferred by the Bayesian 
statistical analysis \cite{Bernhard:2019bmu}.

\begin{figure}[htb]
\vspace*{0.2cm}
\centering
\includegraphics[width=0.6\linewidth]{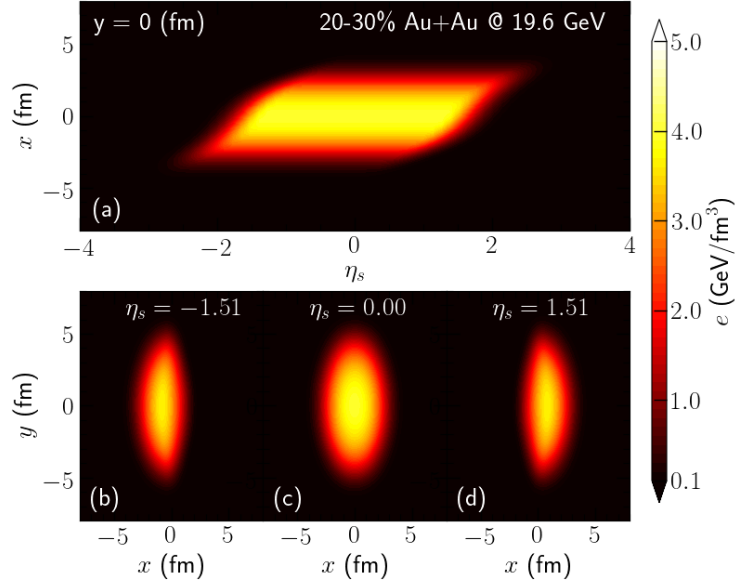}
\caption{Contour plot for the local energy density distribution at $\tau = 1.8\,{\rm fm}/c$
in 20-30\% Au+Au collisions at 19.6 GeV \cite{Shen:2020jwv} in space time rapidity and one
transverse direction (for $y=0$) (a), and in the transverse ($x-y$) plane for three 
different space time rapidities (b)-(d).
\label{fig:initial-simple-model-1}}
\end{figure}

Fig.~\ref{fig:initial-simple-model-1} shows projections of the 3D initial energy density
distribution in 20-30\% Au+Au collisions at 19.6 GeV. In Panel (a), the energy density is
shifted to positive $\eta_s$ for $x > 0$, which is a consequence of longitudinal momentum
conservation. Along the impact parameter direction (positive $x$), the local nucleus
thickness function of the projectile nucleus is larger than that of the target, which 
leads to a positive net longitudinal momentum in the $x > 0$ region. Panels (b-d) 
illustrate the shape of the energy density in the transverse direction for three different
space-time rapidities. The fireball becomes more eccentric in the forward and backward
directions compared to the energy density profile at mid-rapidity. The dipole-deformation 
of the fireball is odd in the space-time rapidity, correlated with the direction of net
longitudinal momentum $P_z$.

\begin{figure}[htb]
\centering
\includegraphics[width=0.478\linewidth]{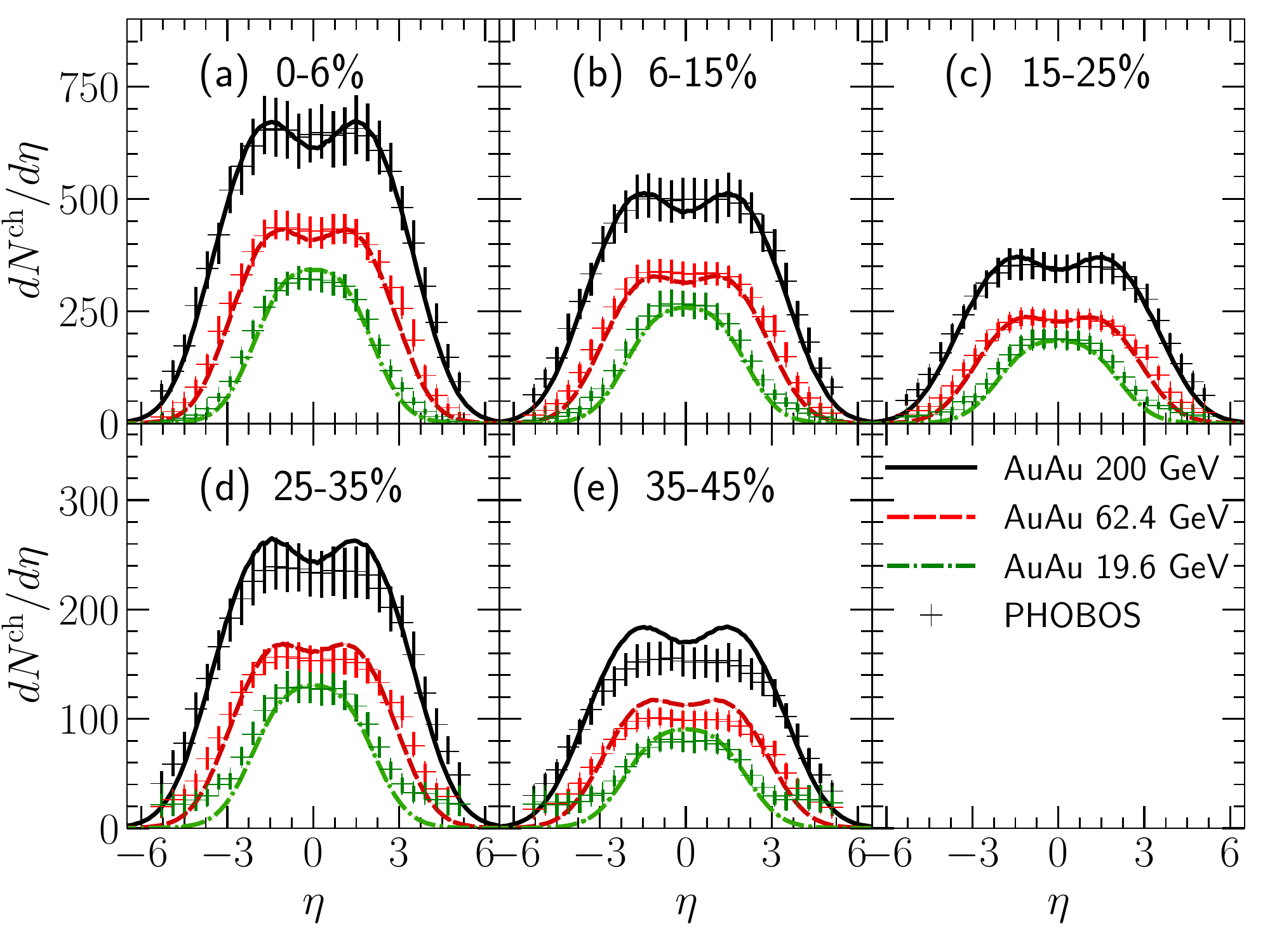}
\includegraphics[width=0.49\linewidth]{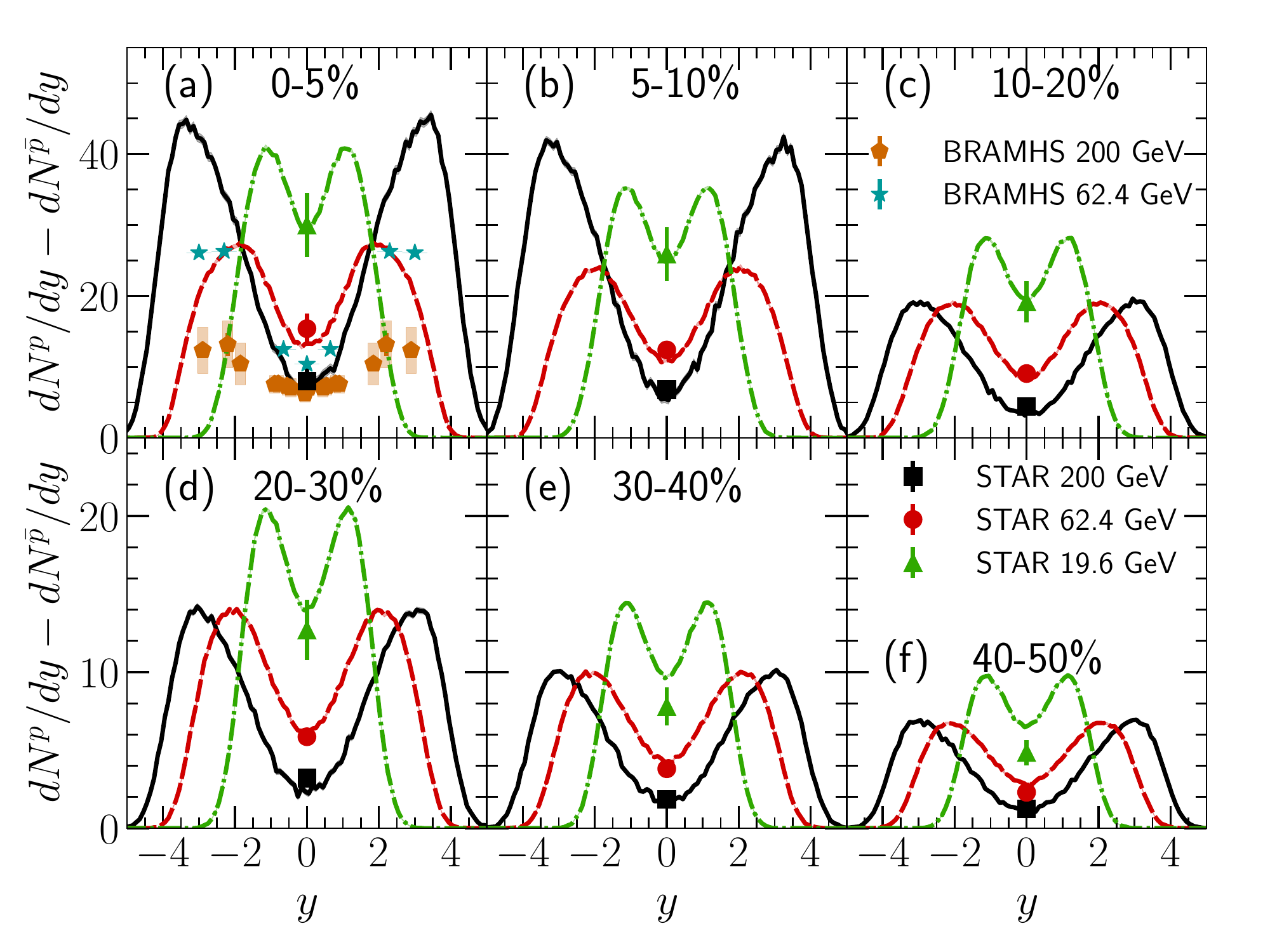}
\caption{The (pseudo-)rapidity distributions of the produced charged hadrons (left) and 
net protons (right) compared with the RHIC measurements for Au+Au collisions at 19.6, 62.4,
and 200 GeV \cite{Back:2005hs, STAR:2017sal, BRAHMS:2003wwg, STAR:2008med}. The figure was taken from Ref.~\cite{Shen:2020jwv}.
\label{fig:initial-simple-model-spectra}}
\end{figure}

Fig.~\ref{fig:initial-simple-model-spectra} shows that the collision-geometry-based initial
state model with hydrodynamics + hadronic transport simulations can achieve a good
description of the (pseudo-)rapidity distributions of the produced charged hadrons and net
protons measured at RHIC. Note that this model was calibrated only with the data in
the most-central collisions in panel (a). The results in other centrality bins were model
predictions. The rapidity evolution as a function of collision centrality was well captured
by this model. For the net proton rapidity distribution,  this model gives a good
description of the experimental measurement at mid-rapidity as a function of centrality,
while the rapidity dependence still has room for improvements.

\subsection{Dynamical initial condition}
\label{sec:dynIn}

\begin{figure}[t]
\centering
\includegraphics[width=0.49\linewidth]{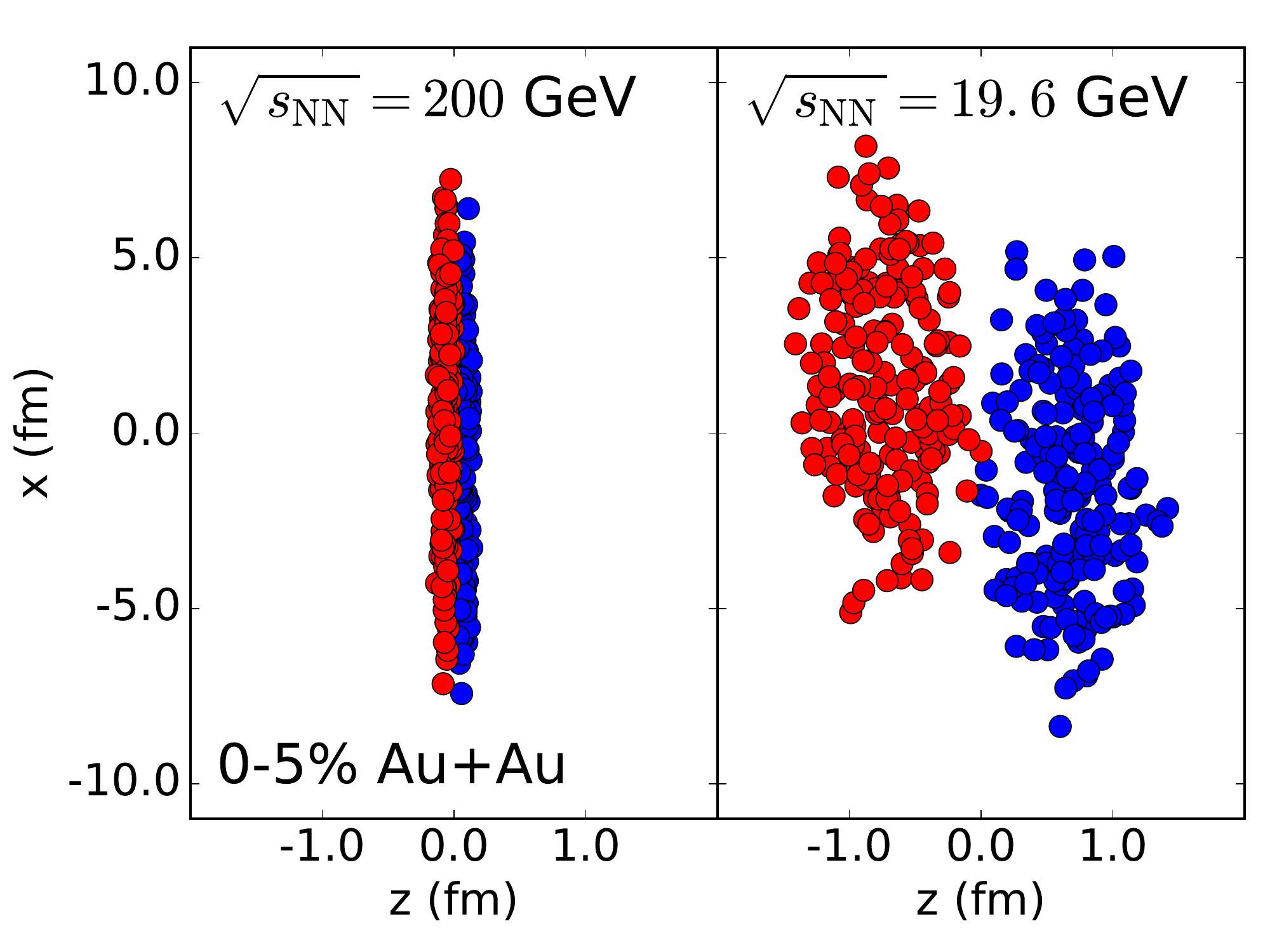}
\includegraphics[width=0.48\linewidth]{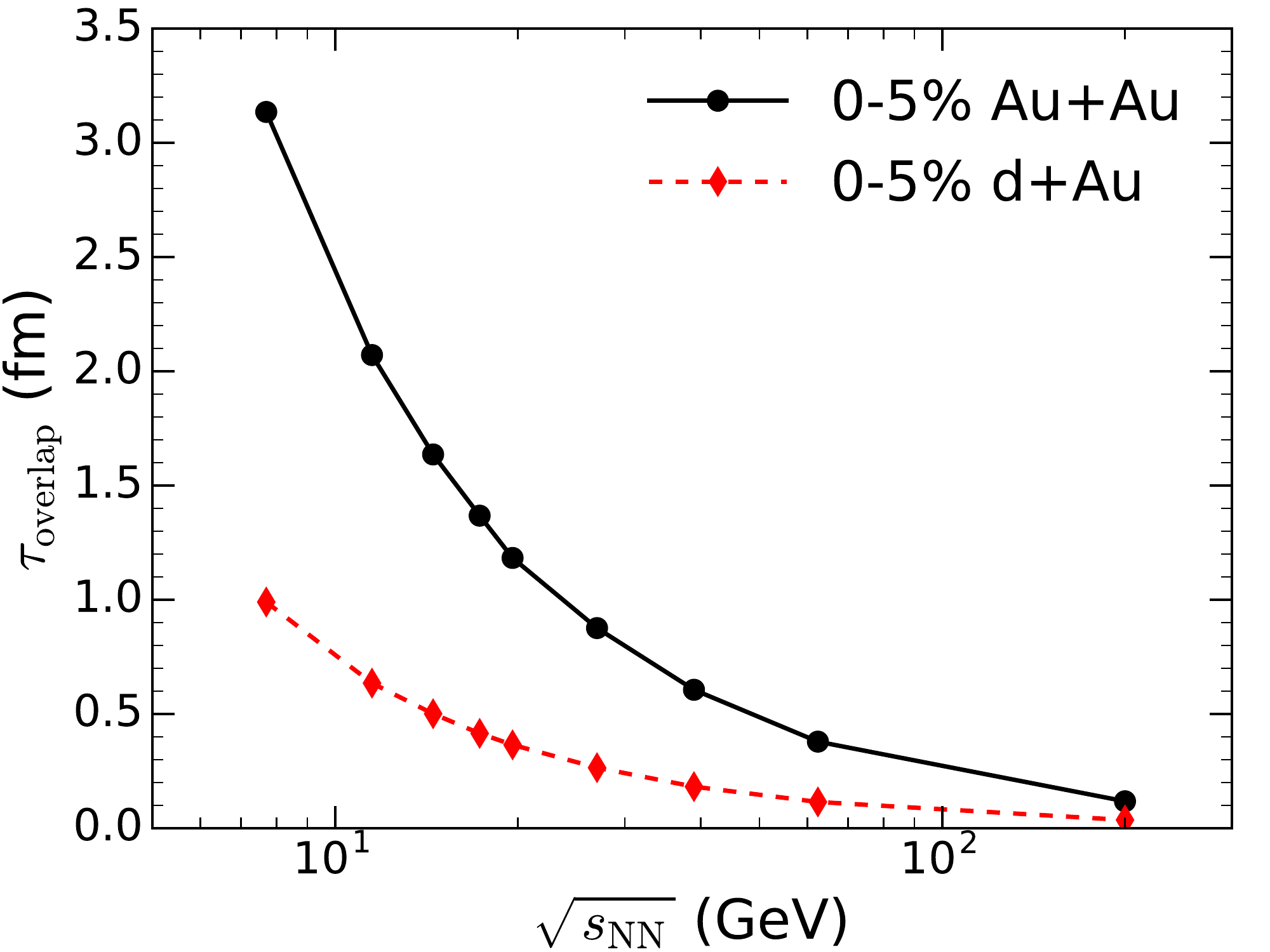}
\caption{Left: Visualization of the Lorentz contraction of incoming nuclei at the time of 
the first nucleon nucleon collision for $\sqrt{s}=200\,{\rm GeV}$ and $19.6\,{\rm GeV}$ 
Au+Au collisions. Right: The overlap time as a function of center of mass energy for Au+Au 
and d+Au collisions. Figure from \cite{Shen:2017bsr}.
\label{fig:initial-dynamic-intro}}
\end{figure}

As already noted, when the collision energy is decreased to $\mathcal{O}(10)$\,GeV, the relativistic Lorentz
contraction factors of the colliding nuclei along the beam (longitudinal) direction are 
no longer large. The overlap time required for the two colliding nuclei
to pass through each other becomes significant compared to the total lifetime of the 
system, which is of the order $10\,{\rm fm}/c$, see Fig.~\ref{fig:initial-dynamic-intro}. 
The nucleon-nucleon collision pairs that collide early will produce energy-momentum 
currents that evolve (possibly hydrodynamically) before the rest of the nucleons collide 
with each other. 

To deal with this situation, a new dynamical framework which connects the
pre-equilibrium stage of the system to hydrodynamics on a local 
collision-by-collision basis was proposed \cite{Shen:2017bsr}. The hydrodynamic
evolution starts locally at a minimal thermalization time after the first
nucleon-nucleon collision. The sequential collisions between nucleons that occur 
later contribute dynamically as energy and net-baryon density sources to the
hydrodynamic simulations.

For nucleon-nucleon collisions we consider energy loss of the valence quarks, whose 
initial momentum fraction is sampled from nuclear parton distribution functions. They
go through a classical string-deceleration model, which generates correlations between
space-time and momentum. 

Going beyond the BEST-developed model discussed in \cite{Shen:2017bsr}, more recent
developments take into account that the energy and momentum deposited in the medium are 
equal to the energy and momentum lost in the deceleration process, resulting in exact 
energy and momentum conservation in the model. Furthermore, baryon number is propagated 
with some probability along the string towards midrapidity, following the idea of baryon
junctions first put forward in \cite{Kharzeev:1996sq}. Consequently, the model includes
spatial fluctuations of the net baryon density and energy density, and thus fluctuations 
of where in the phase diagram the hydrodynamic evolution begins for every position in space.

\begin{figure}[t]
\centering
\includegraphics[width=0.48\linewidth]{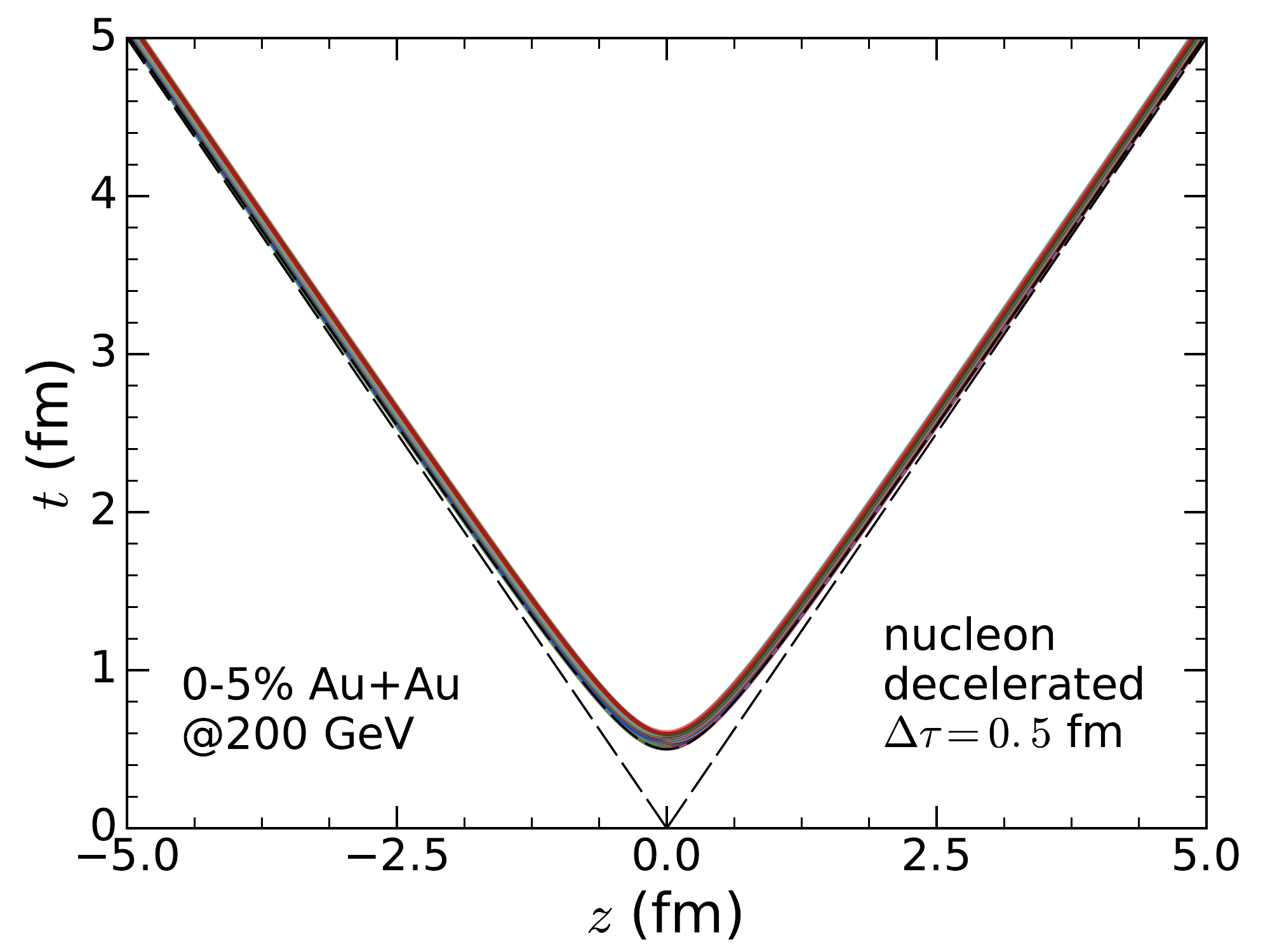}
\includegraphics[width=0.49\linewidth]{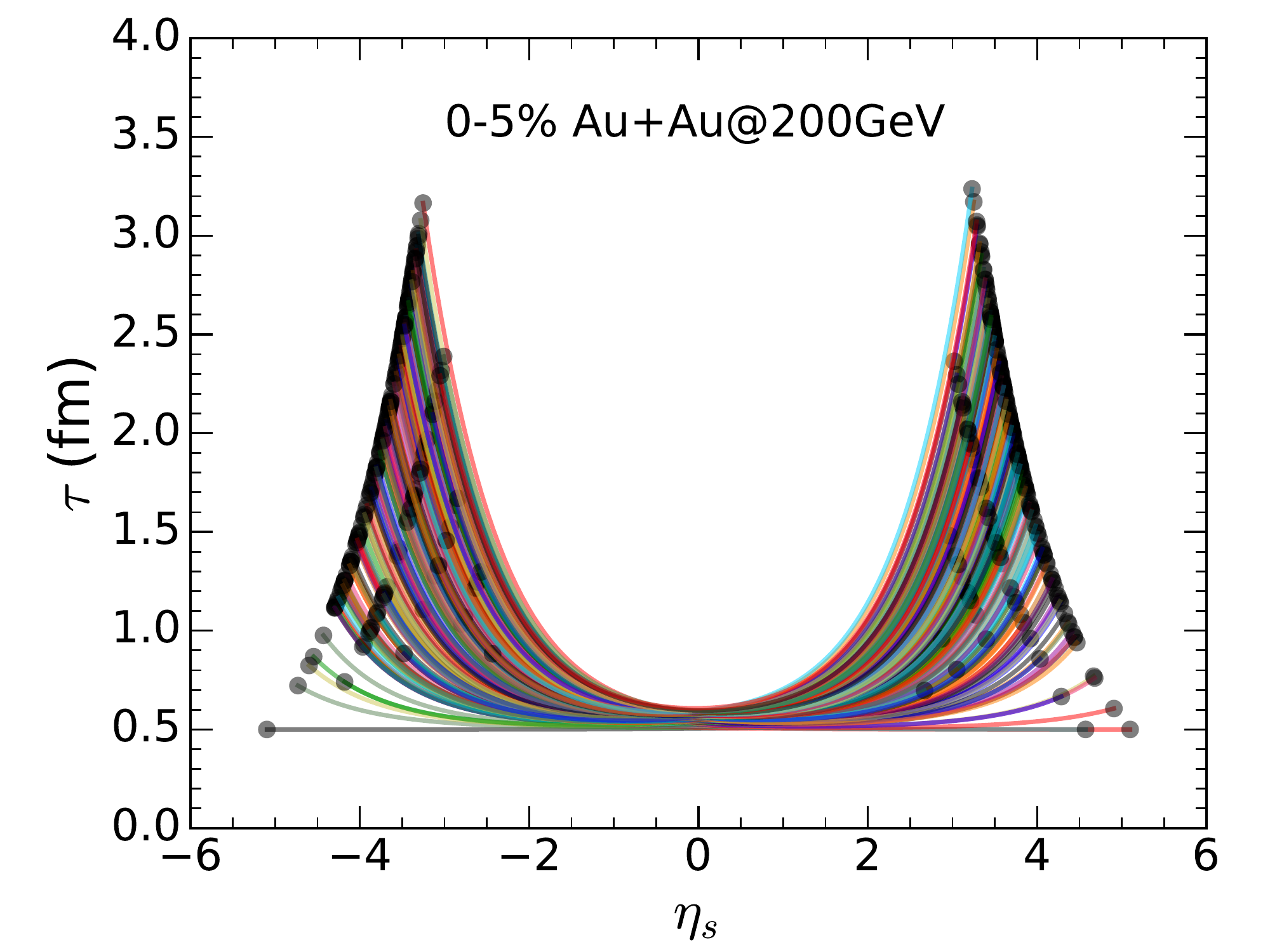}
\includegraphics[width=0.48\linewidth]{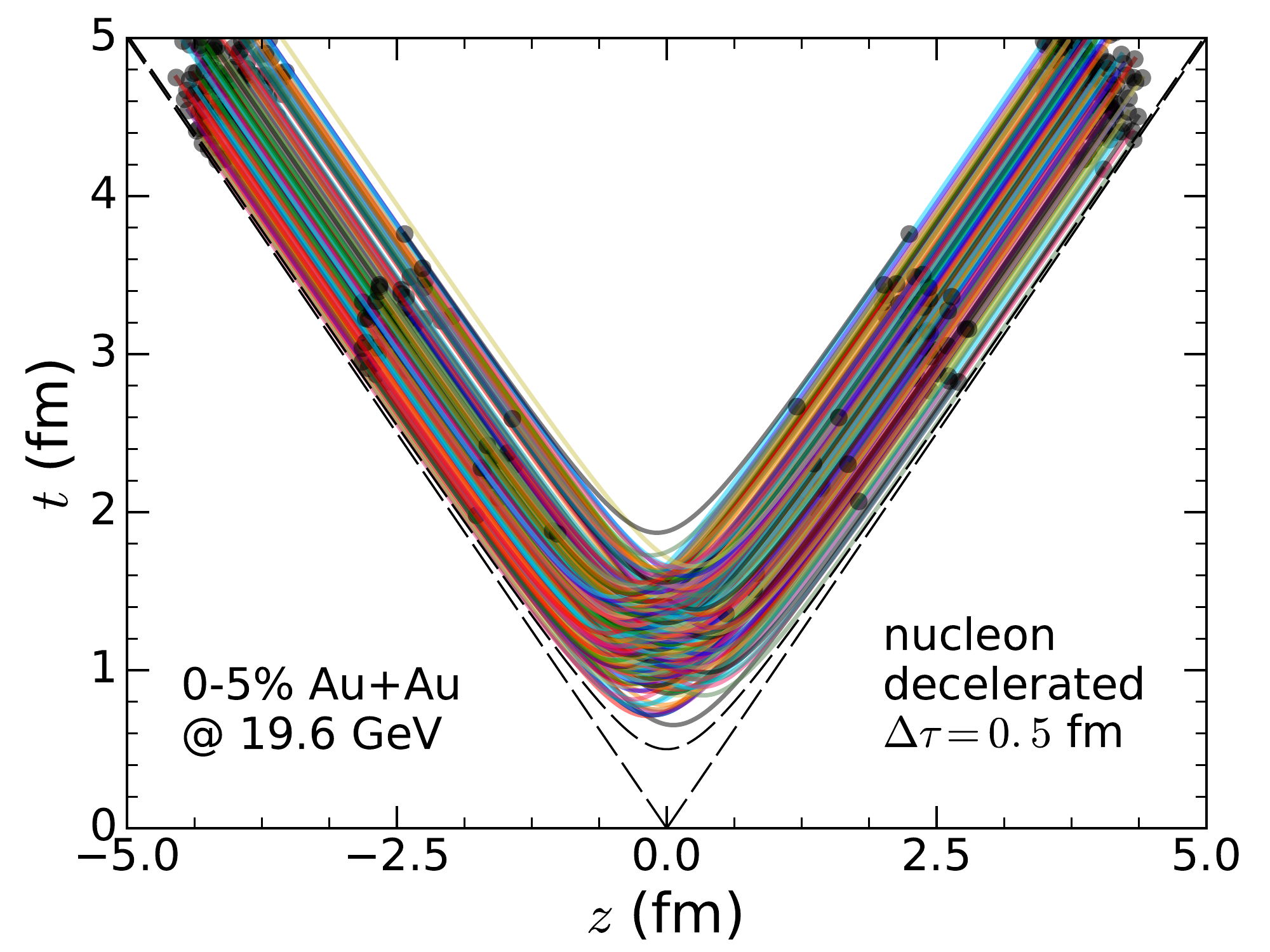}
\includegraphics[width=0.49\linewidth]{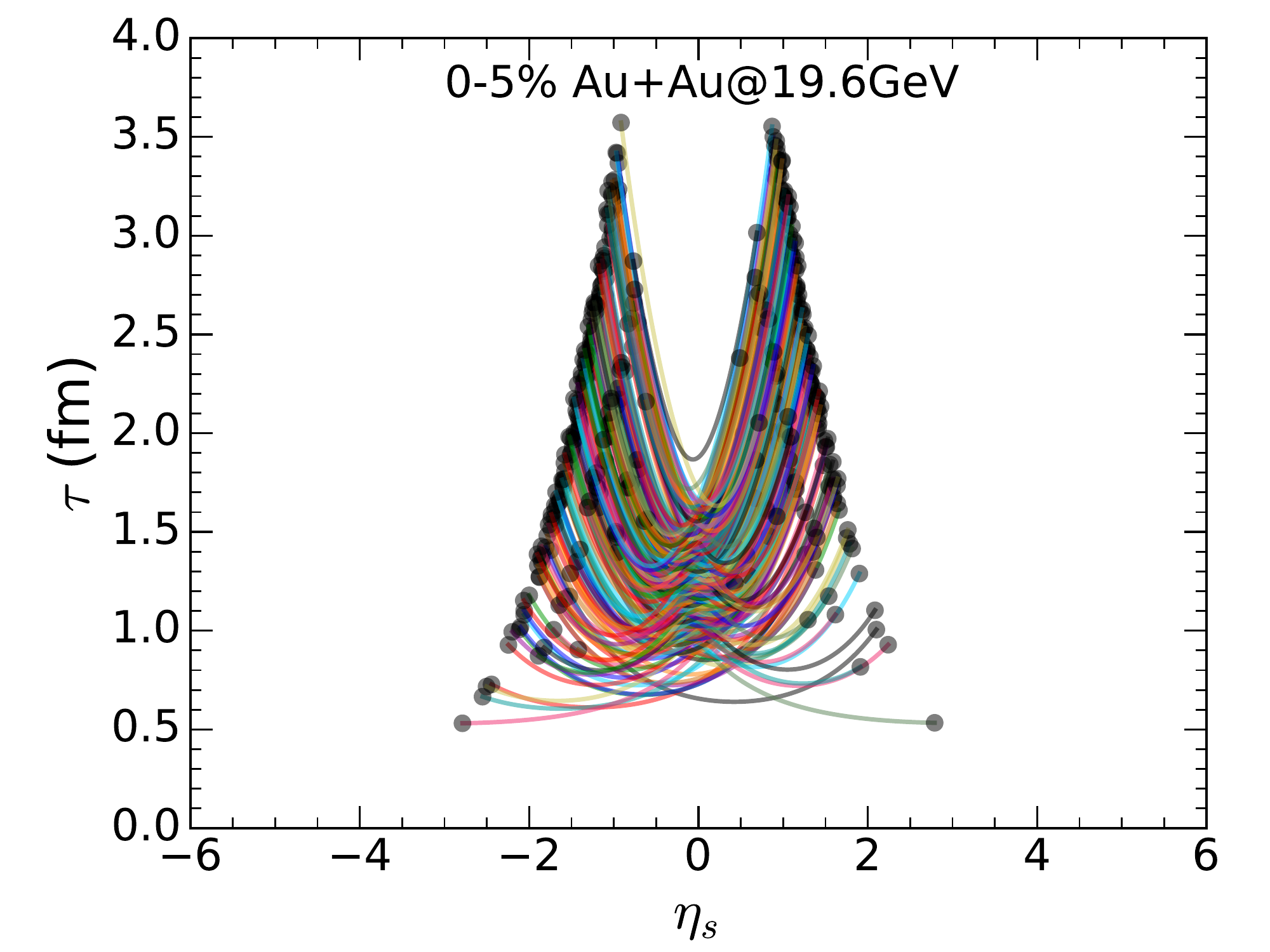}
\caption{The spacetime distribution of strings, which indicate where and when energy 
is deposited into the system. Upper panels show top RHIC energy, lower panels
$\sqrt{s}=19.6\,{\rm GeV}$ collisions. Left panels show $t-z$, right panels $\tau-\eta_s$ 
coordinates. Figure from \cite{Shen:2017bsr}.
\label{fig:initial-dynamic-strings}} 
\end{figure}

In Fig.~\ref{fig:initial-dynamic-strings} we illustrate the space time distribution 
of the sources that enter the hydrodynamic calculation, by showing the distributions 
of strings in the $t-z$ (left panels) or $\tau-\eta_s$ (right panels) plane for
$\sqrt{s}=200\,{\rm GeV}$ (upper panels) and  $\sqrt{s}=19.6\,{\rm GeV}$ (lower panels) 
collisions. While at the higher energy strings are distributed close to what looks 
like a constant $\tau$ surface in the $t-z$ plane, there is a large spread in the $t$
direction for the lower energy collision. Studying the distribution in the $\tau-\eta_s$
plane reveals that while around midrapidity the high energy result is indeed well
approximated by the assumption of a fixed initial $\tau$, at large space time rapidity 
even the high energy collision shows a significant spread in the $\tau$ direction. This
indicates that dynamic sources are relevant for all collision systems, if one is interested
in the physics beyond midrapidity. For low collision energies, there is no way around the
fact that the initial energy deposition takes a significant amount of time, even at
midrapidity. 

Coupling the new dynamic initial state to the hydrodynamic simulation \textsc{music} via
sources as described above and in \cite{Shen:2017bsr}, which in turn is coupled to UrQMD,
which performs hadronic rescattering, we can obtain final particle spectra differential in
rapidity and transverse momentum.  

\begin{figure}[ht!]
\centering
\includegraphics[width=0.6\linewidth]{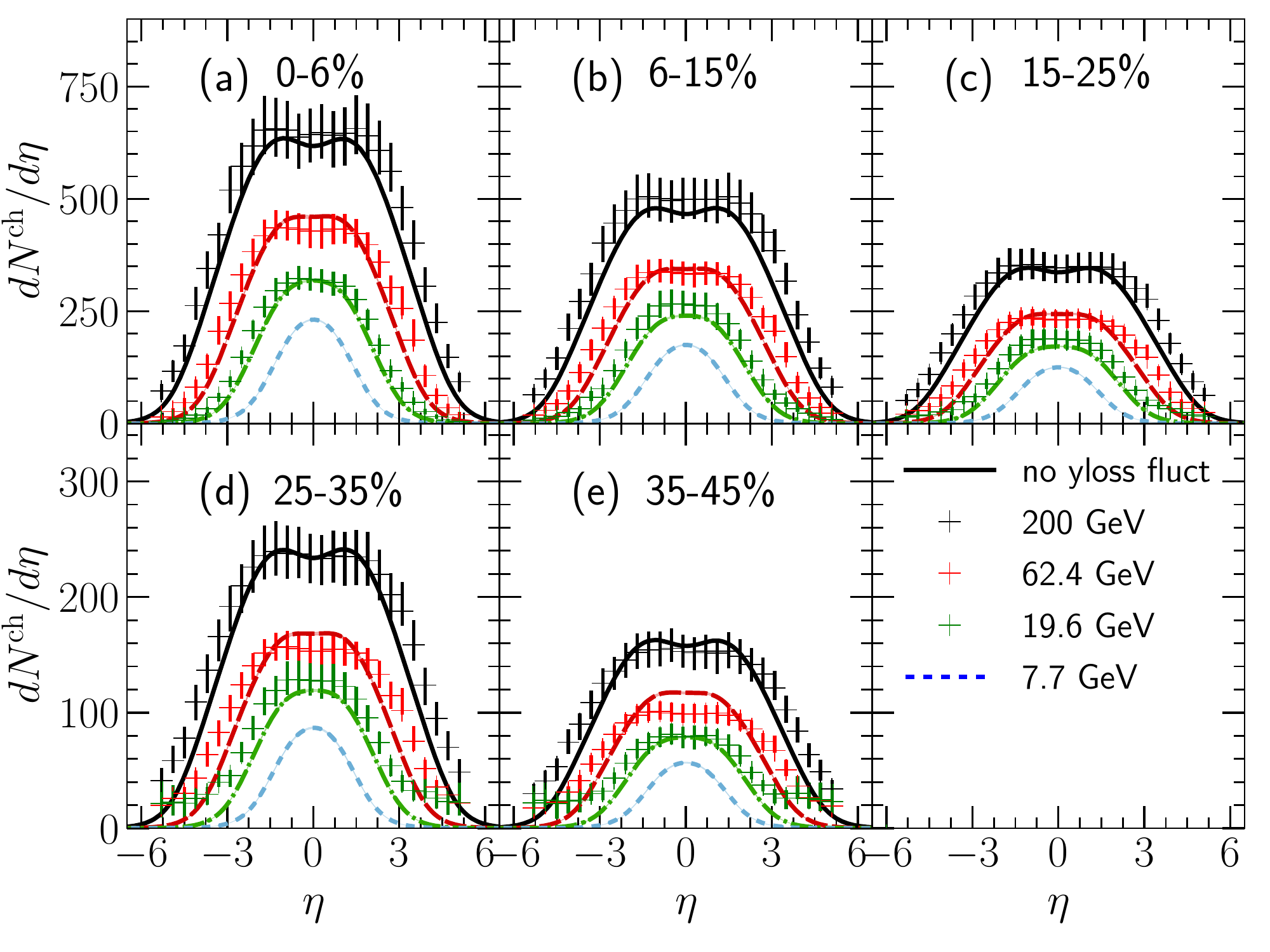}
\caption{The centrality and collision energy dependence of charged hadron pseudo-rapidity
distributions \cite{Shen:2021nbe, Shen:2021Preparation} compared with experimental data from the PHOBOS Collaboration 
\cite{Back:2005hs}.}
\label{fig:BES_dNchdeta}
\end{figure}

\begin{figure}[ht!]
\centering
\includegraphics[width=0.6\linewidth]{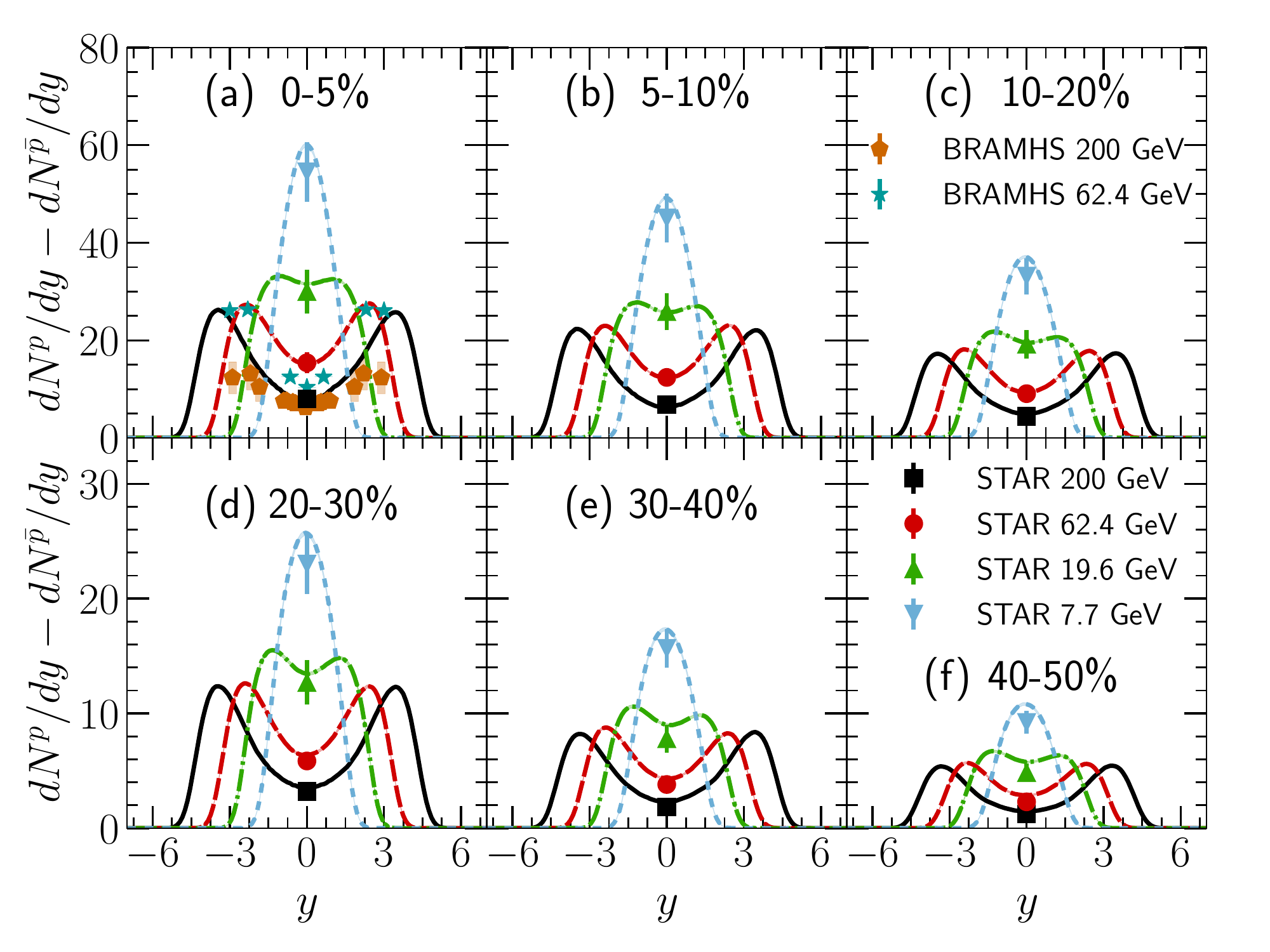}
\caption{The rapidity distributions of net protons for 6 centrality bins 
and different collision energies at RHIC \cite{Shen:2021nbe, Shen:2021Preparation, STAR:2017sal, BRAHMS:2003wwg, STAR:2008med}.
}
\label{fig:BES_netProtondNdy}
\end{figure}

In Fig.~\ref{fig:BES_dNchdeta} we present results for transverse momentum integrated
charged hadron rapidity distributions at various collision energies in different centrality
classes in Au+Au collisions and compare to experimental data from the PHOBOS Collaboration
\cite{Back:2005hs}. The centrality and energy dependence is well described, yet most 
distributions are narrower than the experimental data. A fine tuning of the parameters and
study of the effect of fluctuations in the rapidity loss of the decelerating quarks is still
ongoing and could potentially lead to an improvement of the description of the experimental
data.

In Fig.~\ref{fig:BES_netProtondNdy}, we show net-proton rapidity distributions for
different centrality classes and different collision energies in Au+Au collisions, and
compare to experimental data from the BRAHMS \cite{BRAHMS:2003wwg} and STAR
\cite{STAR:2008med,STAR:2017sal} Collaborations. Except for most central events and 
the highest collision energies, experimental data are only available at midrapidity. Yet, 
the energy and centrality dependence of net-proton production is also well described, with
the rapidity dependence in the 0-5\% bin for 200 and 62.4\,GeV collisions also agreeing
rather well with the data. Unfortunately, rapidity distributions for net protons are not
available for all collision energies and centralities.

It is absolutely essential to describe the baryon stopping as realistically as possible 
when the goal is the extraction of critical fluctuations from net-proton cumulants. Thus,
the proper description of the average net-proton production over a wide range of energies
and centralities is a necessary condition for a model to fulfill. More constraints on the
model, particularly the baryon transport, can be obtained by comparing to experimental data
from asymmetric systems, in particular d+Au collisions, for which data are available at
different collision energies.

\section{Hydrodynamics}
\label{sec:hydro}

\subsection{Introduction and summary}

Prior to the BEST Collaboration, the collision energy dependence of radial and elliptic 
flow in heavy-ion collisions was investigated using boost-invariant simulations
\cite{Kestin:2008bh,Shen:2012vn}. At intermediate collision energies, the heavy-ion 
collisions  break the assumption of longitudinal boost-invariance. Such non-trivial 
longitudinal dynamics was first studied in a simplified 1+1D simulation \cite{Monnai:2012jc}. 
Later, the first 3D hydrodynamic + hadronic transport simulations with the UrQMD
transport-based initial conditions found that the effective specific shear viscosity 
increases as collision energy is lowered \cite{Karpenko:2015xea}. 

The viscous hydrodynamic treatment of heavy ion collision, which has been successfully 
applied to top RHIC and LHC energies, requires several essential extensions in order 
to be applicable for the energies relevant for the BES. At the lower 
energies the net-baryon density does not vanish and thus the theoretical framework needs 
to be able to propagate all the conserved currents, baryon number, strangeness, and electric
charge. At low collision energies $\mathcal{O}(10)$ GeV, the finite longitudinal extension of
the colliding nuclei has to be taken into account, which leads to a substantial overlapping
time $\tau_\mathrm{overlap} \sim 1 - 3$ fm/c during which the two nuclei pass through each
other. The pre-equilibrium dynamics during this overlapping time may play an important role 
in understanding baryon stopping and density fluctuations along the longitudinal direction of
collisions. In order to quantitatively model the dynamics of heavy-ion collision at the RHIC
Beam energy scan energies, the following ingredients are essential:

\begin{itemize}
\item Pre-equilibrium dynamics during the stage when the two colliding nuclei 
pass through each other.
    
\item An equation of state at finite baryon density based on lattice QCD calculations, 
combined with a critical point controlled by a set of adjustable parameters.  

\item Fluid dynamic equations for all conserved charges, including dissipative effects.
\end{itemize}

 We will describe progress on these issues below. For general reviews of hydrodynamic
modeling at RHIC and LHC, and for definitions of hydrodynamic observables we refer 
the reader to Ref. \cite{Heinz_2013,Braun_Munzinger_2016,romatschke2019,Shen:2020mgh}.

\subsection{Conventional relativistic hydrodynamics}

\subsubsection{Results from full 3D dynamical simulations}

Solving the equations of motion of hydrodynamics at intermediate and low collision energies
requires an equation of state (EoS), which describes the thermodynamic properties of nuclear
matter at finite baryon density. Current lattice QCD techniques cannot directly compute
such an EoS because of the sign problem \cite{Ratti:2018ksb}. However, at vanishing net
baryon density, or $\mu_B = 0$, higher-order susceptibilities have been computed by lattice
QCD \cite{Bazavov:2018mes}. These susceptibility coefficients were used to construct 
a nuclear matter EoS at finite baryon densities through a Taylor expansion
\cite{Monnai:2019hkn,Noronha-Hostler:2019ayj, Parotto:2018pwx, Monnai:2021kgu}. These EoS are reliable 
within the region where $\mu_B/T \lesssim 2$ in the phase diagram as shown in 
Sec.~\ref{sec:PD}.

\begin{figure*}[ht!]
\includegraphics[width=0.9\linewidth]{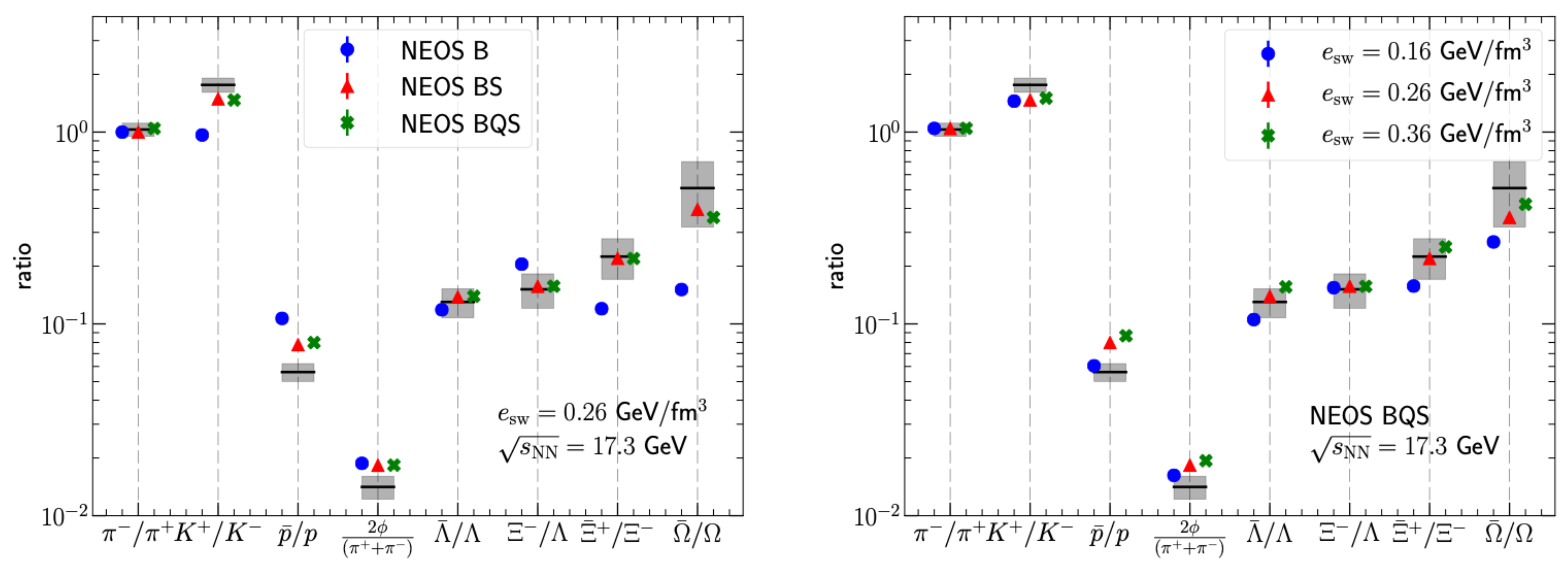}
\caption{(color online) Left Panel: Identified particle ratios for the three different 
equations of state from the same calculation compared to the NA49 experimental data (compiled in \cite{NA49compiled}). Right 
Panel: Identified particle ratios for the three different switching energy densities. The 
figure was taken from Ref.~\cite{Monnai:2019hkn}.}
\label{fig:hydro:pid_ratio}
\end{figure*}

Let us now discuss the phenomenological impacts of various model ingredients, such as
strangeness neutrality and presence of a critical point. Because the colliding nuclei do 
not carry any net strangeness, the strangeness density in nuclear collisions vanishes
on average, $n_s=0$. This condition leads to 
$\mu_s \sim \mu_B/3$ in the QGP phase \cite{Lee:1986mn, Lee:1992hn, Sollfrank:1995bn, Monnai:2021kgu}. Fig.~\ref{fig:hydro:pid_ratio}
shows the effect of the strangeness neutrality on identified particle yields at 17.3 GeV. 
The relative yields of  multi-strangeness baryons increase and agree well with the NA49
measurements once the strangeness neutrality condition is imposed. Further imposing $n_Q =
0.4 n_B$ introduces a small difference between the $\pi^+$ and $\pi^-$ yields. The right panel
shows the dependence of relative particle yields on the choice of switching energy density,
at which fluid cells are mapped to individual hadrons. A lower switching energy density
yields smaller ratio of anti-baryons to baryons. A switching energy density $e_\mathrm{sw} 
\sim 0.2$\,GeV/fm$^3$ is preferred by the NA49 measurements at the top SPS collision 
energy.

\begin{figure*}[ht!]
\begin{tabular}{cc}
\includegraphics[width=0.45\linewidth]{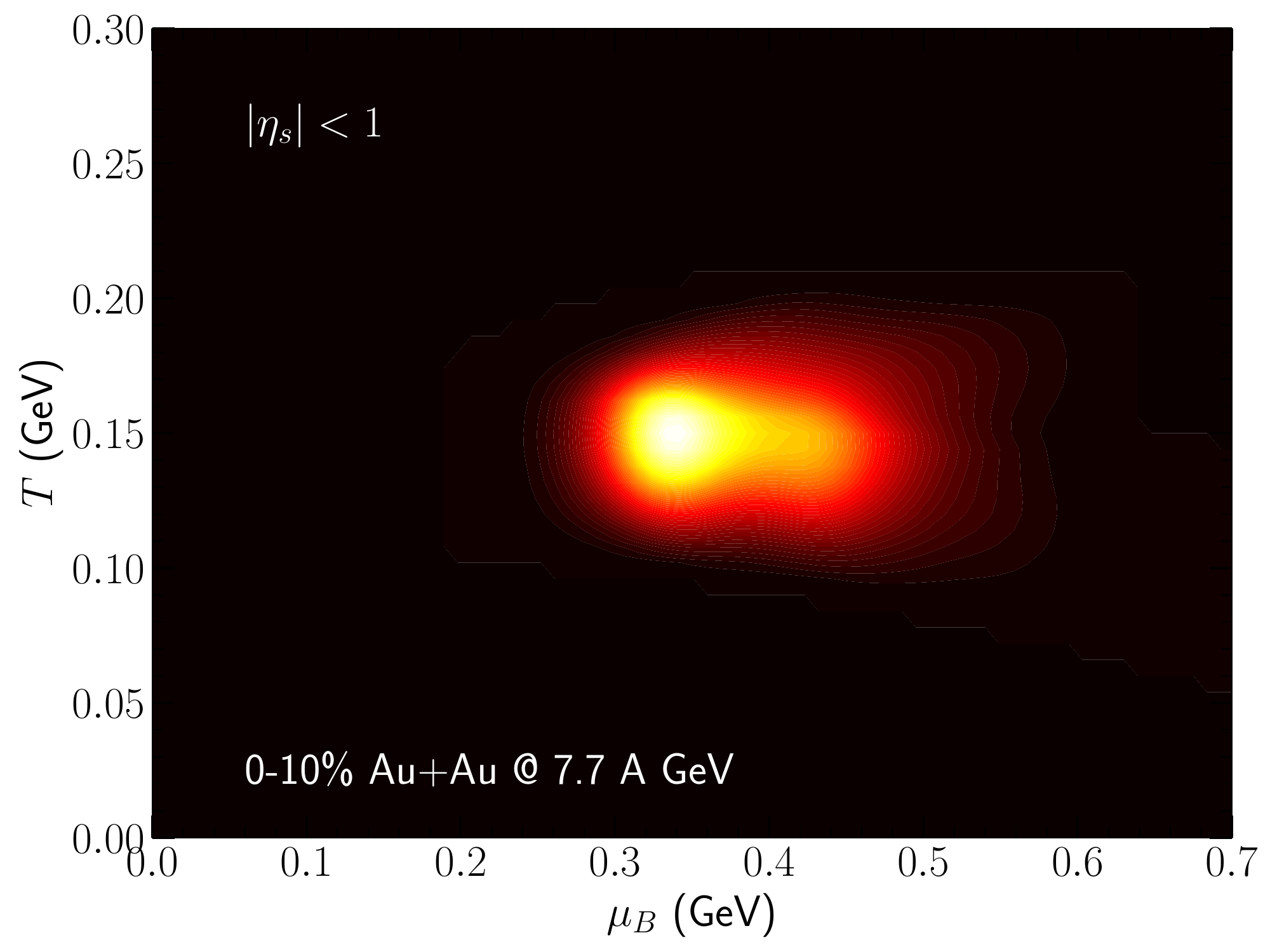} &
\includegraphics[width=0.45\linewidth]{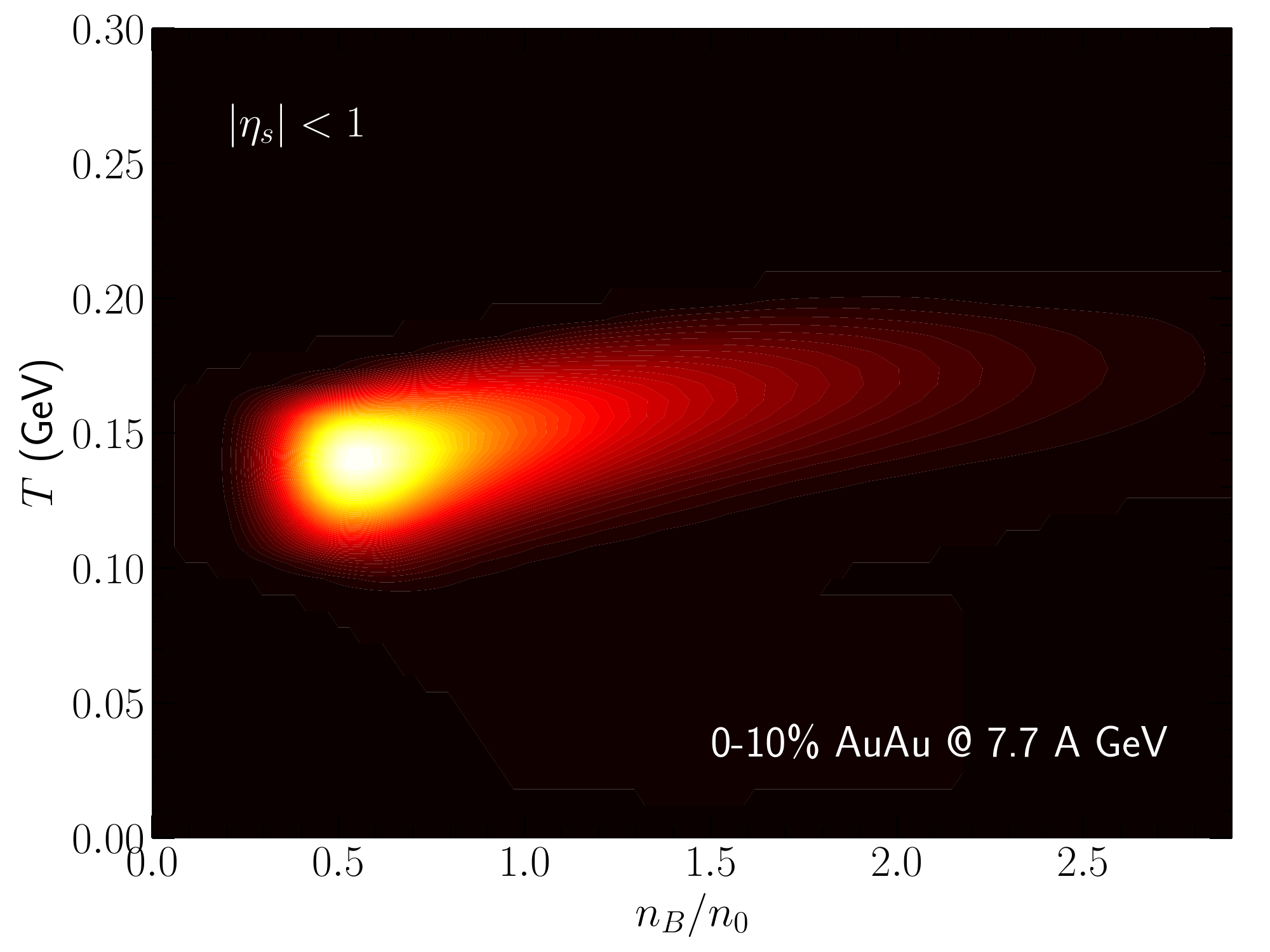}
\end{tabular}
\caption{(color online) Fireball trajectories in the $T-\mu_B$ phase 
diagram (left) and $T-(n_B/n_0)$ (right) for 0-10\% Au+Au collisions at 7.7 A GeV 
\cite{Shen:2020jwv}. The brightness is proportional to the space-time volume of the 
fireball.}
\label{fig:hydro:phase_diagram}
\end{figure*}

Once the dynamical simulations are calibrated with the particle production measurements, 
they can provide a realistic and detailed space-time evolution of the relativistic 
heavy-ion collisions. Fig.~\ref{fig:hydro:phase_diagram} shows the trajectories of an
averaged 0-10\% Au+Au collision at $\sqrt{s_\mathrm{NN}} = 7.7$\,GeV in the QCD phase
diagrams. These trajectories were analyzed from the hybrid simulations performed in
Ref.~\cite{Shen:2020jwv}, which was calibrated to reproduce the measured net proton yield 
at midrapidity. The 3D dynamical framework allows us to map individual heavy-ion collisions
to the QCD phase diagram. At $\sqrt{s_\mathrm{NN}}=7.7$ GeV, in the central rapidity region $|\eta_s|<1$,
most of the fireball explores regions with $T \in [0.1,0.2]$ GeV and $\mu_B \in [0.25,0.5]$ GeV.
The right panel shows the phase diagram as a function 
of the ratio of local net baryon density to the normal nuclear saturation density $n_B/n_0$
with $n_0 = 0.17$ (1/fm$^3$). At 7.7 GeV collision energy, the majority of the fluid cells 
in the fireball reach about half of the nuclear saturation density, $n_B \sim 0.1$\,fm$^{-3}$.
During the early time of the evolution, the value of net baryon density can 
reach up to 0.5 fm$^{-3}$, about three times normal nuclear density. 

\begin{figure*}[ht!]
\begin{tabular}{cc}
\includegraphics[width=0.6\linewidth]{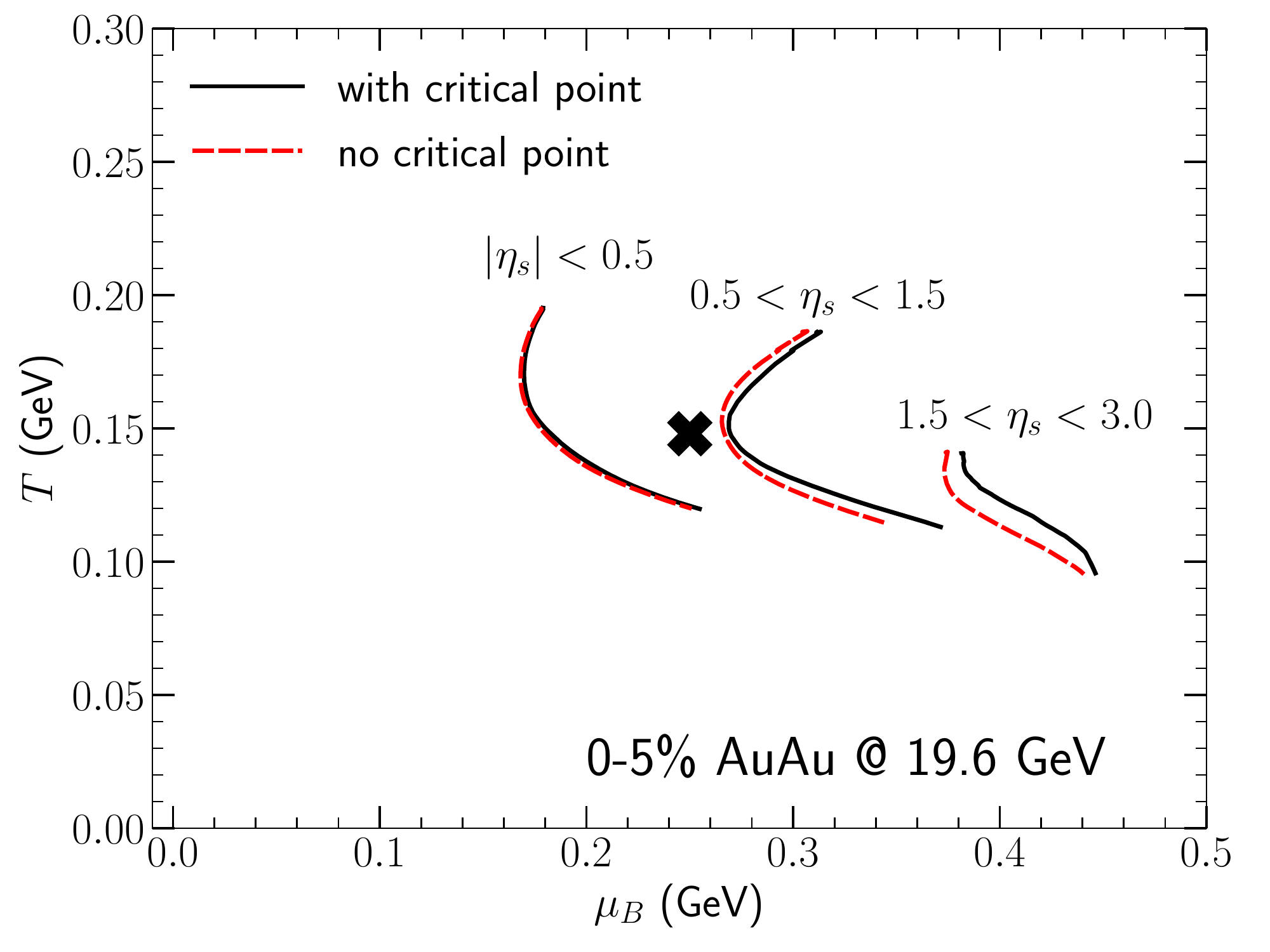}
\end{tabular}
\caption{(color online) The effect of the presence of a critical point on the averaged 
fireball trajectories in three space-time intervals in 0-5\% Au+Au collisions at 19.6 GeV. 
The critical point is located at $T_c = 148$ MeV and $\mu_{B,c} = 250$ MeV, as indicated 
by the cross. }
\label{fig:hydro:phase_diagram_BEST_EoS}
\end{figure*}

In Fig.~\ref{fig:hydro:phase_diagram_BEST_EoS}, we explore the effects of the presence 
of a critical point in the equation of state \cite{Parotto:2018pwx} on the averaged 
fireball trajectories. The critical point and the line of first-order phase transitions
emerging from it for $\mu_B > \mu_{B,c}$ distort the adiabatic ($s/n_B=$const) 
expansion trajectories \cite{Cho:1993mv, Hung:1997du, Parotto:2018pwx}. We consider 
a critical point at $T_c = 148$ MeV and $\mu_{B,c} = 250$ MeV and study how it
influences the time evolution and the final state observables of a heavy-ion 
collisions at 19.6 GeV. Fig.~\ref{fig:hydro:phase_diagram_BEST_EoS} shows 
that the fireball trajectories shift to slightly larger $\mu_B$ values compared to 
whose from the simulations without the critical point.
This effect is consistent with the effect of critical point on constant $s/n_B$ trajectories shown in Ref.~\cite{Parotto:2018pwx}.
The effect is larger at forward rapidities, where the fireball crosses the first-order phase transition. 
The two phases are connected by a Maxwell construction \cite{Parotto:2018pwx}.
This means that we are not trying to simulate nucleation or spinodal decomposition.
Because the fireball trajectories are averaged over a distribution of $s/n_B$ values from individual fluid cells, the trajectory discontinuities at phase transition boundary are smeared. 
We find that the two EoS with and without critical point result in very similar final
particle spectra and flow observables, indicating that these observables have limited
sensitivity to the existence and location of a critical point. But note that the 
implementation of a first-order phase transition at large $\mu_B$ into the dynamical 
simulations has been somewhat rudimentary so far and requires further study.

\begin{figure*}[ht!]
\begin{tabular}{cc}
\includegraphics[width=0.45\linewidth]{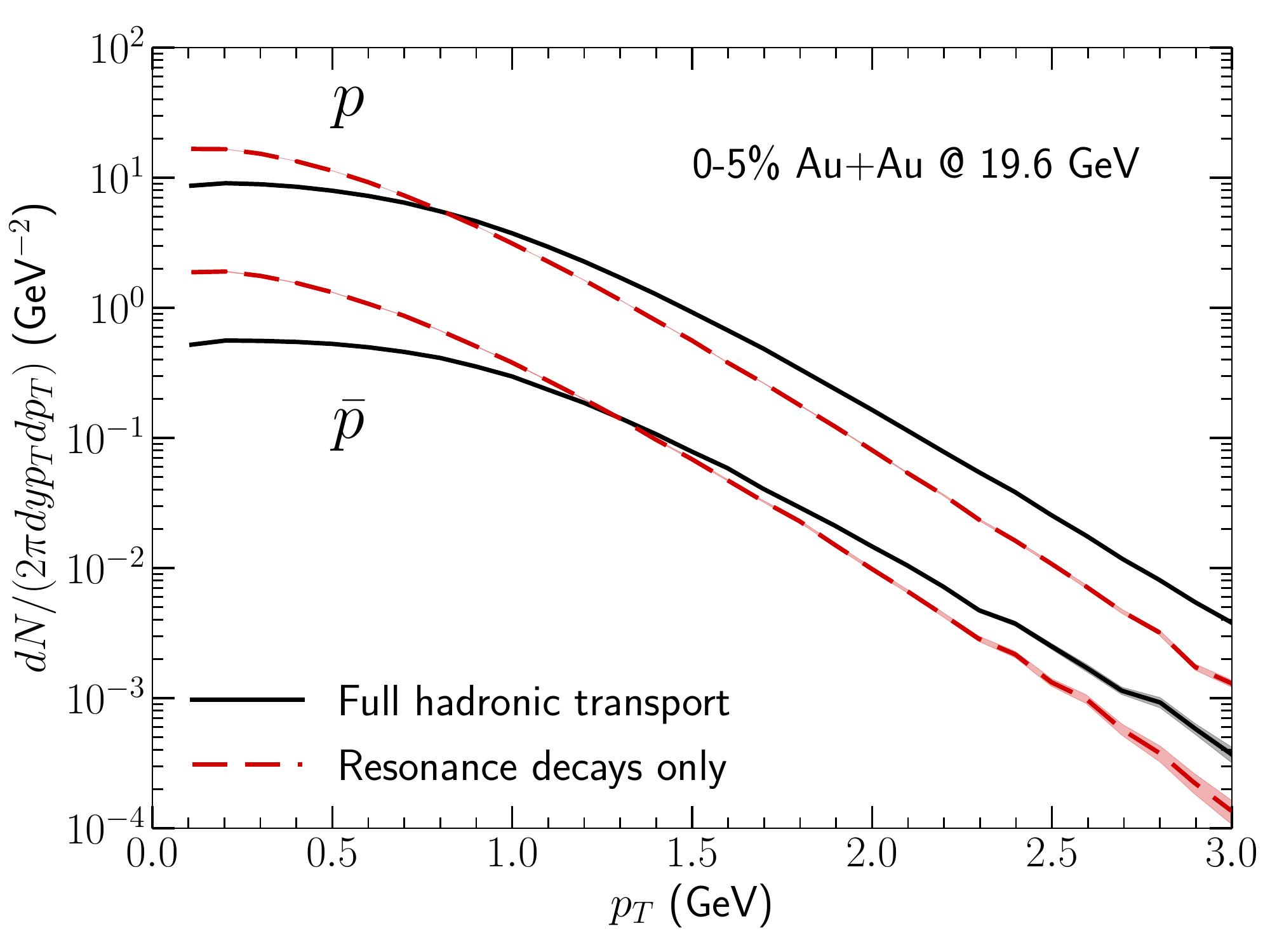} &
\includegraphics[width=0.45\linewidth]{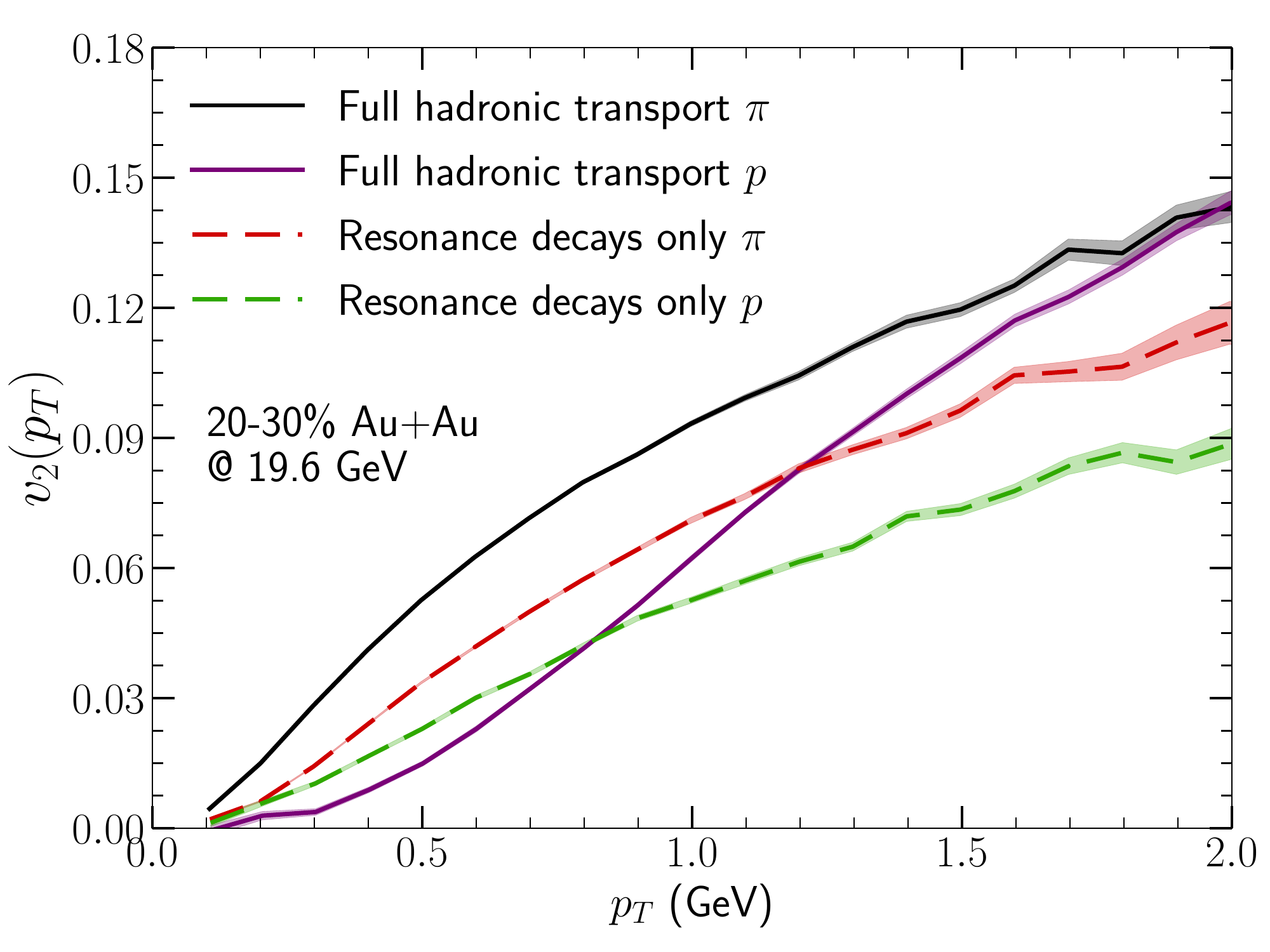}
\end{tabular}
\caption{(color online) Left Panel: The effects of hadronic transport on the transverse 
momentum spectra of protons and anti-protons for 0-5\% Au+Au collisions at 
$\sqrt{s_\mathrm{NN}} = 19.6$\,A\,GeV. Right Panel: The effects of hadronic transport 
evolution on pion and proton $p_T$-differential $v_2(p_T)$ in 20-30\% Au+Au collisions 
at $\sqrt{s_\mathrm{NN}} = 19.6$\,A\,GeV. Results from Ref.~\cite{Denicol:2018wdp}.}
\label{fig:hydro:afterburner}
\end{figure*} 

As the collision energy decreases, the hadronic dynamics becomes more and more important
in describing the dynamics of relativistic heavy-ion collisions. 
Fig.~\ref{fig:hydro:afterburner} demonstrates the effects of hadronic scattering on 
the shape of $p_T$-spectra and $v_2(p_T)$ of identified particles at $\sqrt{s_\mathrm{NN}} 
= 19.6$\,GeV. Hadronic interactions flatten both proton and anti-proton spectra because 
of scatterings with fast moving pions, known as the ``pion wind''. The baryon rich
environment at 19.6 GeV also results in a more prominent annihilation of anti-protons
compared to what is observed at 200 GeV \cite{Song:2011hk, Monnai:2019hkn}. 
The right panel shows that the elliptic flow coefficients of pions and protons continue
to increase in the hadronic phase. The remaining spatial eccentricity continues to generate
momentum anisotropy of particles in the hadronic transport phase, increasing the elliptic
flow of pions and protons at high $p_T$. The low $p_T$ protons' $v_2$ receives a blue shift
from the ``pion wind'', which increases the splitting between pion and proton $v_2$ during
the hadronic evolution. 

\subsubsection{Baryon diffusion}

At intermediate collision energies, the non-vanishing net baryon density $n_B$ forms a conserved particle number current in the hydrodynamic evolution,
\begin{equation}
    \partial_\mu j_B^\mu = 0 \qquad \mbox{with} \qquad j^\mu_B = n_B u^\mu + q_B^\mu.
    \label{eq:hydro:baryondiffusion}
\end{equation}
Similar to the energy-momentum tensor, the evolution of net baryon current involves
dissipative effects, controlled by the net baryon diffusion current $q_B^\mu$. Note 
that in the presence of a conserved baryon charge there are different frames that can
be used to define the fluid velocity $u^\mu$. In the Landau frame the fluid velocity 
is defined by the condition that $T_{\nu\mu}u^\mu=e u_\nu$, so that there is no 
dissipative contribution to the energy flux. The baryon diffusion current $q_B^\mu$
then characterizes baryon diffusion relative to the energy current.
In the Navier-Stokes limit, this diffusion current $q_B^\mu$ is proportional to local gradients of the ratio 
of net baryon chemical potential over temperature, $q_B^\mu \propto \kappa_B \nabla^\mu
(\frac{\mu_B}{T})$, where $\kappa_B$ is the baryon diffusion constant \cite{Denicol:2012cn, Denicol:2018wdp}.

\begin{figure}[ht!]
\begin{tabular}{cc}
\includegraphics[width=0.45\linewidth]{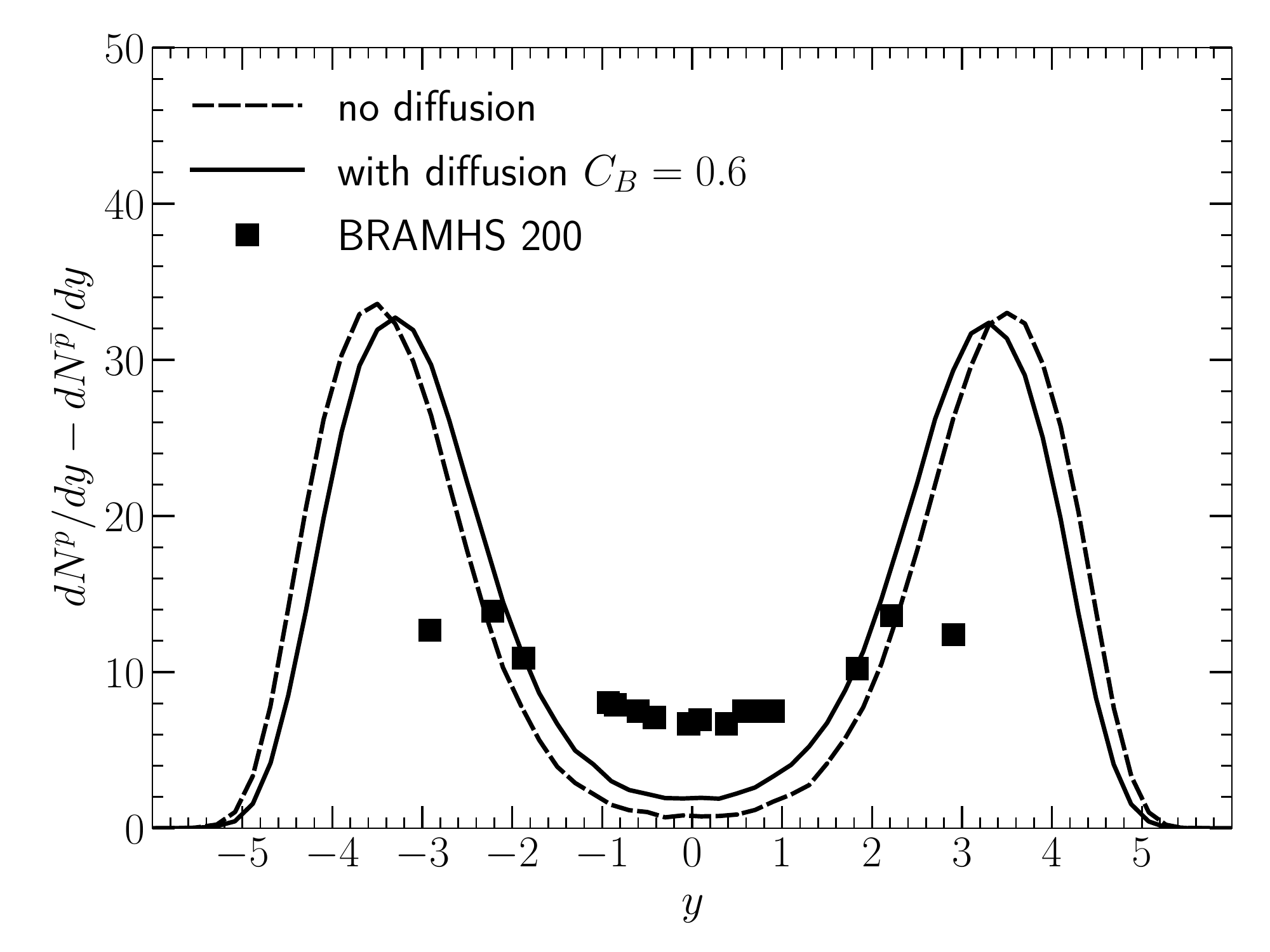} &
\includegraphics[width=0.45\linewidth]{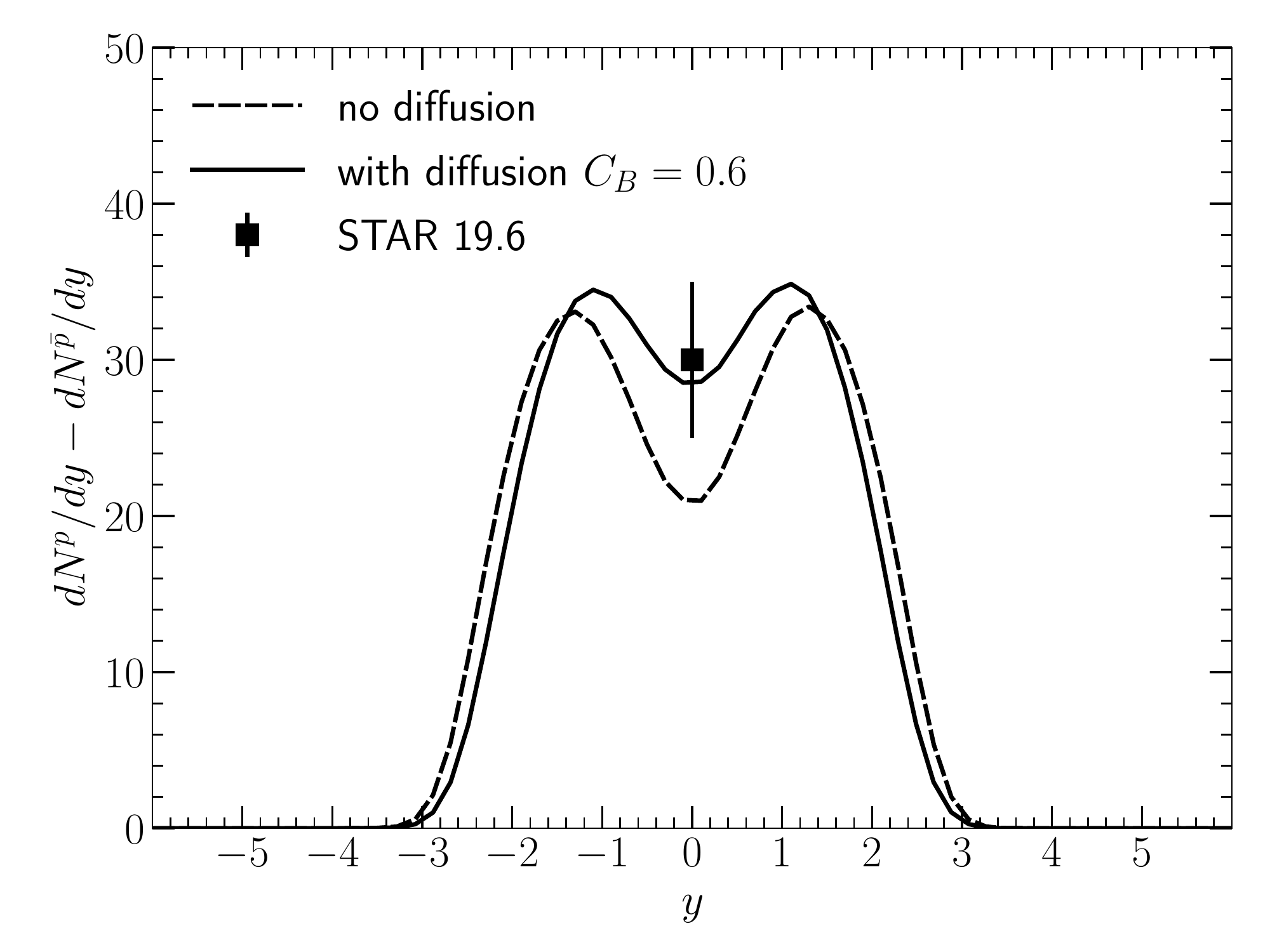}
\end{tabular}
\caption{The effects of baryon diffusion on net proton rapidity distributions. Results 
from Ref.~\cite{Denicol:2018wdp}. The experimental data are from \cite{STAR:2017sal, BRAHMS:2003wwg}.
}
\label{fig:hydro:baryonDiff}
\end{figure}

The effects of net baryon diffusion on phenomenological observables were systematically
studied in Ref.~\cite{Denicol:2018wdp, Du:2021zqz}. The $p_T$-differential observables for proton 
and anti-protons show strong sensitivity to the out-of-equilibrium corrections at
particlization. Fig.~\ref{fig:hydro:baryonDiff} highlights the most prominent 
effect of baryon diffusion currents, which is a change in the rapidity distribution 
of net protons in heavy-ion collisions. A non-zero diffusion constant $\kappa_B$, proportional 
to the parameter $C_B$ in Fig.~\ref{fig:hydro:baryonDiff}, causes a shift of net baryon
number from forward rapidities to the mid-rapidity region. However, the effect of baryon
diffusion during the hydrodynamic evolution alone is not enough to transport enough 
baryon charges to mid-rapidity in Au+Au collisions at 200 GeV. Therefore, the BRAMHS
measurements suggested that there must be substantial baryon stopping during the early
pre-equilibrium evolution of the relativistic heavy-ion collisions at 200 GeV. To extract 
the net baryon diffusion coefficient from the experimental measurements, we need to
disentangle the initial state baryon stopping from the baryon diffusion during the
hydrodynamic phase. Additional experimental observables, such as the charge balance
functions, may help us to set a better constraint on unknown QGP transport coefficient 
in a baryon rich environment. 

\subsubsection{Code validation for different hydrodynamic frameworks}

Relativistic hydrodynamic simulations in full (3+1)D require developing large-scale 
numerical code packages. It is essential to have open-source code packages and standardized
benchmark tests among different implementations.  

\begin{figure*}[ht!]
\includegraphics[width=0.9\linewidth]{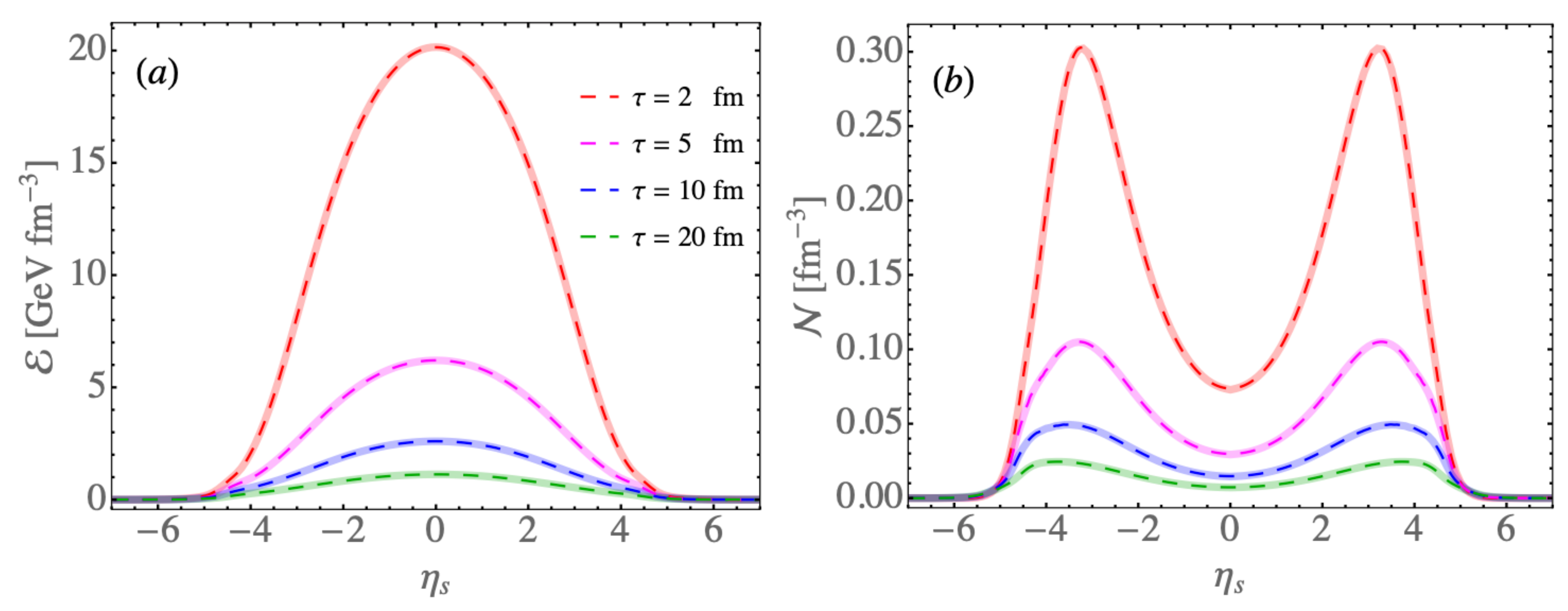}
\caption{(color online) Comparison between the numerical results from BEShydro (dashed 
lines) \cite{Du:2019obx} and the MUSIC simulations \cite{Denicol:2018wdp} of the (1+1)D 
setup described in Ref.~\cite{Monnai:2012jc} (continuous lines): (a) energy density; 
(b) net baryon density. The figure was taken from Ref.~\cite{Du:2019obx}.}
\label{fig:hydro:codecomp}
\end{figure*} 

Within the BEST Collaboration, we performed numerical code validation among two independent
implementations of (3+1)D hydrodynamic simulations with the propagation of net baryon 
current and its diffusion. Fig.~\ref{fig:hydro:codecomp} highlights such a code validation
between MUSIC and BEShydro for the propagation of net baryon current with diffusion in a
simplified 1+1D longitudinal expansion \cite{Du:2019obx}. The two numerical code packages independently
implemented the equations of motion for hydrodynamic fields. The results agree with each
other very well. A variety of additional code validation protocols for (3+1)-dimensional dissipative hydrodynamic codes are described in \cite{Du:2019obx}.

\subsection{Anomalous hydrodynamics}
\label{sec:hydro_anomal}

\subsubsection{Phenomenological simulations of CME}

Before the establishment of the BEST Collaboration, the CME and CMW signals in heavy-ion
collisions have been investigated using ideal chiral 
hydrodynamics~\cite{Hongo:2013cqa,Hirono:2014oda} which evolves non-dissipative 
chiral currents on top a of viscous hydrodynamic background~\cite{Yee:2013cya,Yin:2015fca}. 
A next step towards a more self-consistent treatment of anomalous transport must take 
into account the non-equilibrium correction to both the bulk background and the vector 
and axial vector currents. This is achieved by the Anomalous-Viscous Fluid Dynamics 
(AVFD) simulation package~\cite{Jiang:2016wve,Shi:2017cpu} which solves the evolution of
vector and axial current, including dissipation effect, as linear perturbation on top 
of the viscous hydrodynamic background.

In heavy-ion collision experiments, the CME-induced charge separation is measured by the
charge-sensitive two-particle correlators, known as $\gamma$ and $\delta$. They are 
defined as
\begin{align}
\gamma^{\alpha\beta} \equiv&\; 
    \langle \cos(\phi_\alpha + \phi_\beta - 2\Psi_\text{RP}) \rangle \, ,\\
\delta^{\alpha\beta} \equiv&\; 
     \langle \cos(\phi_\alpha - \phi_\beta) \rangle \, , 
\end{align}
where $\Psi_\text{RP}$ is the reaction plane of the collisions. The indices $\alpha$ 
and $\beta$ label the electric charge, $\alpha,\beta=\pm$. To highlight the CME signal,
experimentalists further consider the difference between opposite-sign and same-sign
correlators, $\Delta\gamma\equiv \gamma^{+-} - (\gamma^{++}+\gamma^{--})/2$ and similarly 
for $\Delta\delta$. Such correlators measure the fluctuation of the charge separation 
vector, and contain not only the CME signal, but also the non-CME background. 
The major sources of such non-CME background come from the effect of Global Momentum 
Conservation (GMC) and Local Charge Conservation (LCC). The GMC and LCC effects imply 
non-vanishing multi-particle correlations, which makes it non-trivial to implement these 
effects in the numerical simulation of freeze-out process.

\begin{figure}[!hbt]
\includegraphics[width=0.45\textwidth]{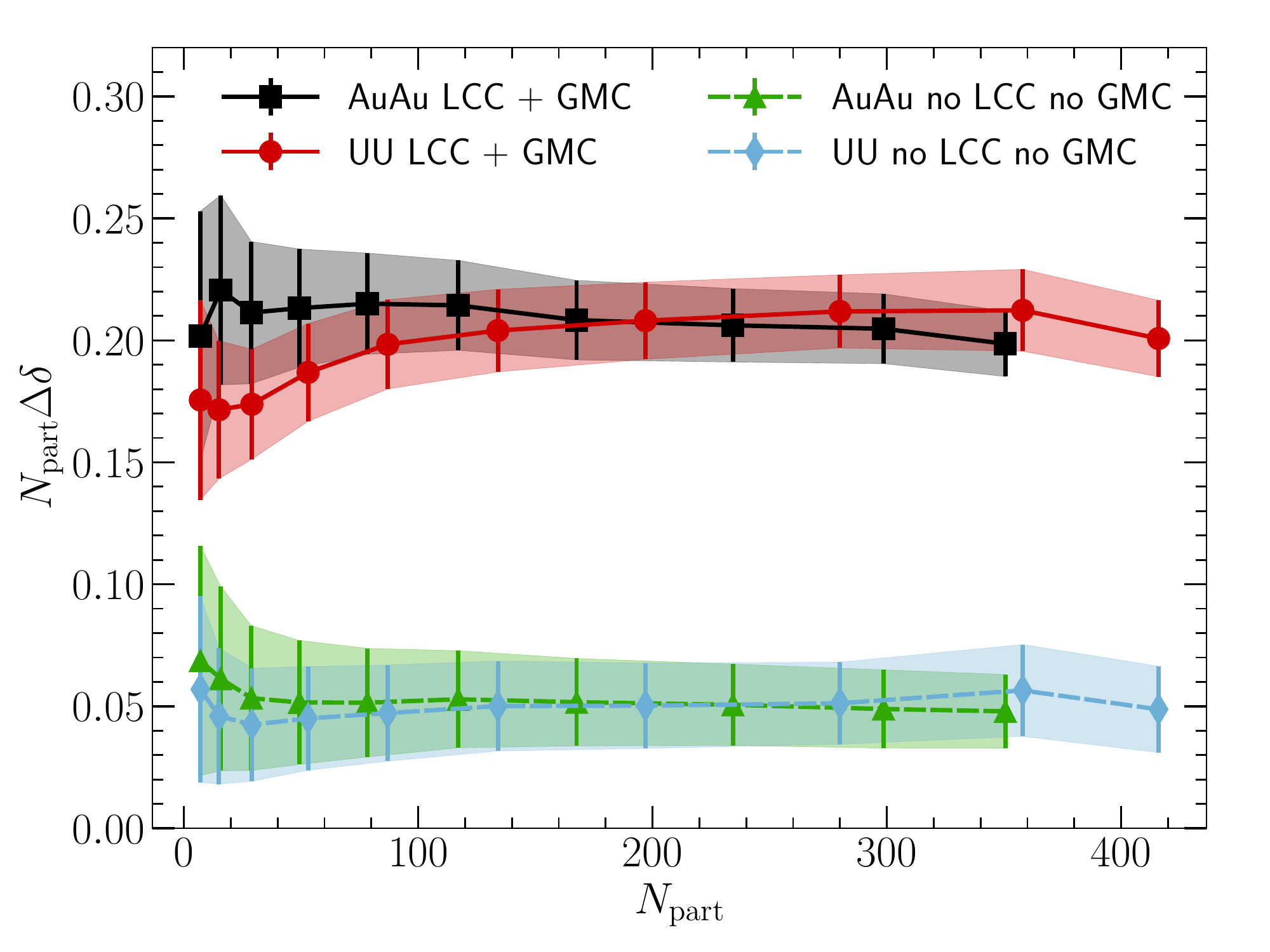}
\includegraphics[width=0.45\textwidth]{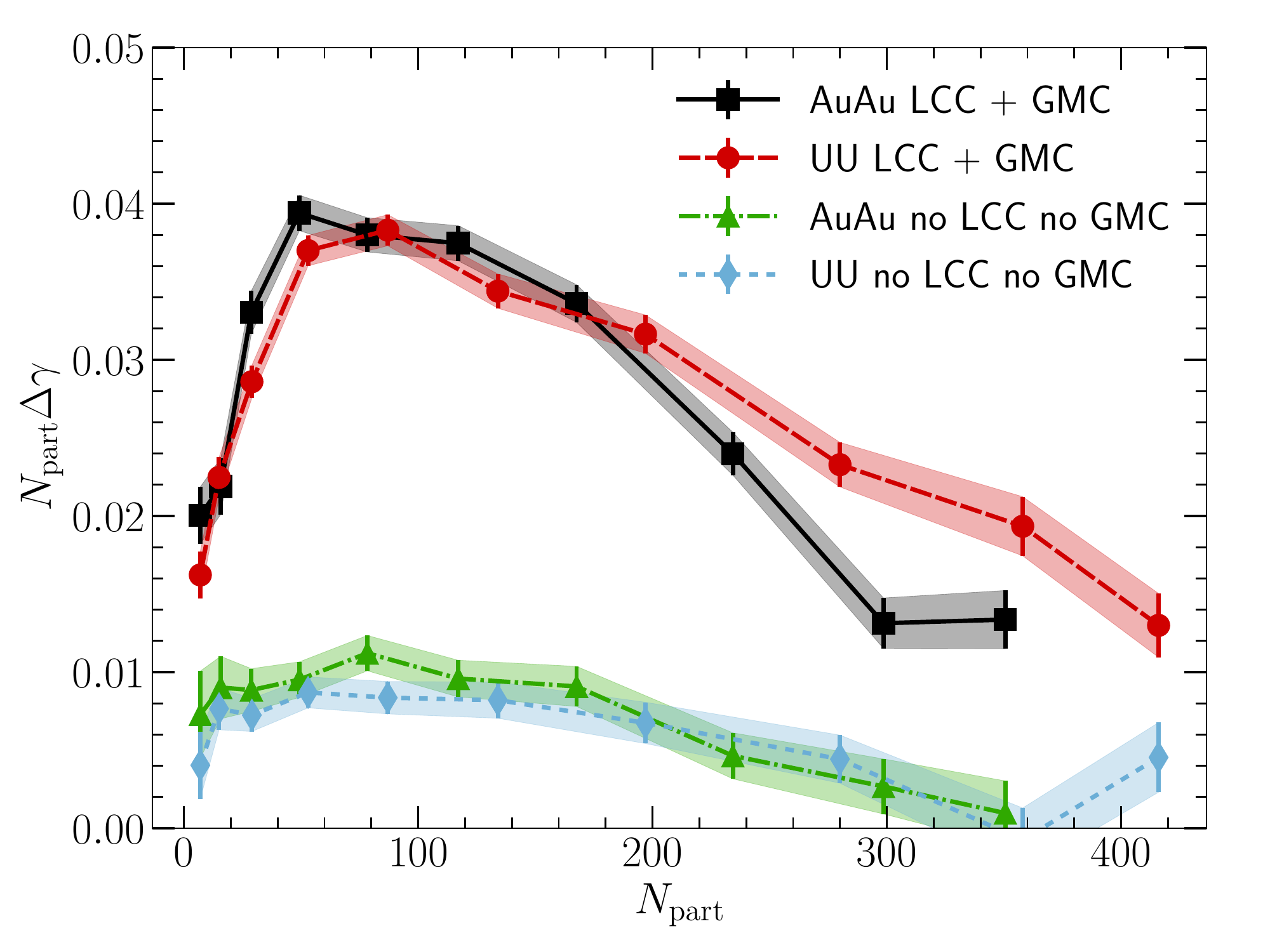}
\caption{(color online, reproduced from \protect{\cite{Schenke:2019ruo}} with permission) 
The difference between opposite-sign and same-sign $\delta$ (left) and $\gamma$ (right) 
correlation functions scaled by number of participants $N_\mathrm{part}$ in Au+Au and 
U+U collisions. Results with and without imposing local charge conservation (LCC) and 
global momentum conservation (GMC) are shown.
\label{fig.sec4_LCC}
}
\end{figure}

In~\cite{Schenke:2019ruo}, both the GMC and LCC effects are implemented in the freeze-out 
process using the numerical implementation first proposed in~\cite{Bozek:2012en}. In this 
work charged hadron-antihadron pairs are chosen to be produced in the same fluid cell, 
while their momenta are sampled independently in the local rest frame of the fluid cell. 
This procedure implicitly assumes the correlation length to be smaller than the size of 
the cell, hence it provides an upper limit for the correlations between opposite sign 
pairs. In addition, the GMC is imposed by adjusting the momentum of final state hadrons. 
As shown in Fig.~\ref{fig.sec4_LCC}, the LCC effect increases the $\Delta \gamma$ and 
$\Delta \delta$ correlators, compared to the case with only resonance decay. Meanwhile, 
GMC changes the absolute value of same-sign and opposite-sign correlators, but has 
negligible influence on the difference between them. Since then a more sophisticated 
prescription of particlization was developed by the BEST Collaboration
\cite{Oliinychenko:2019zfk,Oliinychenko:2020cmr}, which allows for a more realistic 
estimate of GMC and LCC effects on the $\gamma$ and $\delta$ correlators. As discussed 
in detail in Sec.~\ref{sec:transport}, this new particlization method employs the Markov 
Chain Monte-Carlo algorithm to sample hadrons according to the desired distribution, 
respecting the conservation of energy, momentum, baryon number, electric charge, and 
strangeness, within a localized patch of fluid cells on the freeze-out surface. 

\begin{figure}[!hbt]
\includegraphics[width=0.45\textwidth]{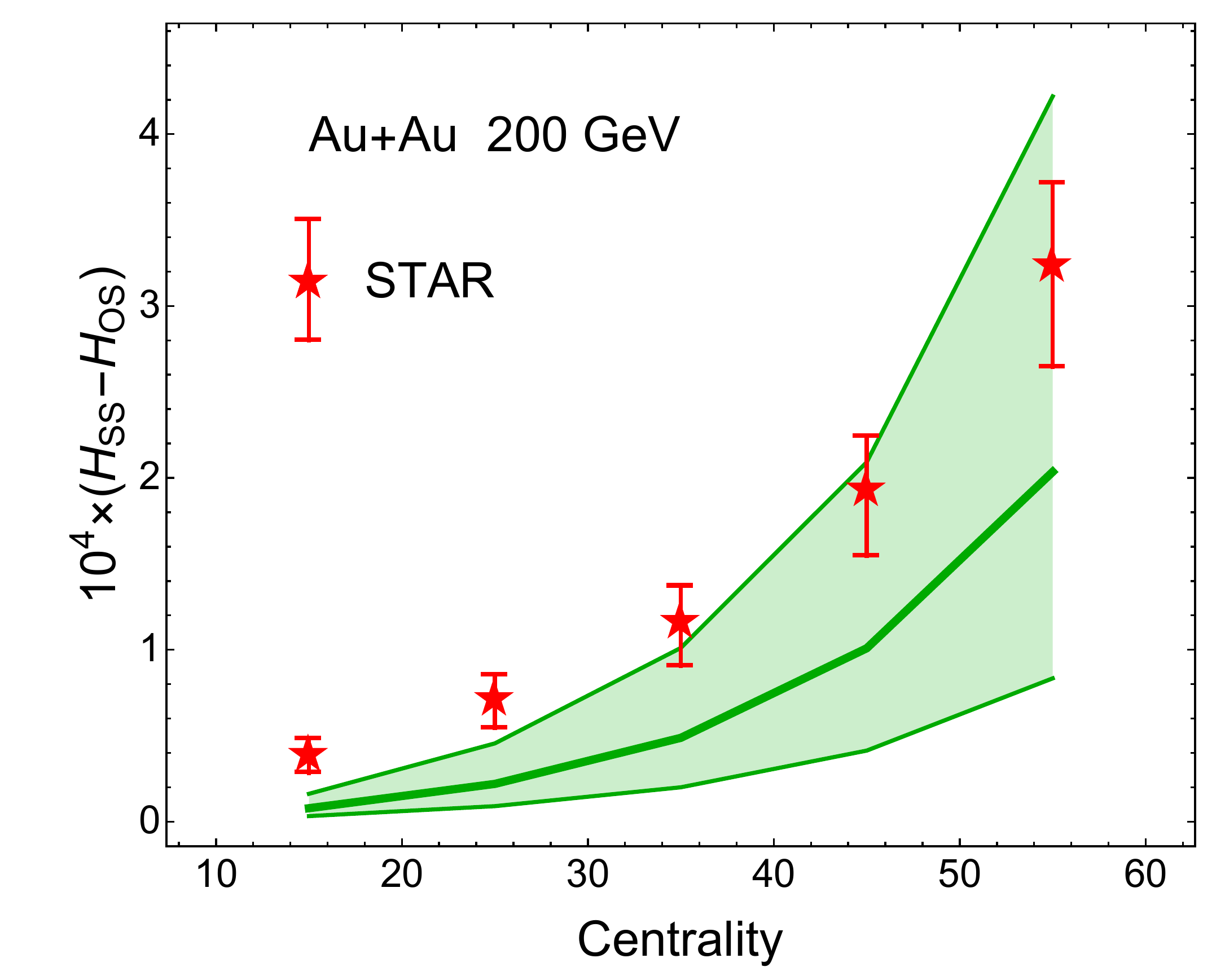}
\includegraphics[width=0.45\textwidth]{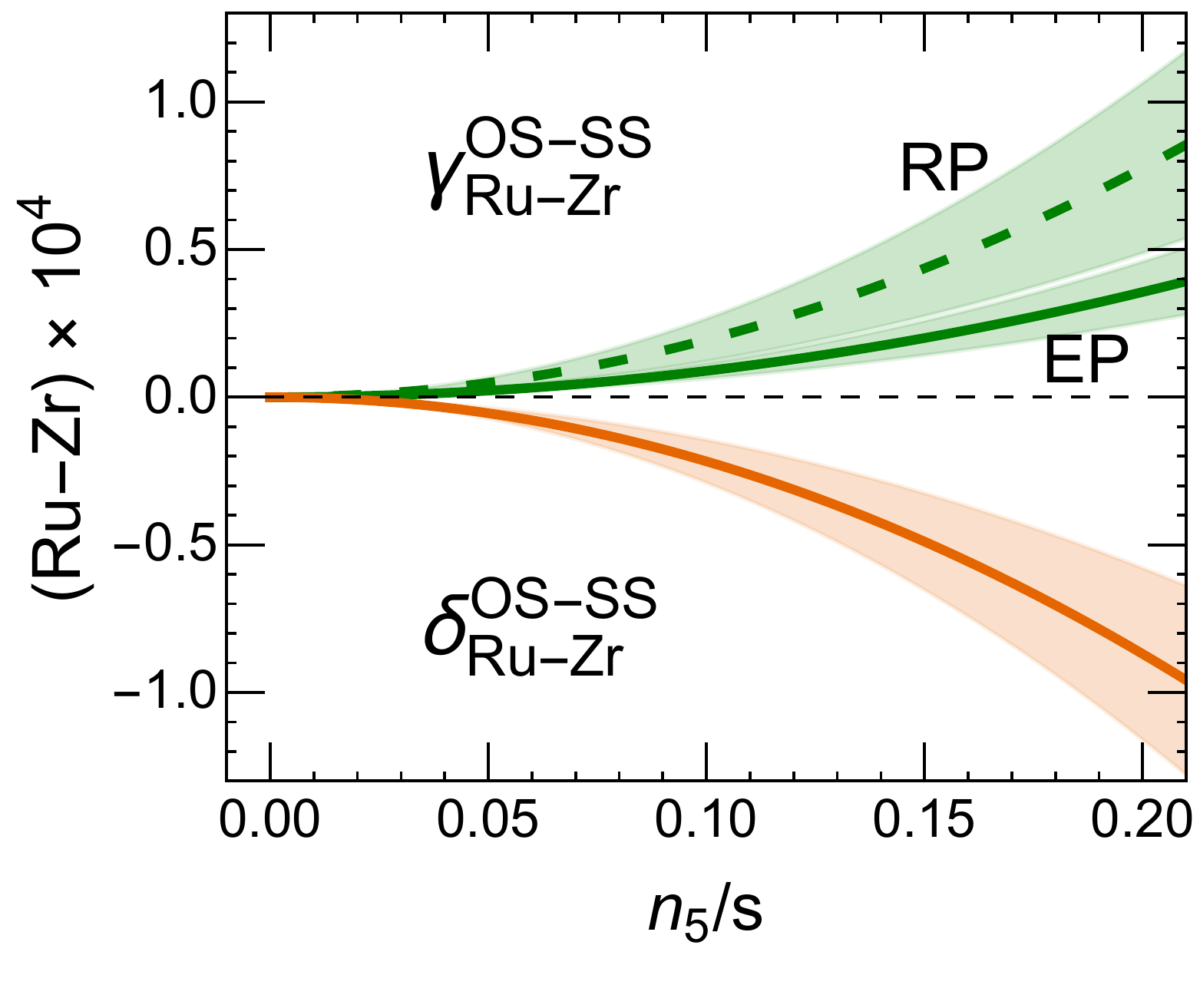}
\caption{(left, reproduced from \protect{\cite{Jiang:2016wve}} with permission) The azimuthal
correlation observable $\left(H_{SS}-H_{OS}\right) $ for various centrality, computed from 
AVFD simulations and compared with STAR data~\cite{Adamczyk:2014mzf}, with the green band 
spanning the range of key parameter from $Q_s^2=1\rm GeV^2$ (bottom edge) to  $Q_s^2= 
1.5 \rm GeV^2$ (top edge).
(right, reproduced from \protect{\cite{Shi:2019wzi}} with permission) 
EBE-AVFD predictions for $\gamma_{Ru-Zr}^{OS-SS}$ (green) and $\delta_{Ru-Zr}^{OS-SS}$ 
(orange) with respect to event-plane (EP) as functions of $n_5/s$.  Error band represents 
the statistical uncertainty from simulations.
\label{fig.sec4_avfd}}
\end{figure}

Over the years the AVFD package has been further improved in essentially three stages:
1) in the first generation~\cite{Jiang:2016wve,Shi:2017cpu}, the simulations start with 
event-averaged initial condition, and tested the sensitivity of the strength of CME charge 
separation with respect to a series of ingredients, particularly the axial charge imbalance 
and the magnetic field lifetime. By using reasonable parameters, the magnitude and centrality 
dependence of possible CME signal can be described, see Fig.~\ref{fig.sec4_avfd} (left).
2) later, a second generation simulation~\cite{Shi:2019wzi} was developed, which takes 
into account the fluctuating initial condition for hydro and magnetic field, and implements 
the LCC effect with prescription of Ref.~\cite{Schenke:2019ruo}. As shown in 
Fig.~\ref{fig.sec4_avfd} (right), a difference of CME signals between the isobaric system 
is predicted, which can be tested in the on-going isobar experiment at RHIC.
3) in a continuing effort of the BEST Collaboration, the AVFD package is upgraded to 
its third generation, and implements the micro-canonical particle 
sampler~\cite{Oliinychenko:2019zfk,Oliinychenko:2020cmr}, followed by the updated 
hadron transport simulation package, SMASH~\cite{Weil:2016zrk}. It provides a global 
description of CME observables for different collision systems, including both the CME 
signal and the non-CME background.

It is worth mentioning that two additional improvements are needed for a more accurate 
description of the CME signal. First, the evolution of non-conserved axial charge requires 
a more careful modeling, to take into account the thermal fluctuations and damping effects.
The second is the time evolution of the electromagnetic field. Current versions of the AVFD 
calculation use a toy-model parameterization for the time revolution, and requires the input 
from a more realistic MagnetoHydroDynamics (MHD) calculation. While the dynamical axial 
charge in under development in an on-going project by the BEST Collaboration, the current 
status of the dynamical electromagnetic field solver will be discussed in the rest of this 
subsections.

\subsubsection{Evolution of Electromagnetic Field}

\begin{figure*}[!hbt]
\includegraphics[width=0.315\textwidth]{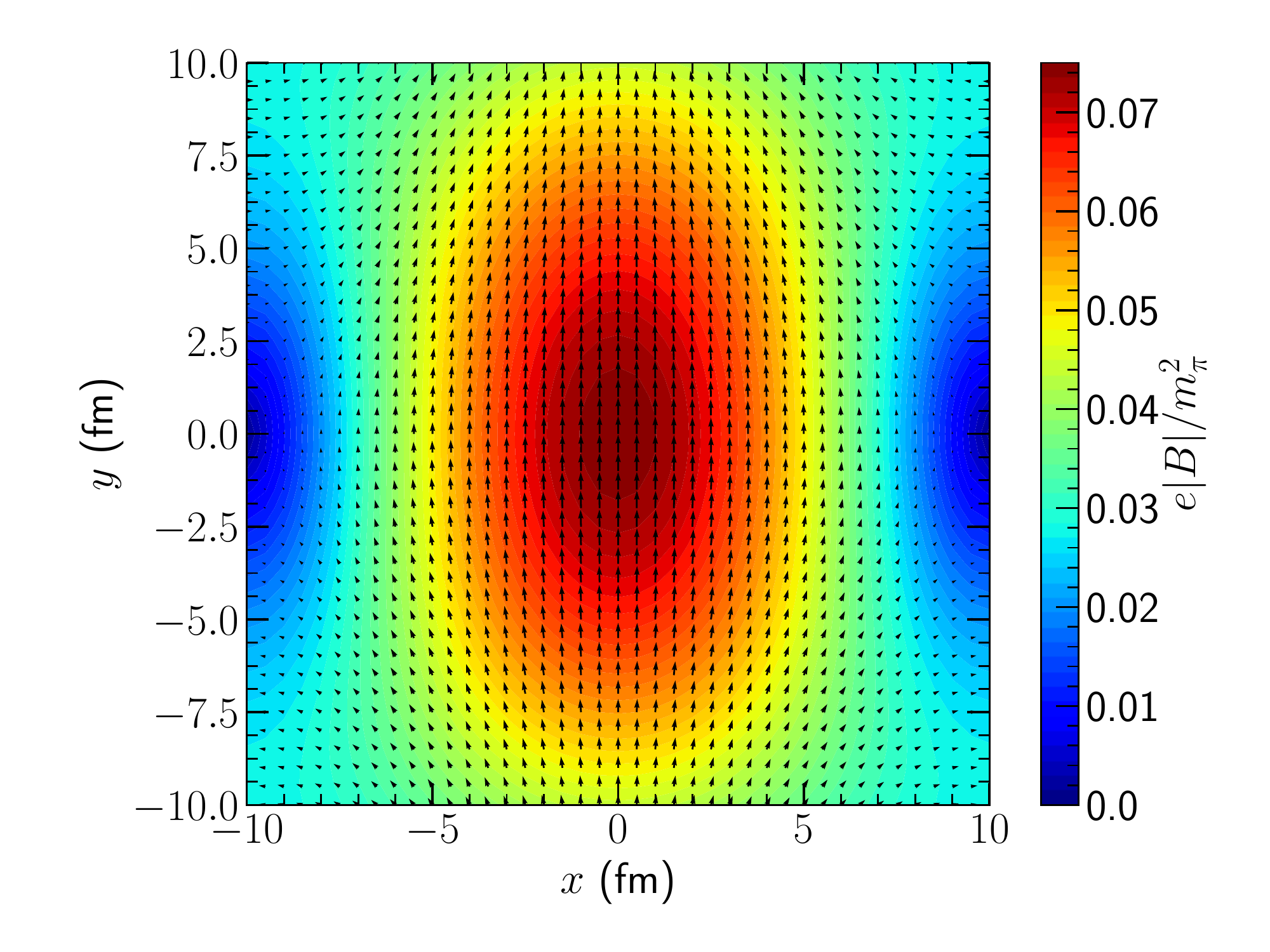}
\includegraphics[width=0.3\textwidth]{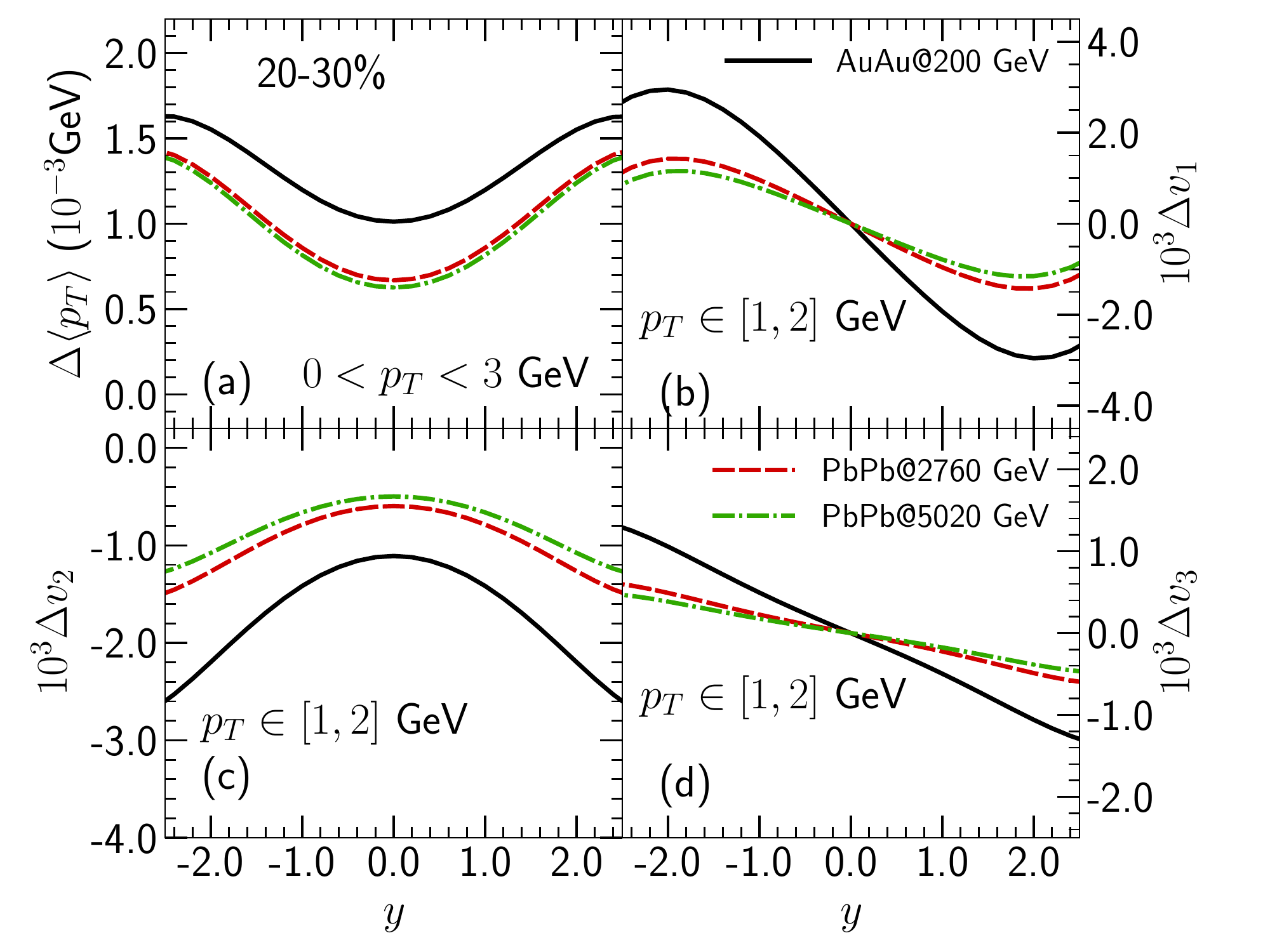}
\includegraphics[width=0.33\textwidth]{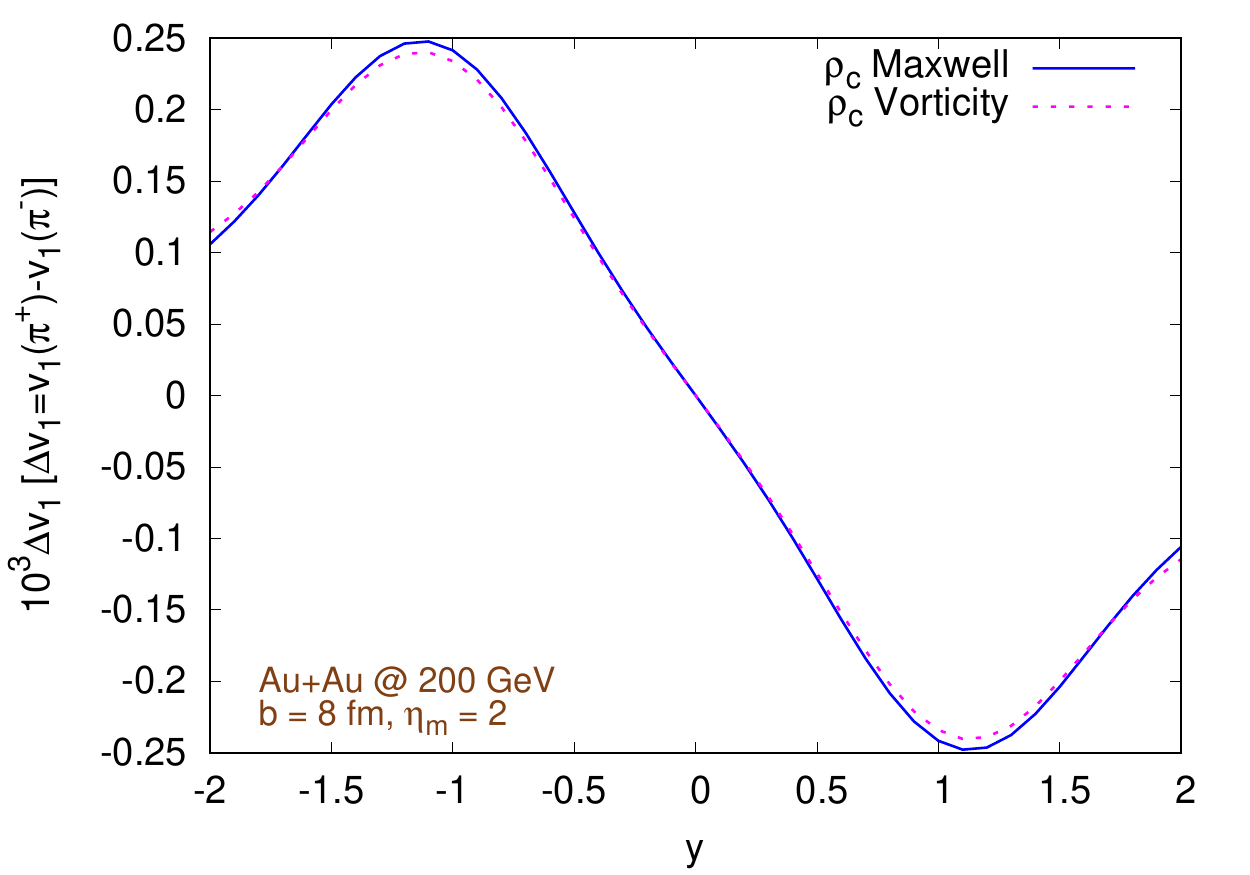}
\caption{(Left, reproduced from \protect{\cite{Gursoy:2018yai}} with permission): The magnetic 
field in the transverse plane at $z=0$  in the lab frame at a proper time $\tau=1$~fm/$c$ after a 
Pb+Pb collision with 20-30\% centrality and with a collision energy $\sqrt{s_{NN}} =2.76$~TeV. 
(middle, reproduced from \protect{\cite{Gursoy:2018yai}} with permission) The collision energy 
dependence of the electromagnetically induced charge-odd contributions to flow observables. The
difference of particle mean $p_T$ and $v_n$ between $\pi^+$ and $\pi^-$ are plotted as a function of
particle rapidity for collisions at the top RHIC energy of 200 GeV and at two LHC collision energies.
(right, reproduced from \protect{\cite{Inghirami:2019mkc}} with permission) Directed flow $v_1$ 
versus rapidity $y$, with an initial energy density distribution non tilted and tilted by using 
different values of the parameter $\eta_m$.
\label{fig.sec4_em}
}
\end{figure*} 

Significant efforts have been made within the BEST Collaboration to numerically solve the
magnetohydrodynamic equations. In \cite{Gursoy:2018yai}, the space-time evolution of the 
electromagnetic field is solved on top of realistic hydrodynamic evolution. This involves
an analytical solution derived in \cite{Gursoy:2014aka} which assumes a constant, 
temperature-independent electric conductivity. The magnetic field profile at $\tau=1$~fm/$c$ in the 
20-30\% Pb+Pb collisions is shown in Fig.~\ref{fig.sec4_em} (left). The EM induced modification 
of particle distribution, i.e. the difference of particle mean $p_T$ and $v_n$ between $\pi^+$ 
and $\pi^-$, are studied as shown in Fig.~\ref{fig.sec4_em} (middle). In particular, a 
charge-odd directed flow, $\Delta v_1$, and a triangular flow, $\Delta v_3$, are found 
to be odd in rapidity. These effects are induced by the magnetic field via the Faraday effect 
and the Lorentz force, as well as by the Coulomb field of the charged spectators. In addition, 
the electric field generated by the QGP with non-vanishing net charge drives rapidity-odd 
radial flow $\Delta \langle p_T\rangle$ and elliptic flow $\Delta v_2$. These studies assume 
that the evolution of the EM field decouples from the hydrodynamic background, neglecting 
the feedback on the medium. 

On the other hand, a study based on ideal magnetohydrodynamics was performed in
Ref.~\cite{Inghirami:2019mkc}. The approach is based on solving the evolution of 
the bulk medium together with the EM field, but it assumes infinite electric conductivity.
The study finds similar behavior to the aforementioned electromagnetic field induced 
modifications, e.g. the charge-dependent direct flow $\Delta v_1$ as shown in 
Fig.~\ref{fig.sec4_em} (right). In both studies, the rapidity slope of $\Delta v_1$ 
are found to be opposite to the ALICE results~\cite{Margutti:2017lup}. Such tension 
reflects the delicate interplay between the Faraday effect and the Lorentz force, which 
contribute oppositely to the sign of $\Delta v_1$, and calls for a more realistic study 
of the evolution of electromagnetic field.
Moreover, a recent study \cite{Giacalone:2021bzr} demonstrated we can the averaged
transverse momentum of the collision system as an experimental handle to manipulate
the magnitude of magnetic fields generated by the spectators.
A new project by the BEST Collaboration, 
aimed towards a more realistic description of the space-time profile of the magnetic 
field in heavy-ion collisions, is in progress. The goal of this project is to solve
the Maxwell equations together with the conservation equation of the electric current, 
taking into account realistic temperature dependent conductivities as perturbations 
to the hot medium. Similar to \cite{Gursoy:2018yai}, the feedback to the medium is
neglected and these equations are solved as perturbations.


\subsection{Fluctuation dynamics}
\label{sec:hydro_flcut}

As discussed in Sec.~\ref{sec:theory} there are two practical approaches for studying the
dynamical evolution of critical fluctuations near a QCD critical point: the deterministic
approach, which solves a relaxation equation for the correlation functions, and the stochastic
approach, which describes the same physics using stochastic equations similar to the Langevin 
equation. The study of critical dynamics in fluids has a long history~\cite{Pomeau:1974hg,
Hohenberg:1977ym}, but until recently many important ingredients were missing in studies of
critical dynamics in heavy ion collisions:

\begin{enumerate}
\item[i] The effects of conservation laws (such as charge conservation) were missing. Since 
the order parameter relevant for the QCD critical point is associated with baryon density, 
charge conservation needs be treated properly for quantitative studies. 
\item[ii] Most studies considered a homogeneous and boost-invariant fireball. Such a set-up is 
far from realistic and does not take into account the effects of advection. 
\item[iii] While the qualitative importance of such out-of-equilibrium effects has been 
well-appreciated, its quantitative relevance has not been fully studied.
\end{enumerate}

In the following, we shall first summarize the simulation results from Hydro+, which employs 
the  deterministic approach, and discuss the progress towards addressing the aforementioned
issues. At the end of this subsection we will summarize the progress and current status of
stochastic hydrodynamics simulations.

\subsubsection{Hydro+ simulations}

\begin{figure*}[!hbt]
\begin{center} 
\includegraphics[width=0.41\textwidth]{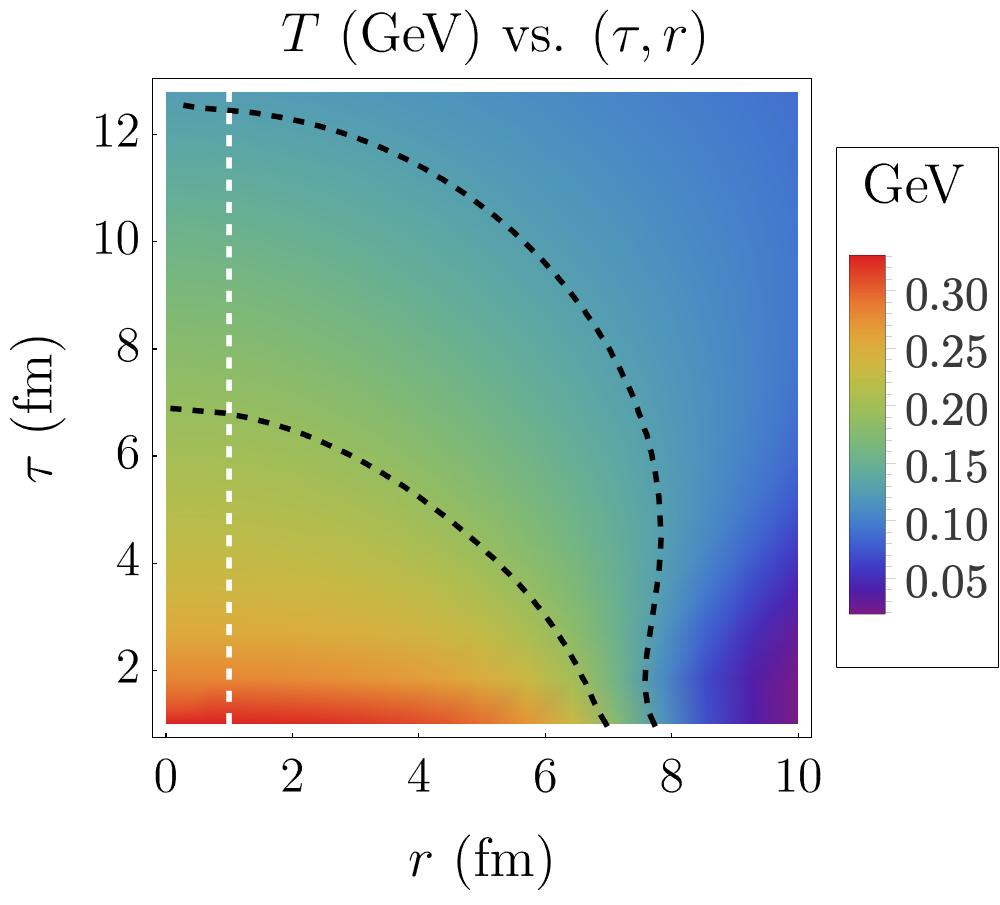} \hspace{0.05in}
\includegraphics[width=0.45\textwidth]{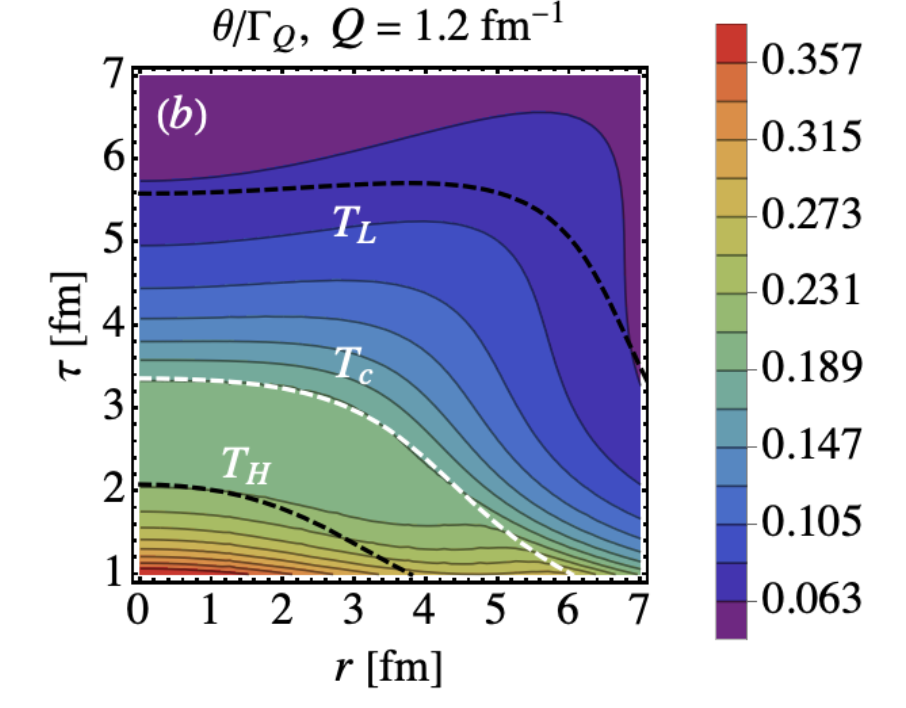} \hspace{0.05in}
\caption{
Temperature profile in the transverse plane for two exploratory studies of Hydro+.
The dashed curves are equal-temperature contours. 
(Left panel): The temperature as a function of $\tau,r$ in the model studied in 
Ref.~\cite{Rajagopal:2019xwg}. 
(right panel): Results for a model based on Gubser flow~\cite{Du:2020bxp}. } 
 \label{fig:hydroplus-profile}
\end{center}
\vspace{-0.1in}
\end{figure*} 

There are two recent simulations within the Hydro+ framework, published in 
Refs.~\cite{Rajagopal:2019xwg} and~\cite{Du:2020bxp}. (See Sec.~\ref{sec:theory_fluct}
for a brief discussion of the ideas underlying Hydro+.) The goal of 
Ref.~\cite{Rajagopal:2019xwg} was to explore Hydro+ in a ``minimal model''. This model
captures the interplay of critical fluctuations and hydrodynamics in a setting where 
the dynamics is similar to that is encountered in a heavy ion collision while the
geometry and equation of state are simplified. Specifically, Ref.~\cite{Rajagopal:2019xwg}
considers a radially and longitudinally expanding fluid, which is boost-invariant and
azimuthally symmetric in the transverse plane, and expands along the temperature axis at 
$\mu_B=0$, with a hypothetical critical point located at small $\mu_B$.
In the exploratory study of Ref.~\cite{Du:2020bxp}, the authors follow the evolution 
of the two-point function of baryon density $\phi(Q)$ (or $\phi_Q$) on top of a 
simplified QCD matter background, known as Gubser 
flow~\cite{PhysRevD.82.085027,Gubser:2010ui}], with large non-zero baryon number.
This set-up allows the authors to clearly distinguish the main effects controlling 
the dynamics of long-wavelength fluctuations and to explore systems with large baryon 
chemical potential. In Fig.~\ref{fig:hydroplus-profile}, we plot the temperature 
profiles in the $(r,\tau)$ plane for both studies. 

\begin{figure*}[!hbt]
\begin{center}
\includegraphics[width=1.00\textwidth]{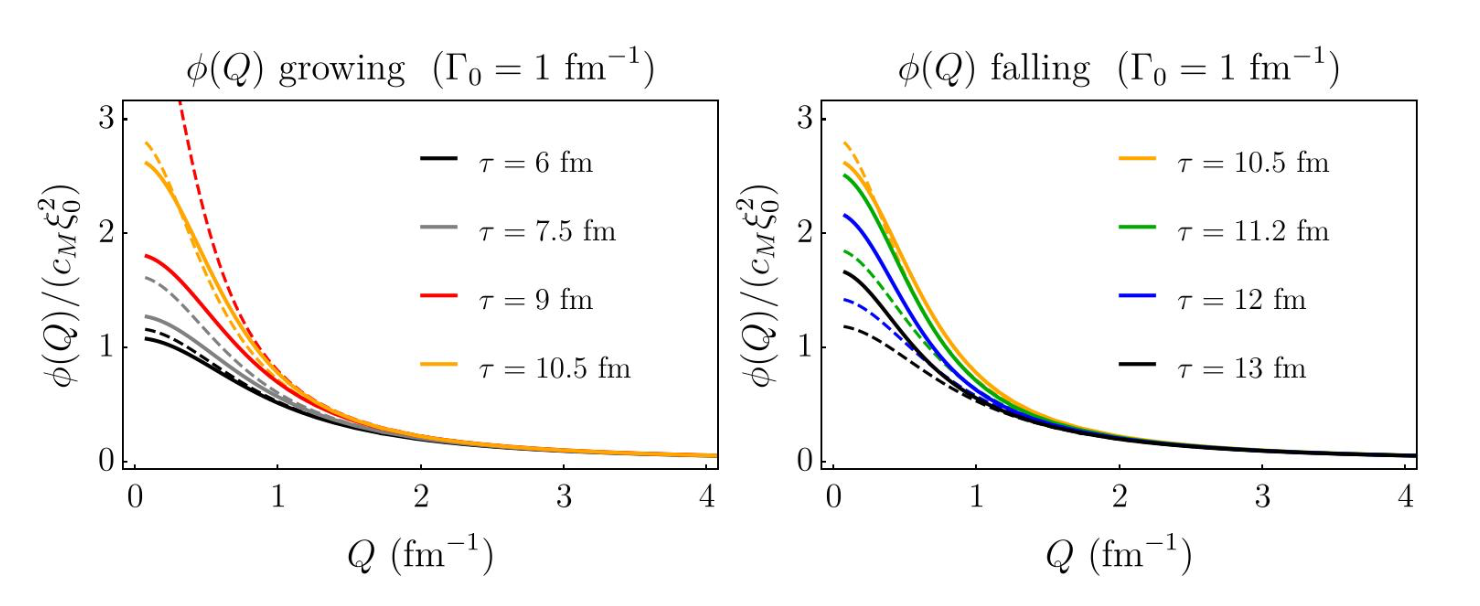} 
\end{center}\begin{center}
\includegraphics[width=0.82\textwidth]{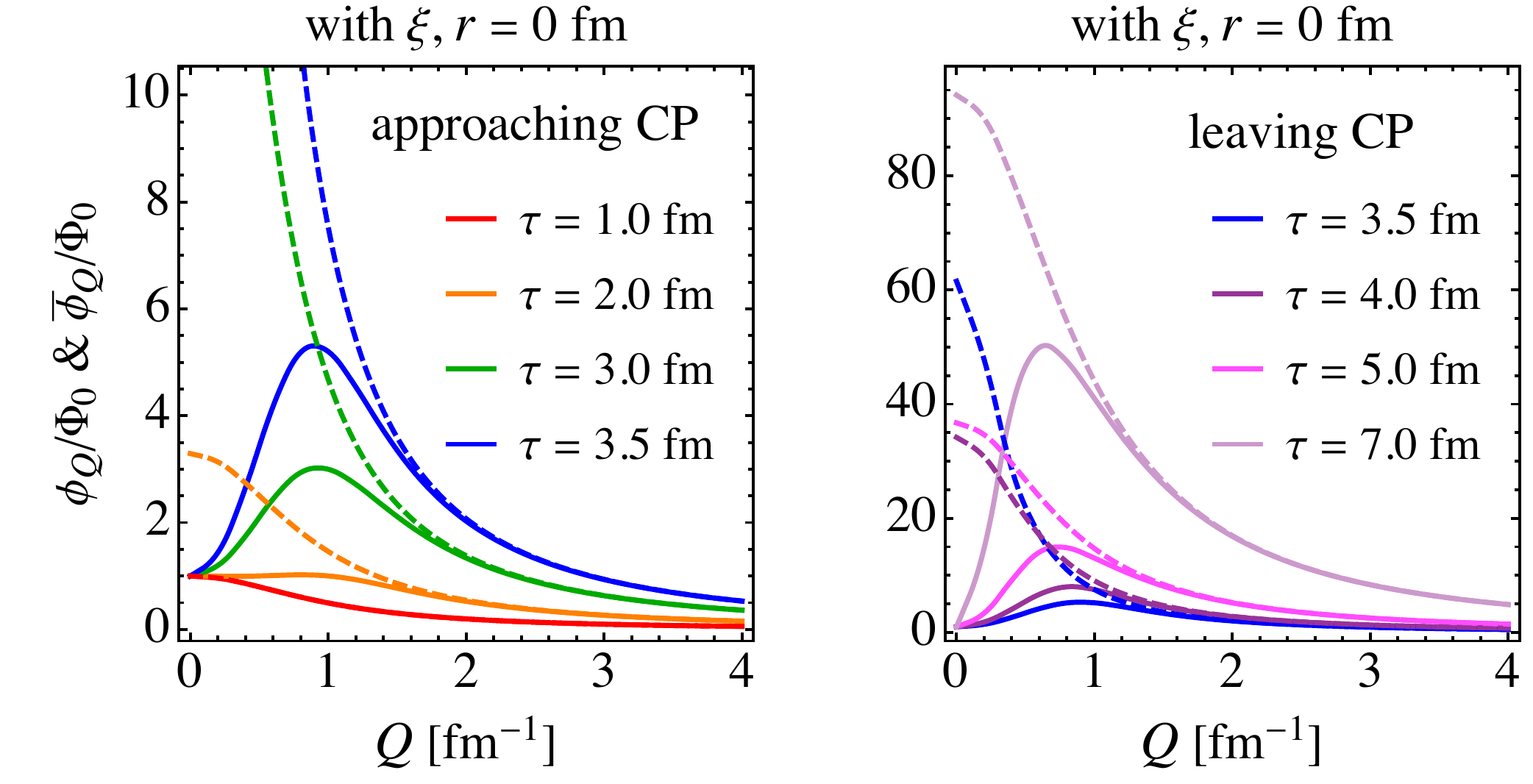}
\hspace{0.3cm}
\caption{
\label{fig:phievo}
The magnitude of the critical fluctuations $\phi(Q)$ (with appropriate normalization) vs 
momentum $Q$ at the representative radial distance for several different values of proper time $\tau$
from Ref.~\cite{Rajagopal:2019xwg} (upper panel) and Ref.~\cite{Du:2020bxp} (lower panel).
The solid curves and dashed curves represent out-of-equilibrium expectation and equilibrium 
results respectively.} 
 \label{fig:memory}
\end{center}
\vspace{-0.1in}
\end{figure*} 

\begin{figure*}[!hbt]
\includegraphics[width=0.85\textwidth]{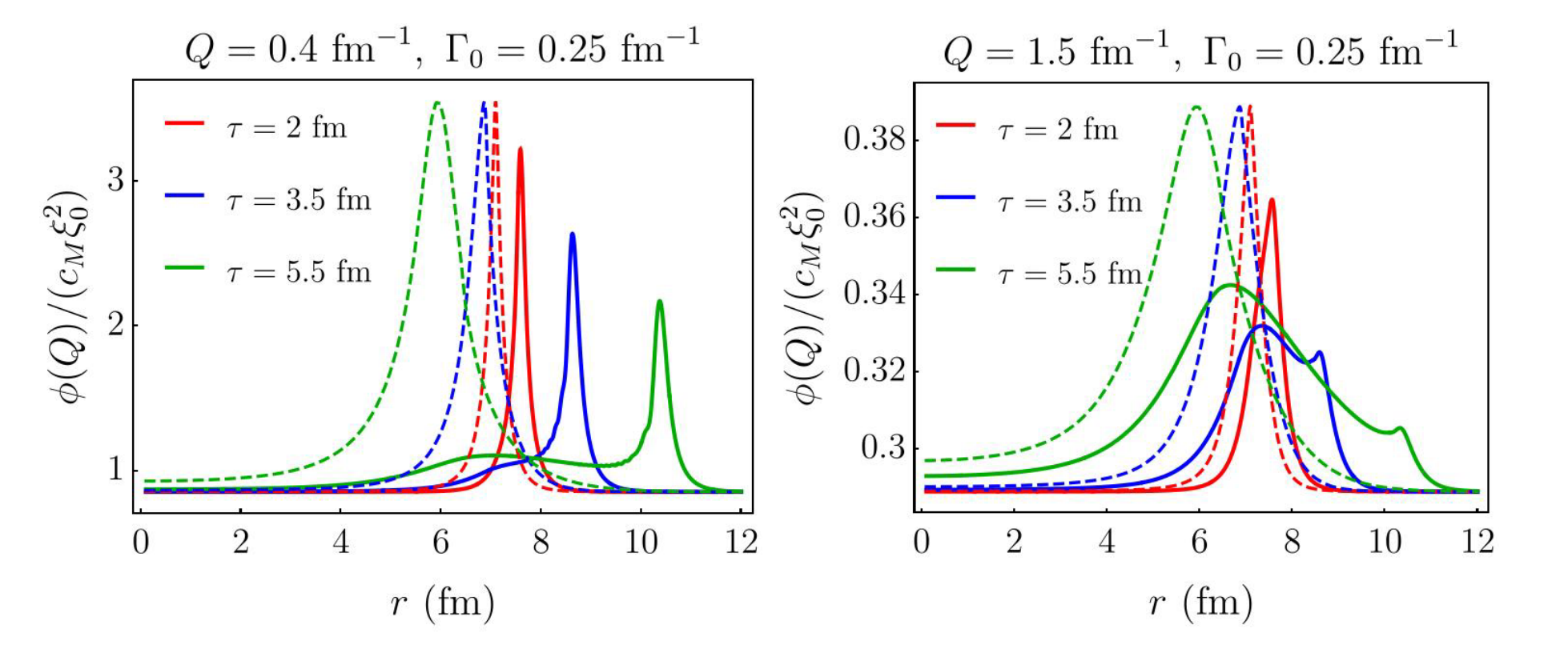} \hspace{0.022in}
\includegraphics[width=0.80\textwidth]{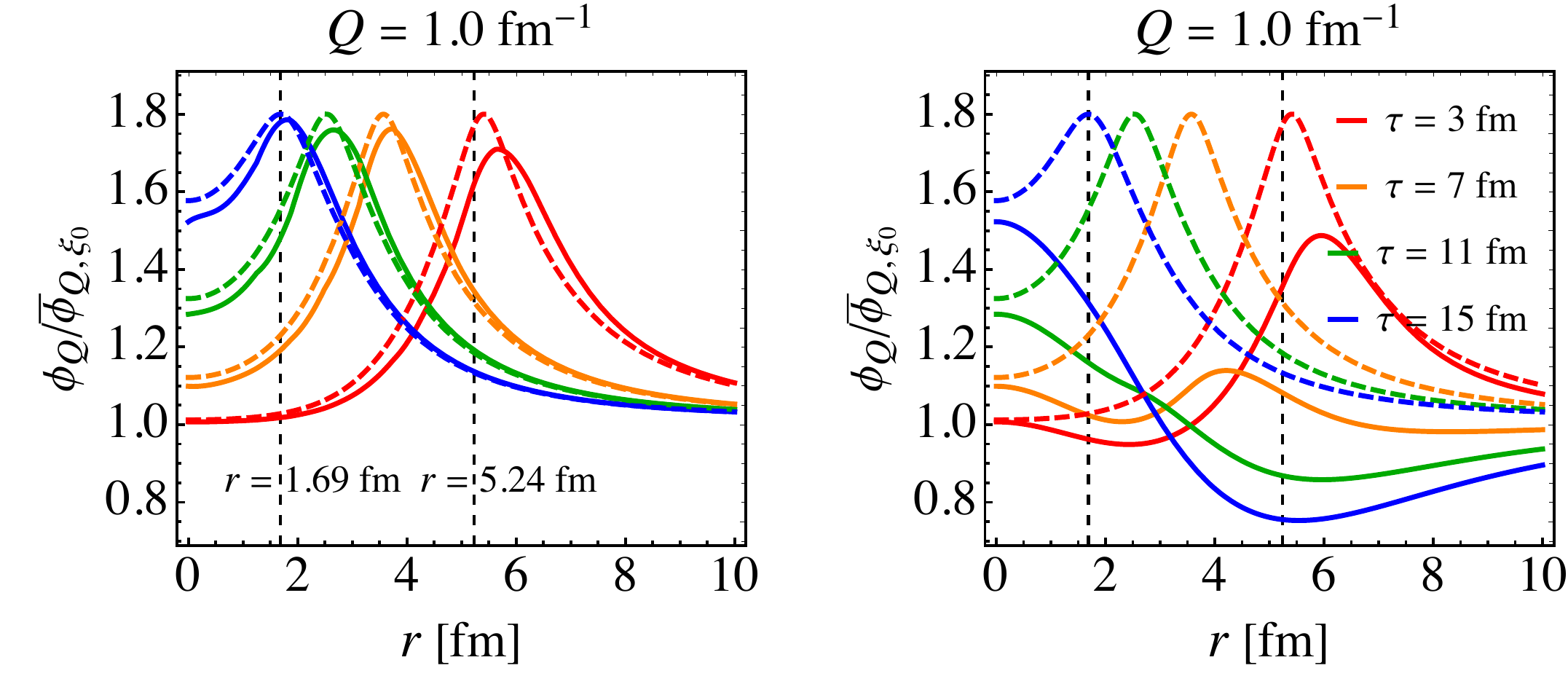} \hspace{0.022in}
\caption{ \label{fig:advection}
The magnitude of the critical fluctuations $\phi(Q)$ (with appropriate normalization) vs radius 
$r$ at a representative momentum $Q$ from Ref.~\cite{Rajagopal:2019xwg} (upper panel) and 
Ref.~\cite{Du:2020bxp} (lower panel)
In all figures, dashed (solid) lines show the equilibrium (nonequilibrium) values. 
}  
\end{figure*} 

We now discuss the main lessons learned in those simulations.  First, both simulations 
demonstrate the need for describing the out-of-equilibrium evolution of fluctuations 
quantitatively. Fig.~\ref{fig:memory} (upper panel) plots the temporal evolution 
of $\phi(Q)$ (as a function of the momentum $Q$) from Ref~\cite{Rajagopal:2019xwg} 
at $r=1$~fm, deep within the interior of the fireball. We observe that, as expected, 
large $Q$ modes stay in equilibrium. Hence we shall focus on small $Q$, i.e. long-wavelength 
modes from now on. We see that the temporal evolution of $\phi(Q)$ at long wavelengths 
falls out of equilibrium in two characteristic stages. First, at earlier times as 
$\bar{\phi}(Q)$ rises as the cooling QGP approaches $T_{c}$ from above,  $\phi(Q)$ 
lags behind Fig.~\ref{fig:phievo} (upper left panel). At later times, as $\bar{\phi}(Q)$ 
drops as the QGP cools away from $T_{c}$ toward lower temperature Fig.~\ref{fig:phievo} 
(upper right panel) $\phi(Q)$ shows a memory effect: the dashed curve remembers where 
the dashed curve used to be (Fig.~\ref{fig:phievo}, upper panel, right). The phenomena 
of lag and memory are also apparent in Fig.~\ref{fig:phievo} (lower panel) where we plot 
the results from Ref.~\cite{Du:2020bxp} at $r=0$~fm. 
Qualitatively, the behavior at small $\mu_B$ \cite{Rajagopal:2019xwg} and large $\mu_B$ \cite{Du:2020bxp}] is similar, but the $\mu_B$-dependence of the transport coefficients leads to significant quantitative differences if (as suggested by lattice QCD and BES experiments) the critical point is located at large $\mu_B$.
To summarize, $\phi(Q)$ encodes information about the criticality, which is why we are working towards describing it
quantitatively. 

As a second lesson we learn that conservation laws play an important role in dynamics.
This can be seen by comparing the behavior of $\phi(Q)$ at a small $Q$ in the two 
studies. Ref.~\cite{Rajagopal:2019xwg} and Ref.~\cite{Du:2020bxp} implement the
dynamical universality class of model A (for a non-conserved order parameter) and 
model B (for a conserved order parameter), in the classification of Hohenberg and 
Halperin \cite{Hohenberg:1977ym}. 
The relaxation rate $\Gamma(Q)$ approaches a constant for the former and vanishes 
as $Q^2$ for the latter. We observe that a conserved order parameter $\phi(Q)$ 
at small $Q$ stays at its initial value. This clearly demonstrates the crucial role of 
the conservation law: $\phi(Q=0)$ corresponds to the fluctuation of the order parameter 
averaged over the whole volume. If the order parameter is associated with conserved 
densities, then the fluctuation at $Q=0$ can not evolve at all. 

The third lesson is that radial flow transports fluctuations by advection, and that
quantitative studies are required to understand this effect. To see this, let 
us first look at Fig.~\ref{fig:advection} (upper panel) where we show $\phi$ vs $r$ at the
representative value of $Q=0.4$~${\textrm{fm}}^{-1}$ from \cite{Rajagopal:2019xwg}.  We
observe that the peak in equilibrium expectation of $\bar{\phi}(Q)$ moves inward as a 
function of time. On the other hand, the spatial dependence of the fully dynamical 
$\phi(Q)$ is determined by the combination of two out-of-equilibrium effects. First, 
the memory and lag effects imply that as the peak in the equilibrium curve (dashed)
moves inward, the actual $\phi$ at its location increases toward it but does not come 
close to reaching it. Second, the peak in the fluctuations is carried outward by 
advection in the expanding fluid. To further illustrative those two effects, let us 
look at Fig.~\ref{fig:advection} (lower panel) where we show $\phi$ vs $r$ from
Ref.~\cite{Du:2020bxp}. In Fig.~\ref{fig:advection} (lower panel, left), the advection 
term in Hydro+ equation is switched off, and we only see the memory and lag effects. 
However, once the advection term is switched on (lower right panel), significant changes
in the evolution of the $r$-dependence of $\phi$ are observed.

A fourth lesson is that the non-equilibrium contributions due to slow modes to bulk 
properties such as entropy and pressure are generally small. In \cite{Rajagopal:2019xwg}
the authors found tiny back-reaction corrections to the background evolution. 
This is further confirmed in Ref.~\cite{Du:2020bxp}. Specifically, the out-of-equilibrium
slow-mode contribution $\Delta s$ to the entropy density is of the order
\begin{align}
    \frac{\Delta s}{s}\sim \mathcal{O}(10^{-5}{-}10^{-4}) \, . 
\end{align}
This can be understood by comparing the phase space volume of out-of-equilibrium critical 
modes $Q^{3}_{_\mathrm{neq}}/(2\pi)^{3}$ with the typical entropy density $s$ :
\begin{equation}
\label{eq:estdeltas}
    \frac{\Delta s}{s} \sim \frac{1}{(2\pi)^{3}}(\frac{T^3}{s}) 
    \left(\frac{Q_\mathrm{neq}}{T}\right)^3 , 
\end{equation}
where $Q_{\mathrm{neq}}\sim \xi^{-1}$ denotes the typical momentum which is not in
equilibrium. Using the value $s=(4\pi^{2}(N^{2}_c-1)+21\pi^{2}N_{f})T^{3}$, which 
corresponds to the entropy density of an ideal QGP at zero baryon chemical potential, 
we arrive at $(\Delta s/s) \sim {\cal O}(10^{-4})$.

 This conclusion is consistent with a study of the critical behavior the 
bulk viscosity near the QCD critical point~\cite{Martinez:2019bsn}. Bulk 
viscosity controls the non-equilibrium contribution to the pressure in an
expanding fluid. In a near-critical fluid, the dominant effect arises from
the lag in the order parameter relative to its equilibrium value. The critical
bulk viscosity is of the form
\begin{align}
    \frac{\zeta}{s}\sim a \, \left(\frac{\xi}{\xi_0}\right)^{3}\, , 
\end{align}
where $a$ is constant of proportionality, and $\xi_0$ is the correlation length 
away from the critical point. Ref.~\cite{Martinez:2019bsn} found that $a$ is 
quite small, $a\lsim 10^{-2}$ on the crossover side of the transition, although it is 
very sensitive to the mapping between the QCD EoS and Ising EoS. Based on the
estimate in \cite{Akamatsu:2018vjr}, we also expect $\xi/\xi_0\lsim 2$. Finally, 
we note that bulk viscosity itself also exhibits critical slowing down, and the 
nonequilibrium contribution to the pressure is smaller in magnitude than the equilibrium expectation
$\delta P = -\zeta \nabla\cdot v$.

From a practical perspective, the smallness of back-reaction effects suggests one 
may neglect the back-reaction in future phenomenological modeling, which will
significantly reduce computational cost. 

\subsubsection{The simulation of stochastic hydro}

A different method to include stochastic effects in hydrodynamics and to consider the 
thermal fluctuations that are demanded by the fluctuation-dissipation theorem is  
to generalize the energy-momentum tensor, $T^{\mu \nu}$,  by including a noise term.
Explicitly, to the usual  ideal and viscous parts, a random fluctuating term 
$S^{\mu \nu}$ is added:
\begin{eqnarray}
T^{\mu \nu}  = T^{\mu \nu}_{\rm ideal} + T^{\mu \nu}_{\rm viscous} + S^{\mu \nu} \, . 
\end{eqnarray}
The introduction of the noise term leads to a feature absent from treatments without
thermal fluctuations. The autocorrelation of noise is proportional to a delta function:
\begin{eqnarray}
\langle S^{\mu \nu} \left(x_1\right) S^{\rho \sigma} \left(x_2\right)\rangle = 2 T 
\left[ \eta \left( \Delta^{\mu \rho} \Delta^{\nu \sigma} 
                 + \Delta^{\mu \sigma} \Delta^{\nu \rho}\right) 
   + \left( \zeta - \frac{2}{3} \eta \right) \Delta^{\mu \nu} \Delta^{\rho \sigma}\right]
   \delta^4 \left( x_1 - x_2\right)
\end{eqnarray}
where $\eta$ and $\zeta$ are the shear and bulk viscosity coefficients,
$\Delta^{\mu \nu} = g^{\mu \nu} - u^\mu u^\nu$, and $x_1$ and $x_2$ are space-time 
four vectors \cite{landau1980statistical,Kapusta:2011gt}. The introduction of thermal
fluctuations turns the hydrodynamical evolution into a stochastic process where the 
noise is sampled over the space of a fluid cell, at each step in proper time $\tau$
\cite{Young:2013fka}. However, the averaged noise will diverge with decreasing cell 
size. This indicates that the theory needs to be renormalized. 
In the perturbative approximation -- where the fluctuation is separated from 
the fluid dynamical background -- equations for the noise and its response decouple 
from the hydro evolution equations \cite{Young:2014pka}.  An approach which goes beyond 
the perturbative limit is discussed next. 

\begin{figure}[hbt]
\centering
\includegraphics[width=0.4\textwidth]{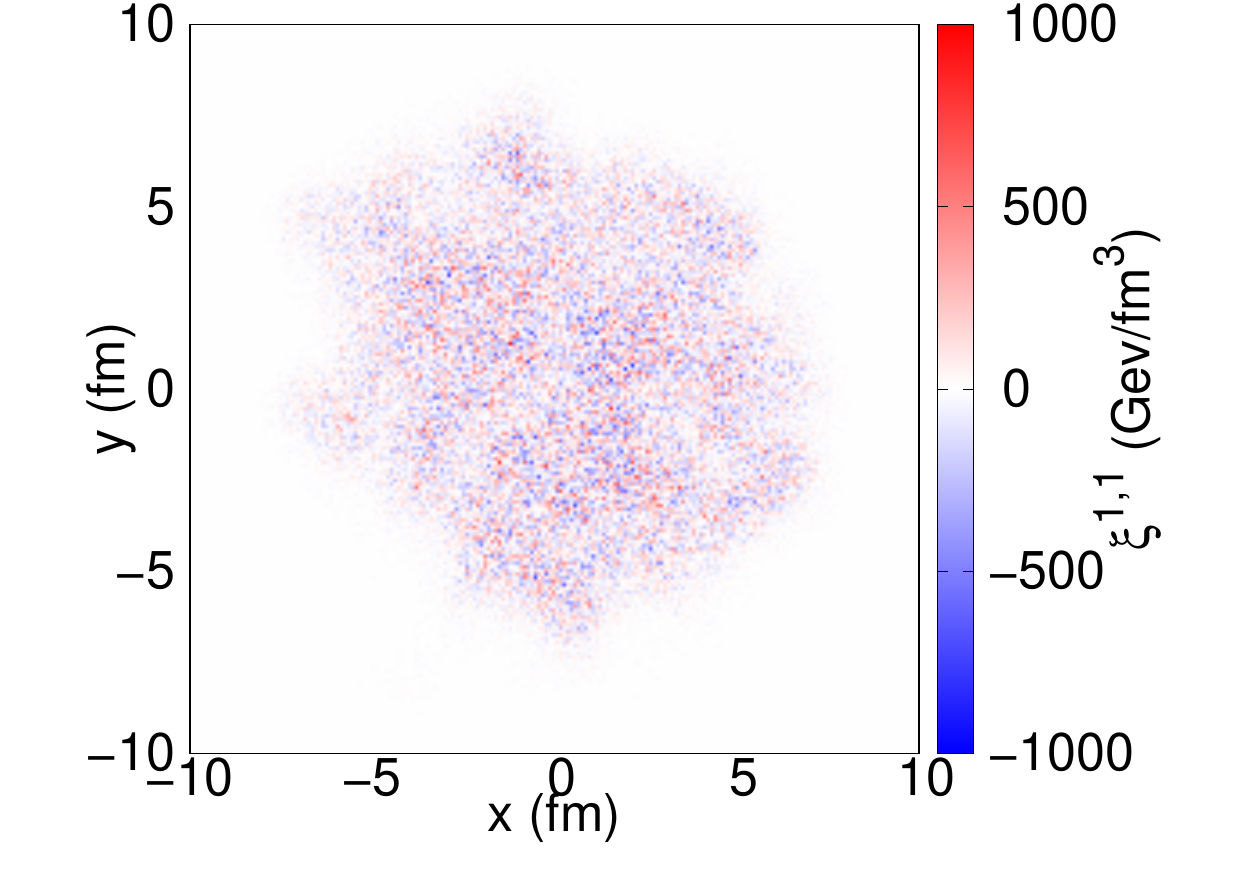}
\includegraphics[width=0.4\textwidth]{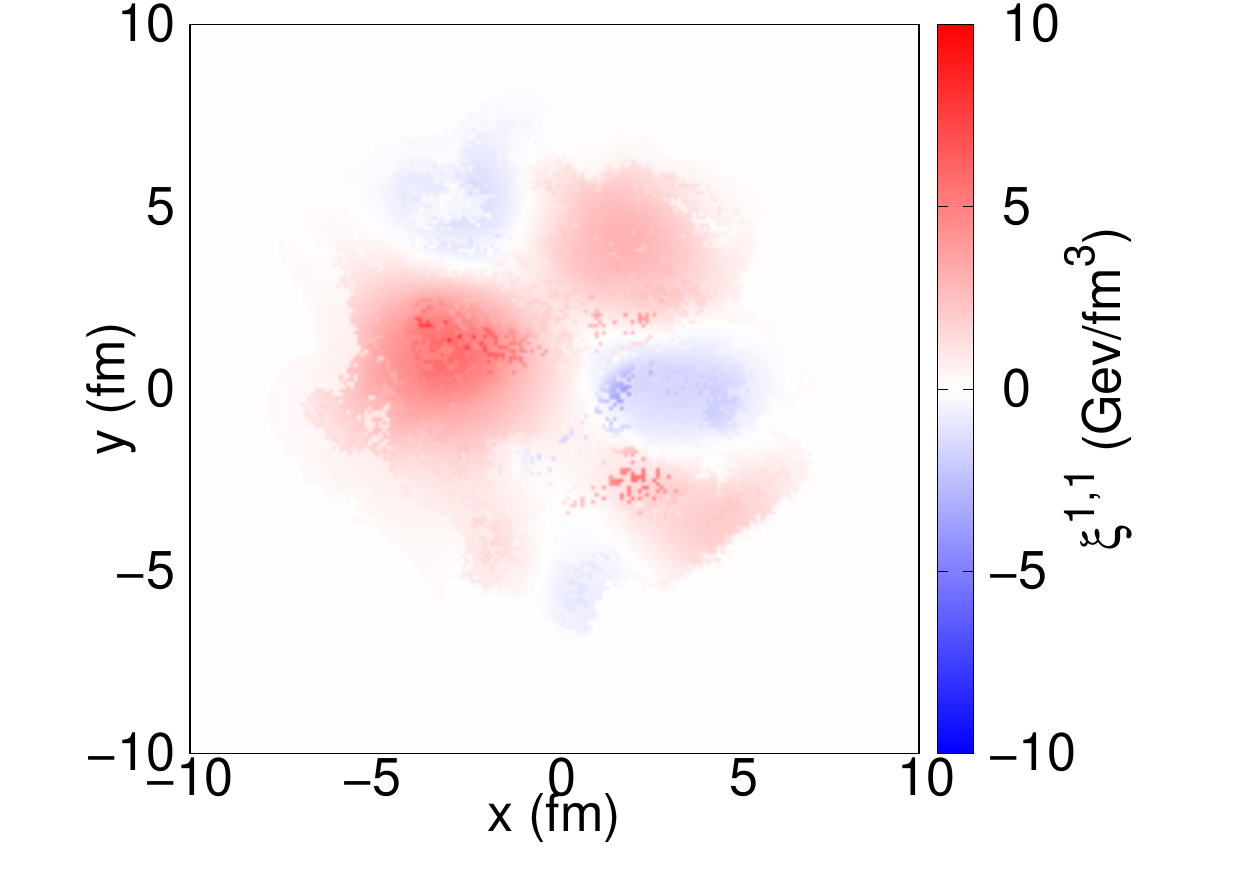}
\caption{The $\xi^{1,1}$ element of the sampled noise tensor \cite{SinghMayankthesis,
Singh:2018dpk}, at midrapidity. Left panel: before removing high wavenumber modes. 
Right: after removing modes larger than $p_{\rm cut} = 0.6/\tau_\pi$}
\label{fig:Singh_before}
\end{figure}

Hydrodynamics is a macroscopic, long wavelength theory. One notes that the finite 
cell size can suppress all wavelengths below $2 \Delta x$, where $\Delta x \sim 
\left( \Delta V\right)^{1/3}$. The discrete grid acts as a low-pass filter allowing 
only modes with wave number less than $\pi/\Delta x$. Consider, as an example, 
fluctuations of shear modes. On physical grounds, one can argue that shear modes 
with a wave number larger than $\sim 1/(\nu\tau_\pi)^{1/2}$ will quickly relax
to equilibrium. Here, $\tau_\pi$ is the decay time for the dissipative stresses 
to relax to the Navier-Stokes form, and $\nu=\eta/(sT)$. These fast modes are in 
thermal equilibrium, and their contribution to physical observables is accounted
for in the equilibrium equation of state and the transport coefficients. 

In typical simulations the cutoff scale set by the inverse cell size is much 
larger than the physical scale set by the relaxation time. In practical applications
it makes sense to remove the fast modes with wave numbers greater than $1/(\nu
\tau_\pi)^{1/2}$ by an additional low pass filter, see also \cite{Murase:2016rhl, Nahrgang:2017oqp, Bluhm:2018plm}.
In \cite{Nahrgang:2017oqp}, the dependence of the hydrodynamical fields and their fluctuations on the lattice spacing $dx$ is demonstrated.
In the following we will use
the kinetic theory relation $\nu\sim \tau_\pi$ to write $p_{\rm cut} = x/\tau_\pi$, 
where $x$ is a parameter of order 1. A procedure to implement a local low pass filter
in relativistic fluid dynamics is described in Ref.~\cite{SinghMayankthesis}.
It is based on boosting fluid cells to the local rest frame, Fourier transforming, 
imposing a wave number cutoff, and then performing the inverse transformation. 
The local coarse-graining limits both noise and ordinary gradients that have the 
potential to invalidate the second-order fluid dynamical treatment. The effect of 
noise filtering is illustrated in Figure \ref{fig:Singh_before}, which displays 
an element of the noise tensor, for a collision of Pb + Pb at $\sqrt{s_{\rm NN}} 
= 2.76$ TeV, in a 0 -- 5\% centrality class, at mid-rapidity, before and after 
the noise-filtering process \cite{Singh:2018dpk}. 

\begin{figure}[hbt]
\centering
\includegraphics[width=0.6\textwidth]{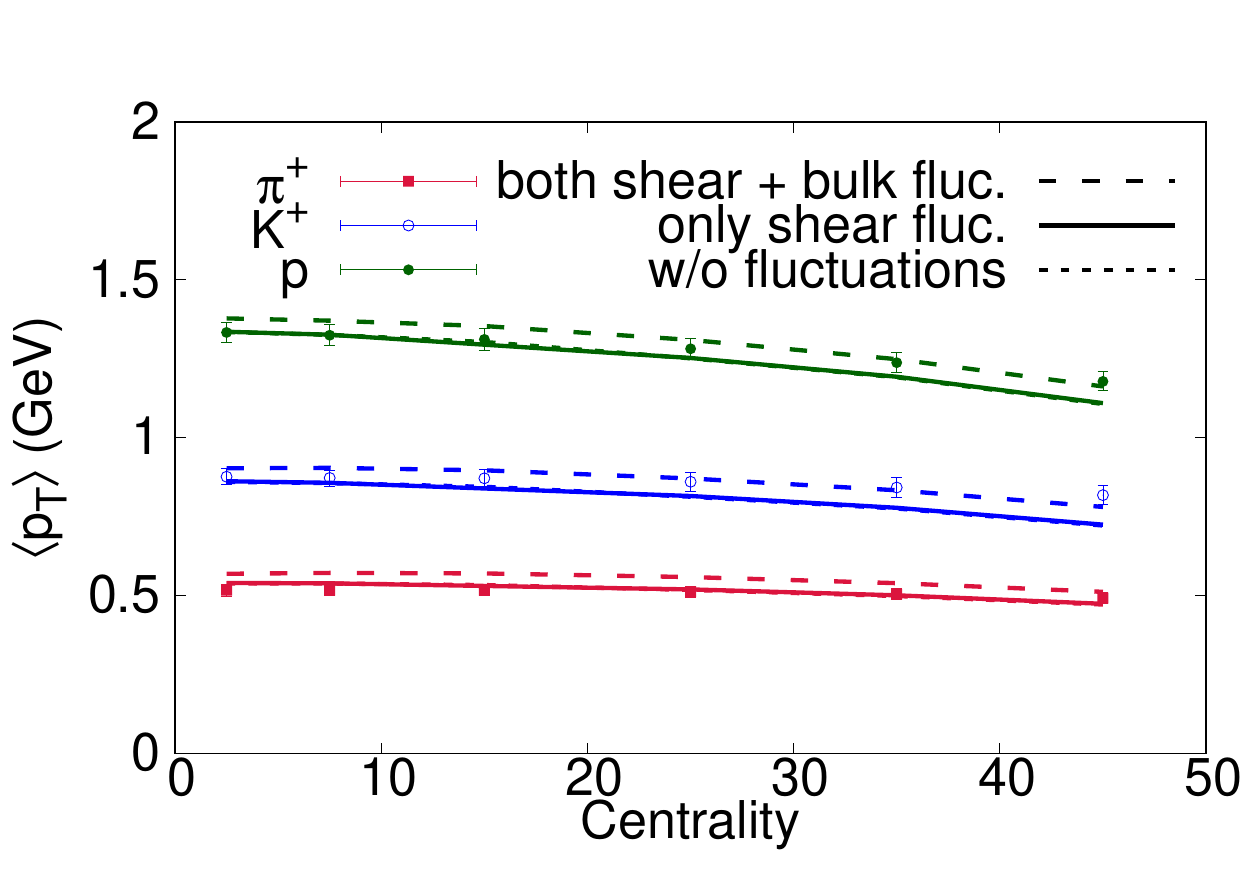}
\caption{The average transverse momentum of $\pi^+$, $K^+$, and $p$, calculated 
with the inclusion of fluctuations associated with the shear viscosity, and with 
both shear and bulk viscosity. Also shown is the results without thermal 
fluctuations. The results are for Pb + Pb with $\sqrt{s_{\rm NN}}= 2.76$ TeV, as 
a function of centrality. The simulations are based on a hybrid model: IP-Glasma +
MUSIC + UrQMD, with $\eta/s = 0.13$, and $\zeta/s$ as modeled in 
Ref.~\cite{Ryu:2015vwa}. }
\label{fig:Singh_fluctuations_pT}
\end{figure}

We leave the details to a forthcoming publication \cite{Singh2021}, but the thermal
fluctuations demanded by the fluctuation-dissipation theorem do have effects on 
observables that are known to highlight the presence of fluctuations, thermal or 
otherwise. The first of those is the class of event-plane correlators, which are 
the correlations between the event planes generated by different harmonic coefficients
$v_{n}$ \cite{Bhalerao:2011yg,Jia:2014jca}, and another are the linear and non-linear
elements of the decomposition of the flow coefficients for higher harmonics ($n \geq 4$), 
as prescribed in Ref.~\cite{Yan:2015jma}:
\begin{eqnarray}
v_n = v^L_n + \sum_{n = p + q} \chi_{n p q} v_p v_q.
\end{eqnarray}
Importantly for the physics pursued by the BEST Collaboration, the inclusion of both 
shear and bulk dissipation modes do have an effect on the phenomenological extraction 
of the transport coefficients of the hot and dense strongly interacting matter. In
particular, the fluctuations associated with the bulk viscosity affect directly the net 
cooling and expansion of the fireball, as seen in Fig. \ref{fig:Singh_fluctuations_pT}
which reports on calculation of the average transverse momentum for different charged
species. The far-reaching conclusion of those studies is that the inclusion of thermal
fluctuations will entail the recalibration of transport coefficients in general, as 
observed in Fig.~\ref{fig:Singh_fluctuations_pT}. A similar conclusion was also reached 
in Ref.~\cite{Young:2013fka}, and it should impact the analysis of heavy-ion collisions 
at all energies, including those performed at the LHC, and the BES runs. 

In what concerns the physics pursued here, the results obtained with stochastic 
hydrodynamics up to now applied to high-energy heavy-ion collisions  need to be 
compared with simulation relying on the Hydro+ framework. The two approaches should be 
complementary, and their comparison will benefit the community as a whole. Then, the
approach to criticality will be considered.  

\begin{figure*}[hbt!]
\includegraphics[width=0.40\textwidth]{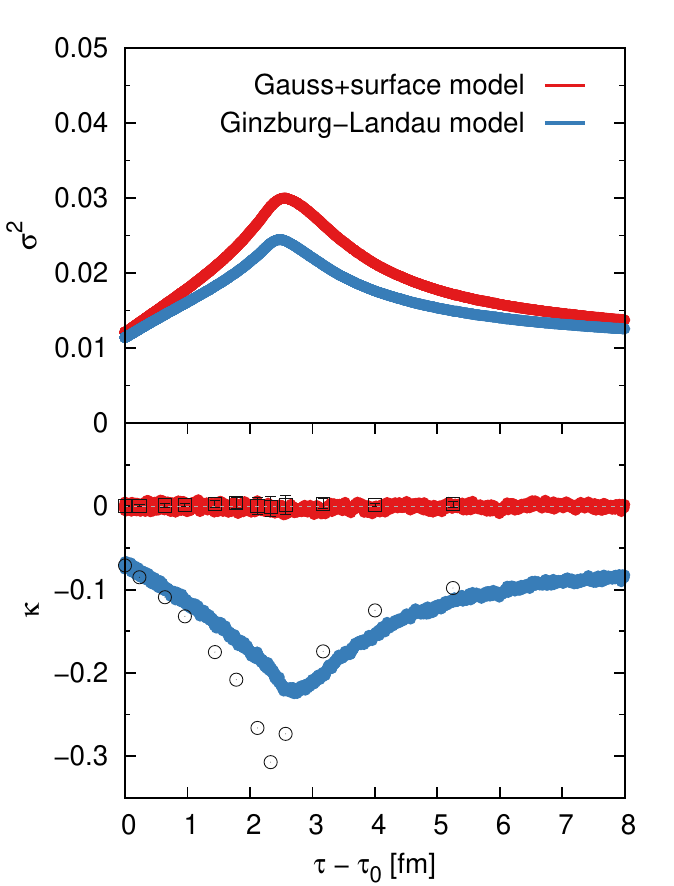}
\caption{The scaled Gaussian cumulants, kurtosis of the net-baryon number as a function 
of scaled temperature $T/T_{c}$ from simulating stochastic diffusive equation with an 
Ising-like equation of state. The equation is solved in a $(1+1)d$ geometry. Here, the 
blue and red curves show the results obtained with two different free energy functionals. 
The first (blue) is a purely Gaussian functional with a surface (gradient) term, and the 
second is a Ginzburg-Landau functional with non-linear terms, see 
Ref.~\cite{Nahrgang:2018afz} for more details. }
\label{fig:2d-K4}
\end{figure*} 

Returning to the physics of the critical point, the authors of Ref.~\cite{Nahrgang:2018afz, 
Nahrgang:2020yxm} consider fluctuations of the net-baryon density near the critical point. They 
solve the stochastic diffusion equation in a finite-size system with Gaussian white noise, using 
an Ising-like equation of state in $1+1$ space-time dimensions. In contrast to earlier
simulations of chiral fluid dynamics (e.g. Ref.~\cite{Herold:2013bi}), where the critical 
mode was identified with the non-conserved sigma field, the critical mode was taken to be 
a conserved density. The effects of charge conservation in a finite system modify the 
equilibrium scaling of the cumulants relative to the expected scaling with the 
correlation length in an infinite system. Typically, the critical growth is reduced, as  
explained in \cite{AgahNouhou:2019reg}. In the exploratory calculation described in 
\cite{Nahrgang:2018afz,Nahrgang:2020yxm} the expected dynamical scaling behavior and 
the impact of critical slowing down are observed. In particular, the results shown 
in Fig.~\ref{fig:2d-K4} demonstrate that Gaussian noise, combined with a non-linear free
energy functional, will generate the expected non-Gaussian cumulants, and that these
cumulants show the effects of memory and lag. Currently, the stochastic diffusion 
equation is studied in expanding systems \cite{Kitazawa:2020kvc, Pihan:2021Preparation} 
and in 3+1 dimensions, where ultraviolet divergencies related to the finite lattice
spacing are more important, and renormalization of the equation of state and the 
transport coefficients has to be taken into account \cite{Attieh:2021Preparation}.
Higher-order cumulants have also been studied in the deterministic approach,
see \cite{An:2020vri}.

\section{Non-critical contributions to proton number fluctuations}
\subsection{Sources of non-critical fluctuations}

 Proton number fluctuation cumulants are one of the primary experimental 
observables to probe the QCD critical point on the phase diagram
\cite{Hatta:2003wn,Athanasiou:2010kw}. The critical point signal in the 
measurements of event-by-event fluctuations of protons should manifest 
itself in deviations of the corresponding measures from the baseline
expectations that do not incorporate any critical point effects. One simple 
choice for the baseline is a Poisson distribution, which would correspond 
to an uncorrelated proton production. However, the event-by-event fluctuations 
of protons in heavy-ion collisions, especially the high-order cumulants, are 
affected by a number of non-critical mechanisms which make the non-critical 
reference distribution considerably more involved than that given by Poisson 
statistics.

\subsubsection{Exact baryon number conservation}

The total net baryon number in heavy-ion collisions is determined by the colliding 
nuclei and conserved throughout the course of the collision. Baryon conservation 
thus introduces correlations between particles. For instance, any newly created 
baryon has to be counterbalanced by an anti-baryon elsewhere in the fireball in 
order to ensure conservation of total baryon number. For this reason alone, cumulants 
of the proton number distribution will show deviations from Poisson statistics.
One can argue that these corrections are small when one measures fluctuations 
in a small part of the whole system, as in this case the small acceptance ensures
the applicability of the grand-canonical ensemble~\cite{Koch:2008ia}. However, the
effects of baryon conservation become large in high-order proton number cumulants 
that are used in the search for the QCD critical point, as first investigated in Ref.~\cite{Bzdak:2012an} in the framework of ideal gas of baryons and anti-baryons.

Recently, a subensemble acceptance method~(SAM) was developed~\cite{Vovchenko:2020tsr}
that allows one to evaluate the effect of global conservation on cumulants measured 
in a subsystem of the full system. To illustrate the effect of global conservation,
consider the ratio of baryon number cumulants $\kappa_n^B$ inside a subvolume of 
uniform thermal system that are affected by global baryon conservation. These are 
given by~\cite{Vovchenko:2020tsr}
\begin{align}
\frac{\kappa_2^B}{\kappa_1^B} & = (1-\alpha) \, \frac{\chi_2^B}{\chi_1^B}, \\
\label{eq:skew}
\frac{\kappa_3^B}{\kappa_2^B} & = (1-2\alpha) \, \frac{\chi_3^B}{\chi_2^B}, \\
\label{eq:kurt}
\frac{\kappa_4^B}{\kappa_2^B} & = (1-3\alpha \beta) \, \frac{\chi_4^B}{\chi_2^B} - 3 
\alpha \beta \left( \frac{\chi_3^B}{\chi_2^B}\right)^2~.
\end{align}
Here $\alpha$ is a fraction of the total volume covered by the subvolume, $\beta = 
1 - \alpha$, and $\chi_n^B$ are the grand-canonical baryon number susceptibilities.
These expressions demonstrate that the effect of baryon conservation disappears in 
the limit $\alpha \to 0$, but that at small finite $\alpha$ the deviations are larger 
for higher-order cumulants.

More recently, the SAM has been extended to the case of multiple conserved charges
\cite{Vovchenko:2020gne}, as well as non-uniform systems, non-conserved quantities
(like proton numbers), and momentum space acceptances \cite{Vovchenko:2021yen}.
The formalism can be used to either subtract the effect of global conservation of 
multiple charges from experimental data, or to include the effect in theoretical 
calculations of proton number cumulants.

\subsubsection{Repulsive interactions}

Another source of particle correlations may come from short-range repulsive interactions 
between hadrons, commonly modeled through the excluded volume~\cite{Rischke:1991ke}. 
The presence of the excluded volume corrections suppresses the variance of particle 
number fluctuations~\cite{Gorenstein:2007ep}. In particular, the HRG model with 
excluded volume effects in the baryon sector leads to an improved description of 
lattice QCD susceptibilities at temperatures close to the chemical freeze-out in 
heavy-ion collisions~\cite{Vovchenko:2016rkn,Vovchenko:2017xad,Karthein:2021cmb}.
As the excluded volume corresponds to purely repulsive interactions, it does not induce 
criticality, thus it is a source of non-critical fluctuations. Incorporating the 
excluded volume effect in heavy-ion collisions is challenging and requires modifications
to the standard Cooper-Frye particlization. Progress in this direction has recently 
been achieved, either through a Monte Carlo sampling of an interacting hadron 
resonance gas at particlization~\cite{Vovchenko:2020kwg}, or an analytic calculation 
of the proton number cumulants~\cite{Vovchenko:2021kxx}.

\subsubsection{Volume fluctuations}

Event-by-event fluctuations of the system volume, which are linked to the centrality
selection and cannot be avoided completely in heavy-ion collisions, comprise an 
additional source of proton number fluctuations~\cite{Gorenstein:2011vq,Skokov:2012ds}. 
The volume fluctuations generally lead to an enhanced variance of fluctuations, whereas 
their effect on the high-order proton cumulants depends on the corresponding cumulants 
of the volume distribution. The volume fluctuations can also be regarded as a 
manifestation of event-by-event fluctuations in the initial state. It is important 
to incorporate this effect in theoretical calculations to match the relevant 
experimental conditions. In some cases, the effect of volume fluctuations can be
removed~(or minimized) from the experimental data, making the theoretical interpretation
of the data easier~\cite{Luo:2013bmi}.

\subsection{Non-critical baseline from hydrodynamics}

An appropriate non-critical baseline for proton number fluctuations can be obtained
within a dynamical description of heavy-ion collisions which incorporates the essential
non-critical contributions. While some non-critical effects, like baryon conservation, 
can be analyzed without dynamical modeling~\cite{Braun-Munzinger:2020jbk}, modeling is
necessary to treat all the different effects simultaneously. This has recently been
achieved in the work~\cite{Vovchenko:2021kxx} in the framework of (3+1)D relativistic
hydrodynamics applied to 0-5\% central Au-Au collision at RHIC-BES energies. The
simulations utilize collision geometry based 3D initial conditions~\cite{Shen:2020jwv}
and viscous hydrodynamics evolution with a crossover-type equation of state
NEOS-BQS~\cite{Monnai:2019hkn} and simulation parameters adjusted to reproduce bulk
observables. The Cooper-Frye particlization takes place at a constant energy density 
of $\varepsilon_{\rm sw} = 0.26$~GeV/fm$^3$, where the cumulants of the (anti)proton
number are calculated in a given momentum acceptance analytically. The calculations 
take into account both the repulsive interactions and global baryon conservation. 
The former are incorporated in the framework of the excluded volume HRG model, in 
line with the behavior of baryon number susceptibilities observed in lattice QCD 
\cite{Karthein:2021cmb}. The effects of global baryon number conservation are taken 
into account using a generalized subensemble acceptance method \cite{Vovchenko:2021yen}.

\begin{figure*}[ht!]
\centering
\includegraphics[width=.49\textwidth]{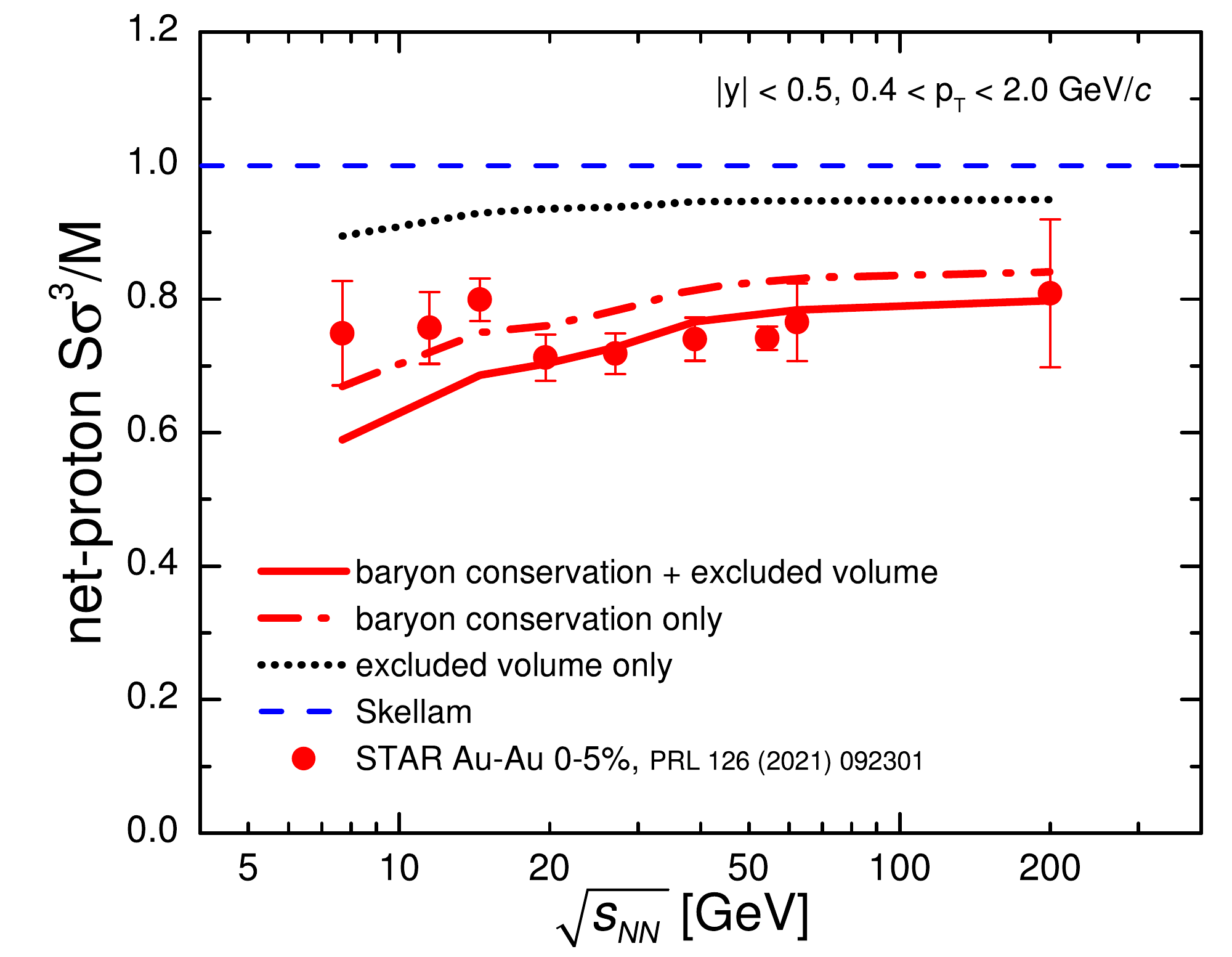}
\includegraphics[width=.49\textwidth]{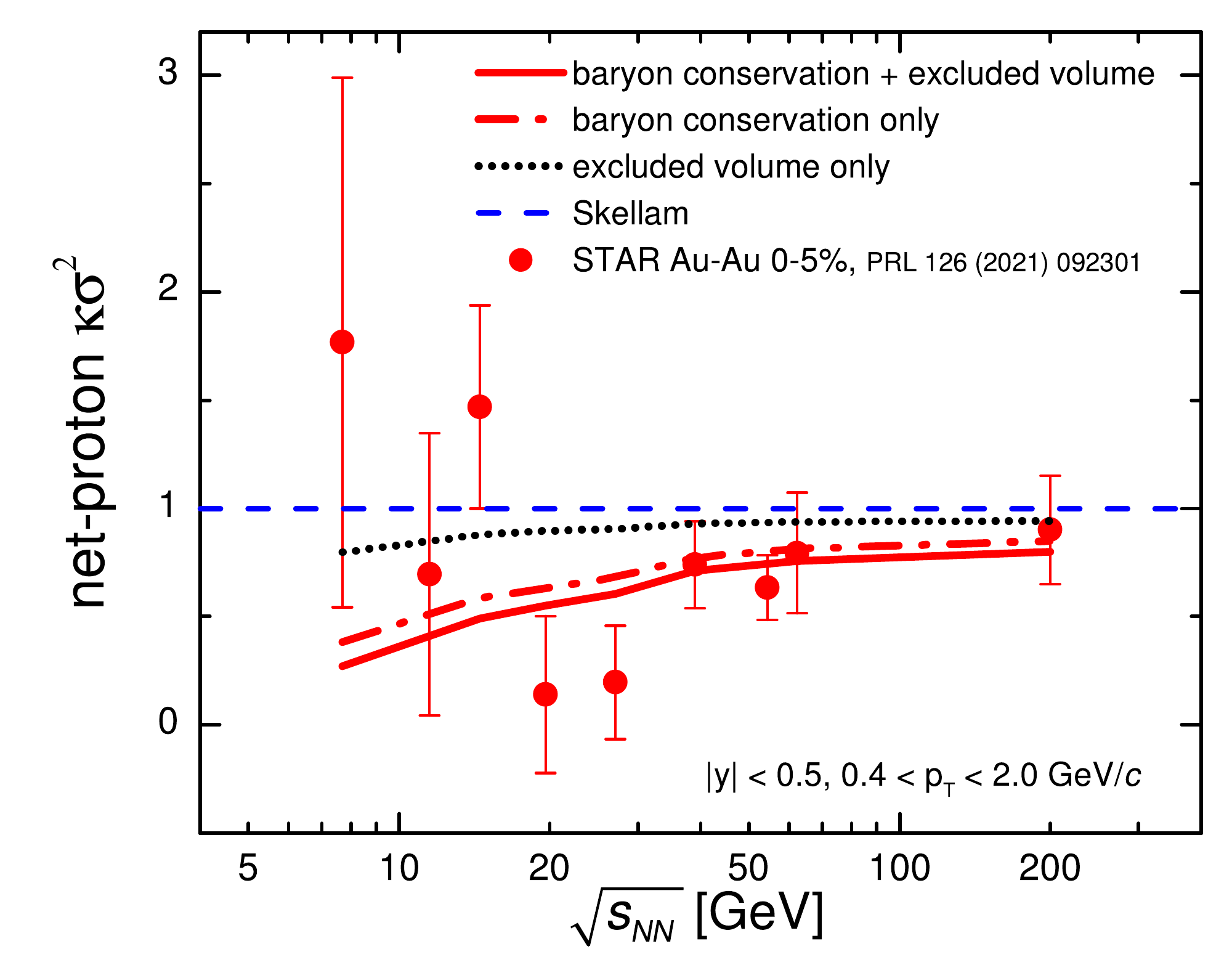}
\caption{
Collision energy dependence of the net-proton cumulant ratios $\kappa_3 / \kappa_1 
\equiv S\sigma^3 / M$~(left) and $\kappa_4 / \kappa_2 \equiv \kappa \sigma^2$~(right) 
in 0-5\% Au-Au collisions at RHIC BES energies in a non-critical scenario compared 
with the STAR measurements~\cite{Vovchenko:2021kxx,STAR:2020tga}.
Calculations use The (3+1)D hydrodynamic medium calibrated in Ref.~\cite{Shen:2020jwv} 
and impose exact global net baryon conservation and excluded volume corrections. 
The figure is adapted from Ref.~\cite{Vovchenko:2021kxx}.
}
\label{fig:BES_NetpCumulants}
\end{figure*}

Figure~\ref{fig:BES_NetpCumulants} shows the collision energy dependence of the net
proton cumulant ratios $\kappa_3/\kappa_1 \equiv S \sigma^3 / M$ and $\kappa_4/\kappa_2
\equiv \kappa \sigma^2$ in comparison with the experimental data of the STAR
Collaboration~\cite{STAR:2020tga}. These ratios are equal to unity in the Skellam
distribution limit of uncorrelated (anti)proton production. Both baryon conservation 
and excluded volume lead to a suppression of these two ratios, which monotonically
increase with collision energy. The stronger effect of baryon conservation at low
energies can be explained by larger fraction of particles ending up at midrapidity
compared to higher energies, while the enhancement of baryon repulsion is due to 
larger baryon densities achieved at lower $\sqrt{s_{\rm NN}}$. It is also clear that
baryon conservation has a larger influence at all energies compared to the excluded 
volume. However, both effects are necessary to obtain a quantitative description of 
the $S \sigma^3 / M$ data at $\sqrt{s_{\rm NN}} \geq 20$~GeV. At lower energies the 
data indicate a smaller suppression of $S \sigma^3 / M$ than predicted by the
calculation. As for the $\kappa \sigma^2$, the STAR data show possible indications 
for a non-monotonic collision energy dependence which is not observed in the baseline
calculation, however, more precise data at the lowest energies are required to make a 
robust conclusion.

\begin{figure}[ht!]
\centering
\includegraphics[width=0.8\linewidth]{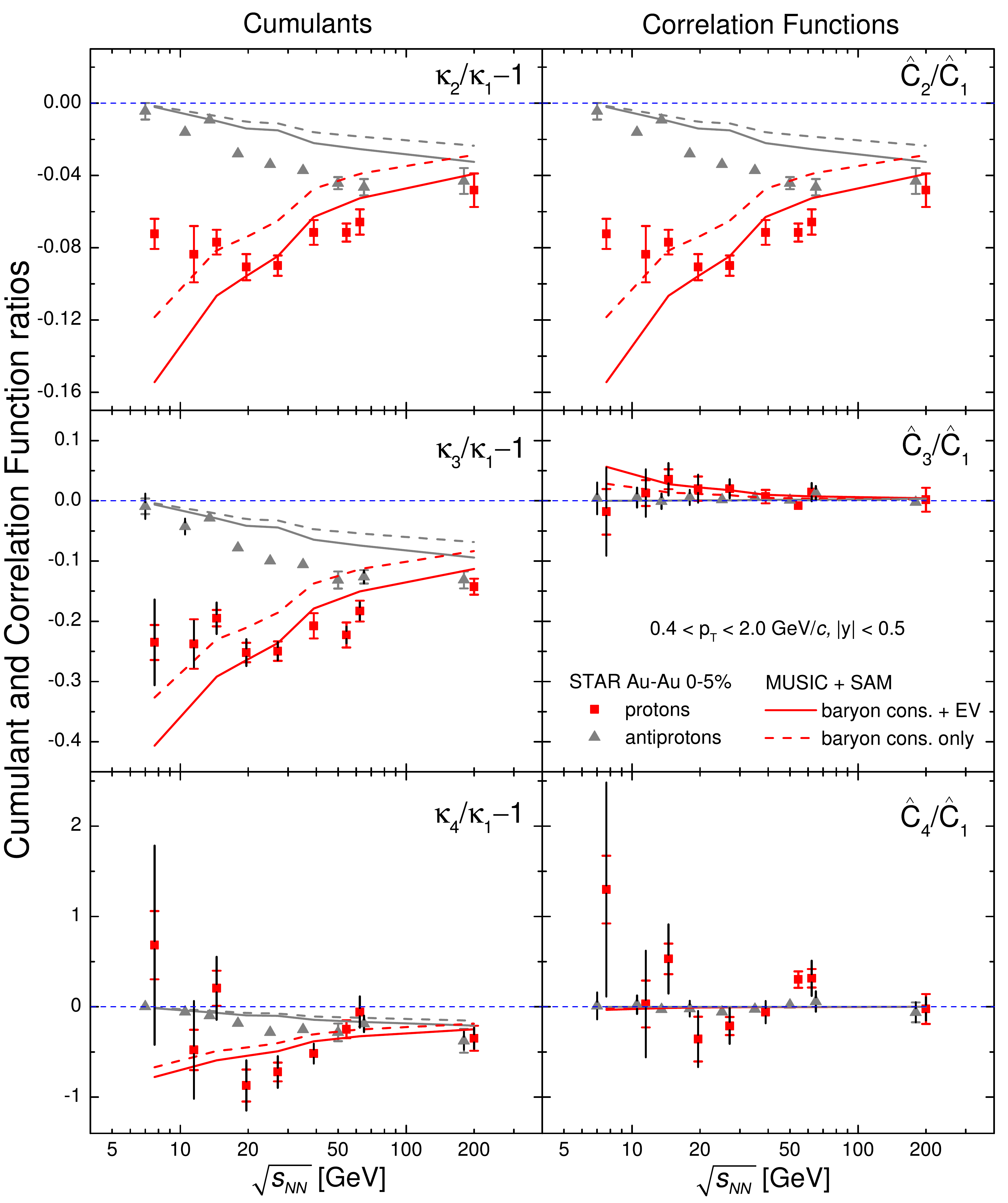}
\caption{
The scaled proton and antiproton cumulants and factorial cumulants (correlation 
functions) in 0-5\% Au+Au collisions at RHIC BES energies in a non-critical scenario 
compared with the STAR measurements~\cite{Vovchenko:2021kxx,STAR:2021iop}.
Calculations use The (3+1)D hydrodynamic medium calibrated in Ref.~\cite{Shen:2020jwv} 
and impose exact global net baryon conservation and excluded volume corrections. 
The figure is adapted from Ref.~\cite{Vovchenko:2021kxx}. }
\label{fig:BES_CorrelationFunctions}
\end{figure}

Additional insights can be gained by analyzing proton and antiproton distributions
separately. In particular, one can study, in addition to the ordinary cumulants, 
the (anti)proton correlation functions~(factorial cumulants) $\hat{C}_k$, which 
probe genuine multi-particle correlations and thus should be sensitive probes of 
the critical behavior. Figure~\ref{fig:BES_CorrelationFunctions} shows the collision
energy dependence of the scaled factorial cumulants $\hat{C}_2/\hat{C}_1$,
$\hat{C}_3/\hat{C}_1$, and $\hat{C}_4/\hat{C}_1$ of protons and antiprotons.
The second factorial cumulants of both the protons and antiprotons indicate negative 
two-particle correlations. The results for protons agree with the experimental data 
at $\sqrt{s_{\rm NN}} \geq 20$~GeV but overestimate the strength of negative 
correlations at lower collision energies. The data for antiprotons are reproduced
qualitatively, however, in contrast to the protons, here the calculation underestimates 
the negative two-particle correlations in the collision energy range 19.6 GeV $\leq
\sqrt{s_{\rm NN}} \leq$ 62.4~GeV.

The high-order factorial cumulants, $\hat{C}_3/\hat{C}_1$  and $\hat{C}_4/\hat{C}_1$, 
exhibit a behavior which is quite different from the corresponding ordinary cumulants.
The calculations indicate the presence of only mild multi-particle correlations among 
protons in the non-critical scenario. The baryon conservation and excluded volume 
effects lead to small positive $\hat{C}_3/\hat{C}_1$, which agrees with the available
experimental data. For the four-particle correlations one obtains $|\hat{C}_4/
\hat{C}_1 | \ll 1$. This also agrees with the available experimental data within error 
bars, although the errors in the data are large for $\sqrt{s_{\rm NN}} \leq 20$~GeV.
If future measurements establish sizable multi-proton correlations, then this result  
would be difficult to describe in a non-critical scenario.

\section{Particlization and kinetic transport}
\label{sec:transport}

\subsection{Introduction and overview}

The goal of this section is to provide an overview over advances made by the BEST Collaboration 
in the area of particlization and kinetic transport in the hadronic phase, in particular:

\begin{itemize}
    \item Microcanonical particlization
    \item Particlization of hydro+
    \item Hadronic transport with adjustable mean-field potentials
\end{itemize}

Dynamical models of heavy-ion collisions typically involve both a hydrodynamic 
stage and a hadronic transport stage. Hydrodynamic description applies in systems 
that are largely equilibrated. However, as the system expands, the matter cools 
and hadronizes. Once hadrons form, it becomes increasingly difficult to maintain 
local equilibrium. A hadron gas consists of hundreds of species with masses 
ranging from 135 MeV/$c^2$ to a few GeV/$c^2$. As the matter cools and becomes 
more dilute, inelastic collisions are too rare to maintain chemical equilibrium,
and chemical equilibration times are much longer than the lifetime of the hadronic
stage \cite{Prakash:1993bt,Rose:2020lfc}. Even local kinetic equilibrium 
is difficult to maintain, in part due to the wide range of masses. A gas of 
non-relativistic particles, with masses much larger than the temperature $T$, 
cools such that $T\sim 1/V^{2/3}$ as the volume $V$ increases, whereas massless
particles cool as $T\sim 1/V^{1/3}$. Thus, without a high collision rate heavy 
particles cool faster than light particles. Further, lighter particles, 
due to their higher thermal velocities, tend to diffuse through their heavier 
neighbors in a phenomenon known as the pion  wind. Once the species no longer 
flow together, a hydrodynamic description is no longer appropriate 
\cite{Pratt:1998gt}. This stage is most suitably described by molecular dynamics 
or Boltzmann approaches. In a molecular dynamics approach one samples representative
particles, which collide and interact with one another by the experimentally known or 
modeled cross sections of hadrons in a dilute environment. Microscopic models of this 
type are sometimes referred to as {\em afterburners}. If one over-samples the 
particles by some factor $N_{\rm sample}$ and assigns each test particle a reduced 
charge of $1/N_{\rm sample}$ while at the same time reducing the cross sections by 
the same factor, one arrives at a test particle representation
of the Boltzmann limit.

Several physical observables are modified during the afterburner stage: 

\begin{itemize}
\item Spectra and flow. Hadronic kinetics typically does not affect pion and kaon yields 
as well as spectra by more than few percent. Proton yields can be changed by around 10\% to 30\%, depending on collision energy, 
mainly due to annihilations. Baryon spectra shift towards higher transverse momenta due 
to the ``pion wind'' effect. Elliptic and radial flow are usually increased by an afterburner
stage. These effects are demonstrated in Fig.\ \ref{fig:hydro:afterburner}. Overall, the impact 
of the afterburner is significant, but at least some of the effects can be taken into account 
by implementing partial chemical equilibrium in the hydrodynamic stage \cite{Motornenko:2019jha, Oliinychenko:2021enj}. Given the uncertainties 
of particlization and the afterburner itself (e.g. unknown cross sections, resonance properties,
difficulty or impossibility of implementing multi-particle reactions, uncertainties of in-medium 
interactions), this can be an efficient approach at the higher energies, however, it has not been
tested in the case of afterburners that include mean-field effects.
    
\item Fluctuations and correlations. The extent to which fluctuations are affected by 
hadronic kinetics is still being investigated. There are indications that diffusion due to 
rescattering in the final stage smears particle distributions and therefore changes fluctuation
observables \cite{Steinheimer:2016cir}. On the other hand, a direct check of net-charge,
net-proton, and net-kaon correlations shows only minor afterburner effects on these 
observables, despite confirming the presence of diffusion and isospin randomization \cite{Dima:tbd}. 
Overall, the role of the afterburner for fluctuations and correlations requires further studies.

\item The role of mean-field potentials in hadronic transport in the energy range explored
by the Beam Energy Scan is not well studied, because the inclusion of mean-field potentials 
is computationally expensive, and additionally the potentials are not well constrained theoretically
at high baryon densities. Mean-field potentials change the equation of state in the low-temperature
high-baryon-density region, where lattice QCD calculations at present cannot provide theoretical 
inputs. Using adjustable mean-field potentials one can explore the sensitivity of observables 
to the equation of state in this region.
\end{itemize}

The interface between the hydrodynamic and microscopic stages is known as {\em particlization}.
During this process the energy, momentum and charge carried by the fluid is transformed into 
a distribution of particles, which on average reproduces the conserved quantities. Building 
such an interface requires confronting several issues. First, the hyper-surface that separates 
the hydrodynamic and microscopic domains, known as the particlization surface, moves relative 
to the fluid. For most of the emission the particlization surface is spacelike, and there is no 
possibility for particles to re-enter the hydrodynamic domain from the hadronic stage. However, 
in some instances the particlization surface may be timelike, similarly as in the case of 
evaporation from a static surface, and then one must consider the effect of trajectories 
reentering the hydrodynamic region, a phenomenon known as backflow \cite{Oliinychenko:2014tqa}. 
Different algorithms handle backflow differently. Another place where algorithms vary is 
in the implementation of viscous corrections. These variations can alter the values of $v_2$, 
especially at high $p_t$. A third challenge facing the interface involves the implementation of 
local charge, energy, and momentum conservation. It is critical to account for such effects when 
analyzing correlations and fluctuations. Many of the hadrons generated at the interface are 
resonances, with large widths $\sim 100$ MeV. Because these widths are not much less than the 
temperature, accounting for the widths is also critical. Finally, the hadronic simulation might 
also incorporate mean fields. These can alter yields or masses, and the interface must be designed 
so that energy and momentum is preserved throughout the particlization process in such a way 
that the particle distributions are thermodynamically consistent. 

Although many of the challenges listed above had been addressed prior to the efforts of the BEST
Collaboration, additional progress was made in several areas. This includes the implementation of
conservation laws and finite resonance widths at the hydrodynamic interface, and the implementation 
of mean fields. For this report we focus on three areas where the BEST contributions are particularly
significant.

\begin{itemize}
\item Microcanonical particlization \cite{Oliinychenko:2019zfk, Oliinychenko:2020cmr}. We have developed a method that takes into account local
conservation of charge and momentum in the particlization process, which is a crucial ingredient 
for a proper description of fluctuation measurements .

\item Particlization of Hydro+. We have shown how to imprint fluctuations from Hydro+ onto 
produced hadrons. This allows us to quantify manifestations of critical behavior in 
final-state measurements.

\item Hadronic transport with adjustable potentials. Mean-field potentials have previously
been implemented at lower (e.g., GSI) energies, where the high baryon densities lead to 
large effects. These effects have been largely ignored at the highest RHIC energies or at the 
LHC. For BES energies, the effects should again be important, but unlike at the lower energies 
one must account for hundreds of hadronic species.
\end{itemize}

In this section we first briefly summarize recent progress in handling particlization in case fluctuations and correlations are of interest. A local microcanonical approach is suitable for fluctuating
hydrodynamics, where fluctuations are realized as fluctuations in a set of hydrodynamical 
simulations. Another approach is required for particlization in hydro+, where second-order 
correlations and fluctuations are available already at the hydrodynamic stage and need to be 
transferred correctly to particles. Then we proceed to discussing the hadronic afterburner 
with adjustable mean-field potentials

\subsection{Local microcanonical particlization of fluctuating hydrodynamics}

\begin{figure}
\centering
\includegraphics[width=0.75\textwidth]{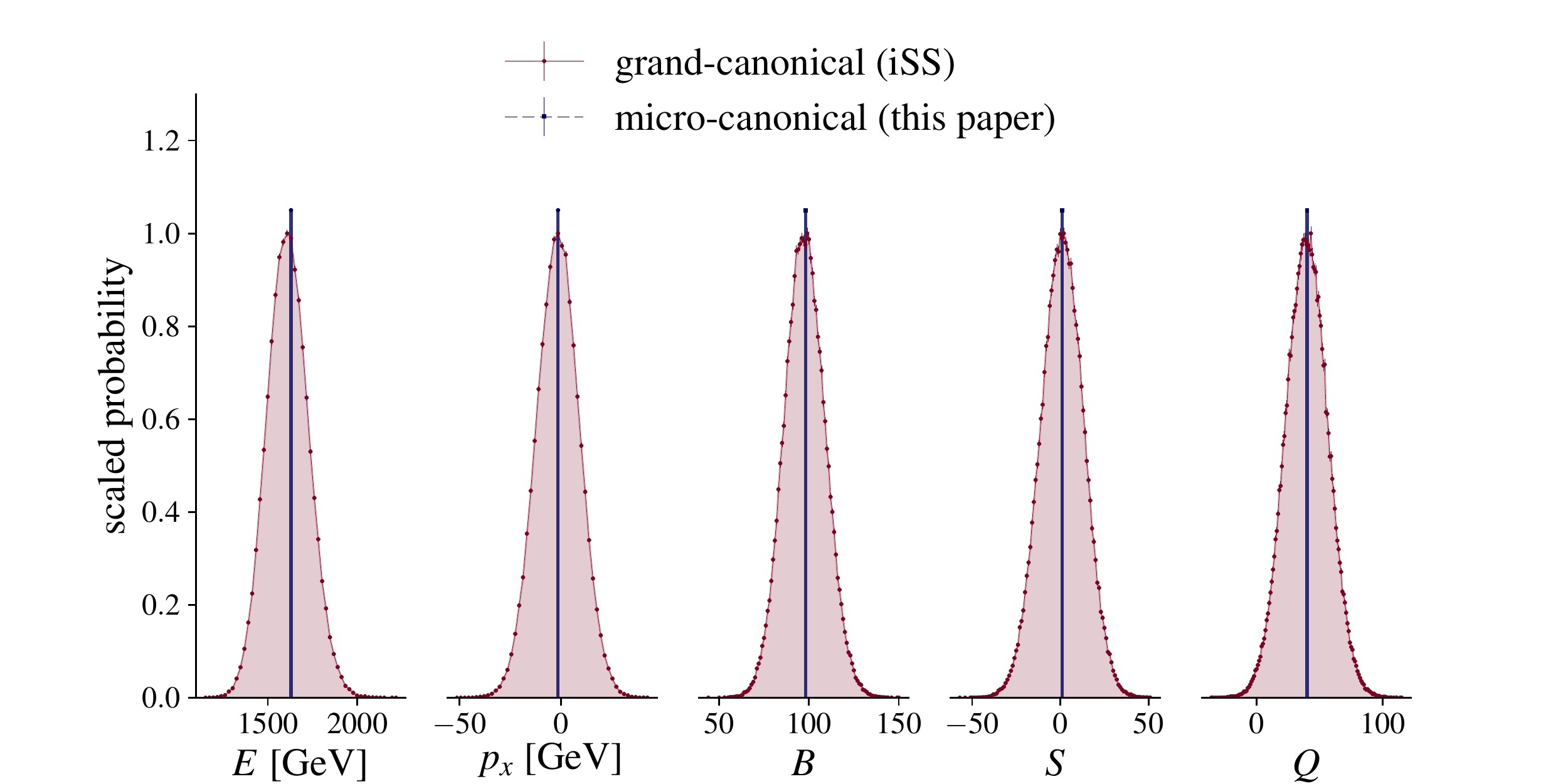}
\caption{Comparing distributions of energy, $x$-component of momentum, baryon number,
strangeness, and electric change from the standard grand-canonical and microcanonical 
particlization. For viewing convenience distributions are scaled to reach maximum of 1 (for 
grand-canonical) or 1.05 (for micro-canonical). Figure taken from \cite{Oliinychenko:2020cmr}.}
\label{fig:microcanonical_particlization}
\end{figure}

 Hydrodynamic approaches with stochastic terms produce an ensemble of events which encode 
the  fluctuations and correlations. For example, consider fluctuations of baryon number $B$ 
in a certain rapidity and transverse momentum window. Suppose that there are $N_{hydro}$ runs, 
and in each of them this baryon number is different. The distribution of $B$ includes both 
thermal and non-thermal (critical, initial state) fluctuations. Therefore, the sampling should 
not introduce additional thermal fluctuations, because they are already present in the ensemble 
of the hydro runs. For this one needs microcanonical sampling.

The concept of microcanonical sampling is shown in Fig.\ \ref{fig:microcanonical_particlization}. 
In contrast to the usual grand-canonical sampling, where energy, net baryon number, net 
strangeness, electric charge are conserved on average over samples, in microcanonical sampling 
they are conserved in each sample. In \cite{Oliinychenko:2019zfk} we have proposed a mathematical 
method for implementing microcanonical particlization and introduced the concept of patches -- 
compact space-time regions on the particlization hypersurface, where conservation laws are enforced.
Some methods to obtain the correct fixed energy, momentum, and charges in every sample were 
suggested previously, see \cite{Schwarz:2017bdg} for overview. However, none of these earlier approaches produces 
the correct microcanonical distribution in the simple limiting case of microcanonical sampling 
in a static, uniform box. In the follow-up work \cite{Oliinychenko:2020cmr} we have tested our 
method in a realistic setup and explored some effects of the microcanonical sampling on heavy-ion observables. The main conclusions of the work are the following:

\begin{itemize}
\item The decomposition of the hypersurface into patches on which the conservation laws are
    enforced can be controlled by a parameter -- the rest frame energy of the patch in our case.
    Even after this parameter is fixed, there is a considerable freedom in selecting the location
    of patches. However, the effect of this additional freedom on observables is significantly
    smaller than the effect of the patch energy parameter. This allows for meaningful applications 
    of our method.
    
\item Fluctuations and correlations are significantly affected by event-by-event 
    conservation laws. Mean values, such as spectra and flow, are only affected in small 
    systems, or in the limit of small patches.
\end{itemize}

We have integrated our open-source microcanonical sampler into a framework with the SMASH hadronic
afterburner and found that the effects of microcanonical sampler on fluctuations and correlations
survive until the end of the afterburner evolution \cite{Oliinychenko:2020cmr}. This means that 
if one studies correlations and fluctuations with an afterburner, then a microcanonical sampler 
is required for consistency. 

\subsection{Freezing out critical fluctuations}
\label{sec:part-crit}

The input data for particlization in Hydro+ is different from that in stochastic even-by-event
hydrodynamics. In stochastic hydrodynamics correlations and fluctuations are embedded in an 
ensemble of hydrodynamic events. In contrast, Hydro+ simulations directly provide the mean
and the two-point correlation functions of the hydrodynamic densities near the critical point. 
In this section we show how particle correlations and fluctuations can be computed in this case,
and we present some initial results, see Ref.~\cite{Pradeep}. These results illustrate how to
translate correlations on a Hydro+ freeze-out hypersurface into predictions for experimental 
observables such as particle multiplicities and their cumulants. Ref.~\cite{Pradeep} proposes 
a freeze-out procedure to convert the critical fluctuations in the hydrodynamic stage into 
cumulants of particle multiplicities. The idea is to introduce  a critical sigma field, so 
that fluctuations of the field are imprinted on the observed hadrons due to the coupling of 
the sigma field to hadrons. 

 As explained in the previous section, the traditional Cooper-Frye procedure~\cite{Cooper:1974mv}
matches only the averages of the conserved densities between the hydrodynamic and particle
descriptions on the freeze-out hypersurface. This is inadequate near a critical point. To ensure 
that the two-point correlation functions describing the critical fluctuations are carried over 
to the particle description, one needs to employ an extended freeze-out prescription near the 
critical point.

\begin{figure}
\includegraphics[scale=0.42]{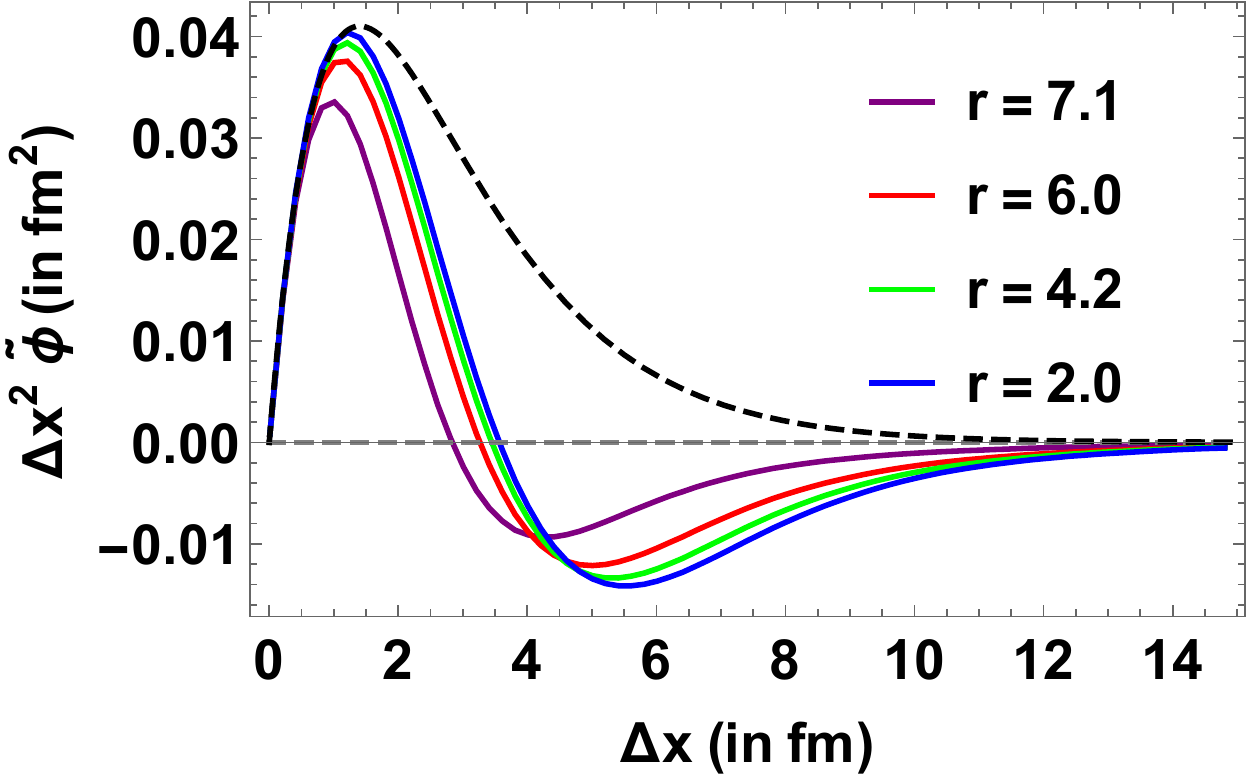}
\includegraphics[scale=0.4]{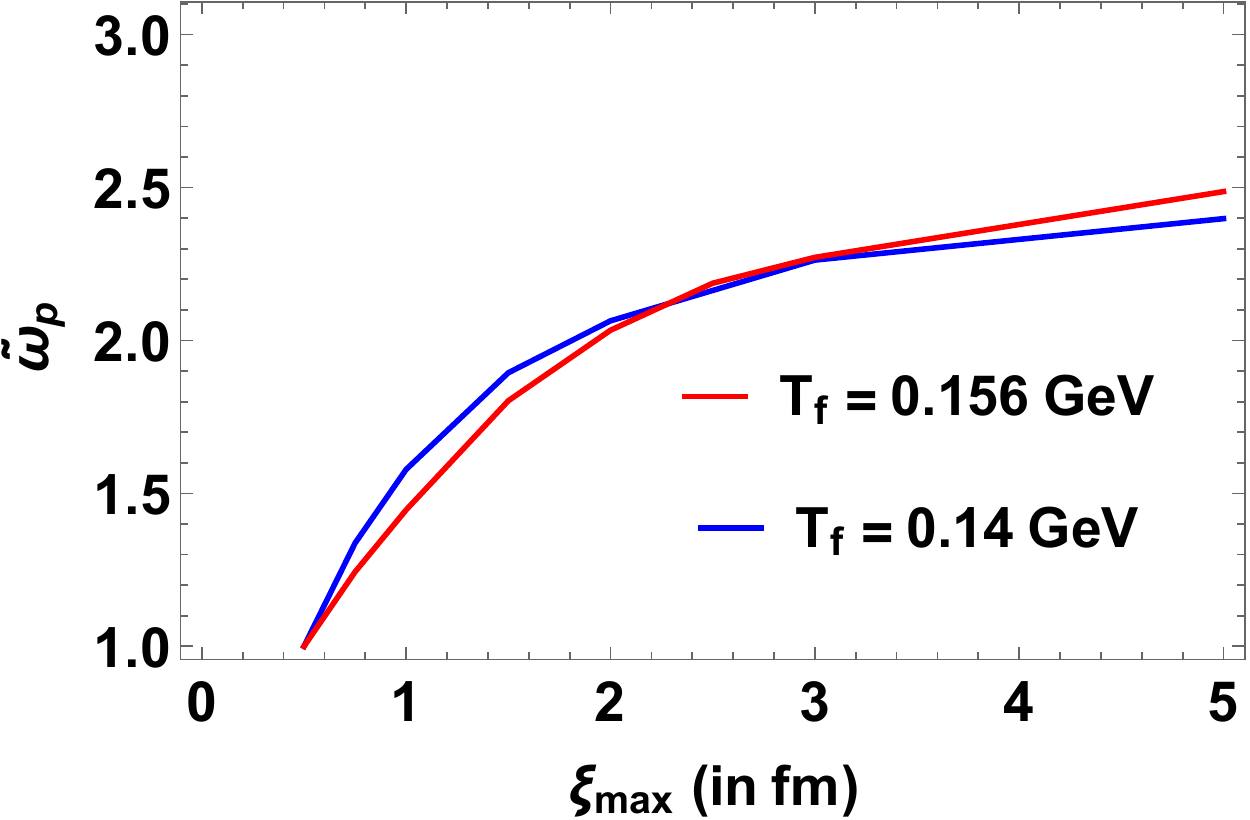}
\includegraphics[scale=0.4]{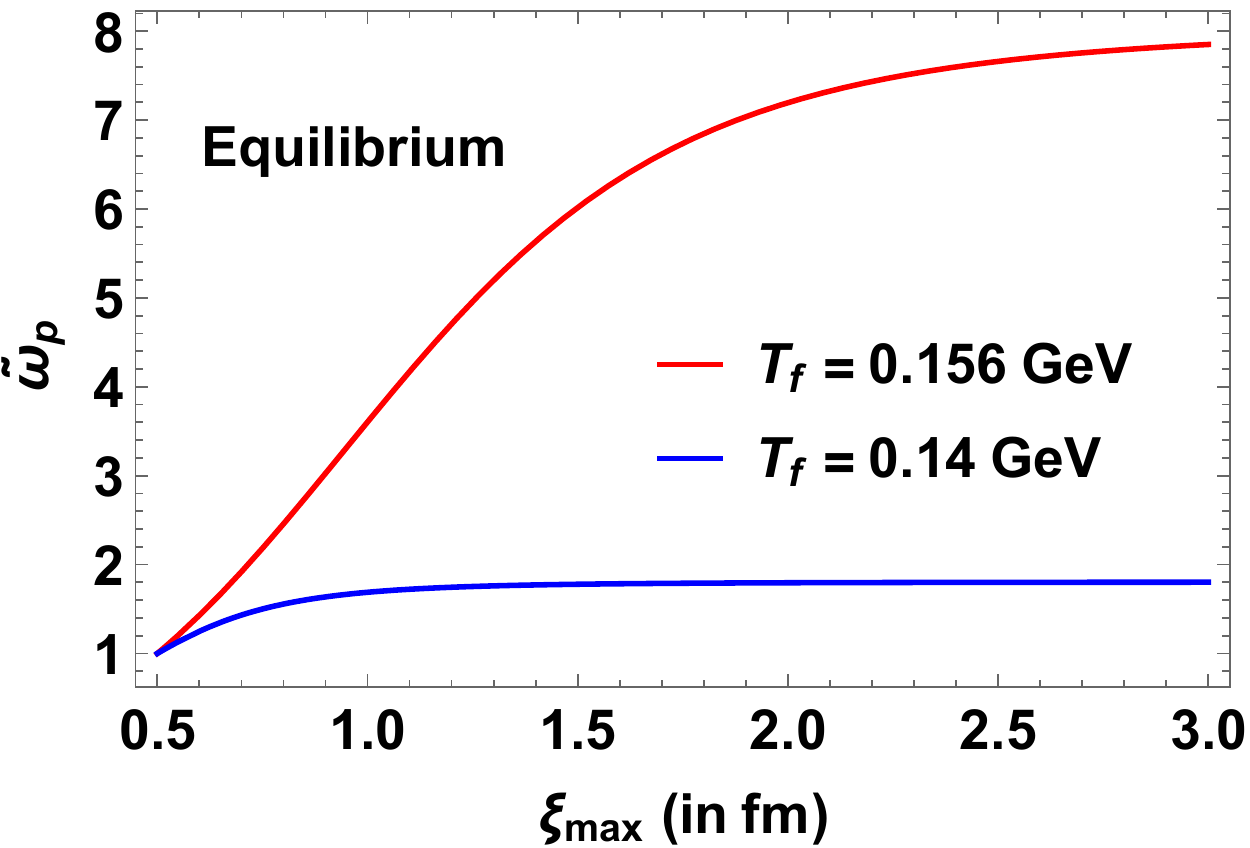}
\caption{\textit{Left}: 
The two-point correlation of $s/n$ in an out-of-equilibrium scenario is plotted 
as a function of the  separation $\Delta x = \sqrt{\Delta x_\perp^2+ (\tau\Delta\eta)^2}$ 
between two points centered on four different locations on the freeze-out hypersurface which 
are labeled by their $r$ values. The dashed black curve in this plot is the equilibrium
expectation. \textit{Center}:  $\tilde\omega$ for protons (defined in
Eq.~(\ref{omega1hydroplosfreezeout})) as a function of the maximum equilibrium correlation length along the system's trajectory, $\xi_{\rm max}$ is plotted for two different isothermal freeze-out scenarios. \textit{Right}:  $\tilde\omega$ for protons if the fluctuations were fully equilibrated at freeze-out is plotted for the same isothermal freeze-out scenarios depicted in the \textit{center} plot. Comparing the \textit{center} and \textit{right} plots,  it can be inferred that the sensitivity of $\tilde{\omega}_A$ to the freeze-out temperature is significantly reduced when the fluctuations are away from equilibrium.}
\label{hydro+particlisationfigure}
\end{figure}

	The critical fluctuations are incorporated in the kinetic (particle) description via an 
effective coupling between the particles and the critical sigma field, $\sigma$, which modifies 
the masses of the particles. The modified particle distribution function is given by:
\begin{eqnarray} \label{fA}
f_A(x,p)=\left<f_A(x,p)\right>
	 + g_A\frac{\partial \left<f_A(x,p)\right>}{\partial m_A}\, \sigma(x)
\end{eqnarray}
Here, $A$ denotes the particle species and $\left<f_A(x,p)\right>$ is the particle distribution
function without including the critical effects, which is taken to be the Boltzmann distribution
function. The coupling $g_A$ measures the strength of the interaction between the particles of
species $A$ and the sigma field. In the preliminary study only pions, nucleons and their
anti-particles were included, but the model can be extended to a full hadron resonance gas in 
a straightforward manner. The field $\sigma$ is a stochastic variable such that:
\begin{eqnarray} \label{eq:hydroplus_sigmafield}
\left<\sigma(x)\right>=0 \, ,\quad 
\left<\sigma(x_+)\sigma(x_-)\right>=Z\,\tilde{\phi} (x,\Delta x)\equiv 
   Z\,\left<\delta \frac{s}{n}(x_+)\delta \frac{s}{n}(x_-)\right> \ ,
\end{eqnarray}	
where $Z$ is an appropriately chosen normalization factor. The two-point correlation function 
of the entropy per baryon, denoted as $\tilde{\phi}$, is obtained from  Hydro+ simulations, 
for example those of Ref.~\cite{Rajagopal:2019xwg}.  The resulting modification to the particle
masses changes the variance of the particle multiplicity distributions, 
 \begin{equation}\label{delNa}
\left<\delta N_{A}^2\right>_\sigma
   = g^2_{A} \int dS_{\mu} J^{\mu}_{A}(x_{+})\int dS^{'}_{\nu} J^{\nu}_{A}(x_{-})
   \left<\sigma(x_+)\sigma(x_-)\right> \, . 
\end{equation}
Here, $dS_\mu$ and $dS^{'}_\nu$ are differential elements on the freeze-out hypersurface 
pointing along the direction of the normal at $x_+$ and $x_-$ respectively. $J^\mu_A$ in
Eq.~(\ref{delNa}) is given by:
\begin{equation} 
J^{\mu}_A=2\,d_A\,\int \,\frac{d^{4}p}{(2\pi)^{3}} 
    \delta(p^2-m_A^2) \, p^{\mu}\,\frac{\partial \left<f_A\right>}{\partial m_A}\ .
\end{equation}
The result $\left<\delta N_{A}^2\right>_\sigma$ in Eq.~(\ref{delNa})  gives an estimate 
of critical effects in the variance of the particle multiplicity. As an exploratory study, this procedure is used 
to freeze out the system generated in a Hydro+ simulation obeying Model H relaxation dynamics 
in an azimuthally symmetric boost invariant background (see Sec. \ref{sec:theory_fluct}). The ratio of the variance defined in Eq.~(\ref{delNa}) to the mean multiplicity is denoted as $\omega_A$,
\begin{eqnarray}\label{omegahidyoplusfreezeout0}
\omega_A=\frac{\left<\delta 
N_A^2\right>_\sigma}{\left<N_A\right>}
\end{eqnarray}
The excess of the critical fluctuations over the non-critical baseline can be quantified, via $\tilde{\omega}_A$  defined below
\begin{eqnarray}\label{omega1hydroplosfreezeout}
\tilde{\omega}_A=\frac{\omega_A}{\omega_A^{\text{nc}}}\ ,
\end{eqnarray}
where $\omega_A^{\text{nc}}$ is the estimate for $\omega_A$ when the correlation length is microscopic and equal to some non-critical value. $\tilde{\omega}_A$  obtained within the Hydro+ framework for the simulation from Ref.~\cite{Rajagopal:2019xwg}
is shown in Fig.~\ref{hydro+particlisationfigure}.

This procedure can be extended to higher moments and employed to calculate higher cumulants 
of particle multiplicity once higher order fluctuations from hydrodynamic simulations are 
available. This work has the potential to quantitatively addresses the effects of critical 
slowing down and conservation laws on particle number cumulants, assuming that these
observables are not substantially modified by the hadronic transport stage. These modifications
could be studied by generating an ensemble of $\sigma$ field configurations satisfying Eq.\ 
(\ref{eq:hydroplus_sigmafield}), and then propagating particles through the kinetic regime.

\subsection{Hadronic afterburner with adjustable mean-field potentials}

Lattice QCD calculations of the equation of state of nuclear matter at finite baryon density
can only reach chemical potentials $\mu_B$ up to about 400 MeV and temperatures down to around 
120 MeV, which corresponds to maximum baryon densities around $\frac{1}{4} n_0$. For higher 
baryon density and lower temperature, an ideal hadron resonance gas equation of state is often
assumed in the simulations. However, in this (vast!) region of the phase diagram there is a 
nuclear liquid-gas phase transition, and the possible first-order QCD phase transition between
hadrons and the quark-gluon plasma. An afterburner with adjustable mean-field potentials can 
provide a versatile equation of state in this region and allows studies of the following 
questions:
\begin{itemize}
\item What region of $(T,\mu_B)$ or $(T,n_B)$ can one probe, in principle, through heavy-ion collisions? Note that while some regions might be reached in practice, there might be no 
observable sensitive enough to signal such an occurrence. 
\item How sensitive are the observables to the equation of state in the high-density 
region? How much do the nuclear liquid-gas and the possible QCD phase transitions 
influence observables at RHIC BES energies?
\end{itemize}
Besides these questions, adjustable potentials are useful to smoothly match the equation of 
state used in hydrodynamics to the one realized in kinetic transport.

Despite these features, afterburners are often run in the cascade mode, in which mean-field interactions between
hadrons are neglected. Although this is done in large part to achieve better numerical efficiency,
another reason for doing so is the fact that the mean-field potentials commonly used in hadronic
transport are only fit to reproduce the behavior of cold nuclear matter, and as such do not contain
information on the possible influence of the QGP phase transition on the nuclear matter EOS. However,
this means that the role of many-body interactions in the hadronic stage of a heavy-ion collision
evolution is largely unexplored, and it is possible that transport simulations are missing important
effects at high baryon densities, where both the mean-fields and the time that the system spends in a
hadronic state are substantial. 

\begin{figure}[ht]
\includegraphics[width=\textwidth]{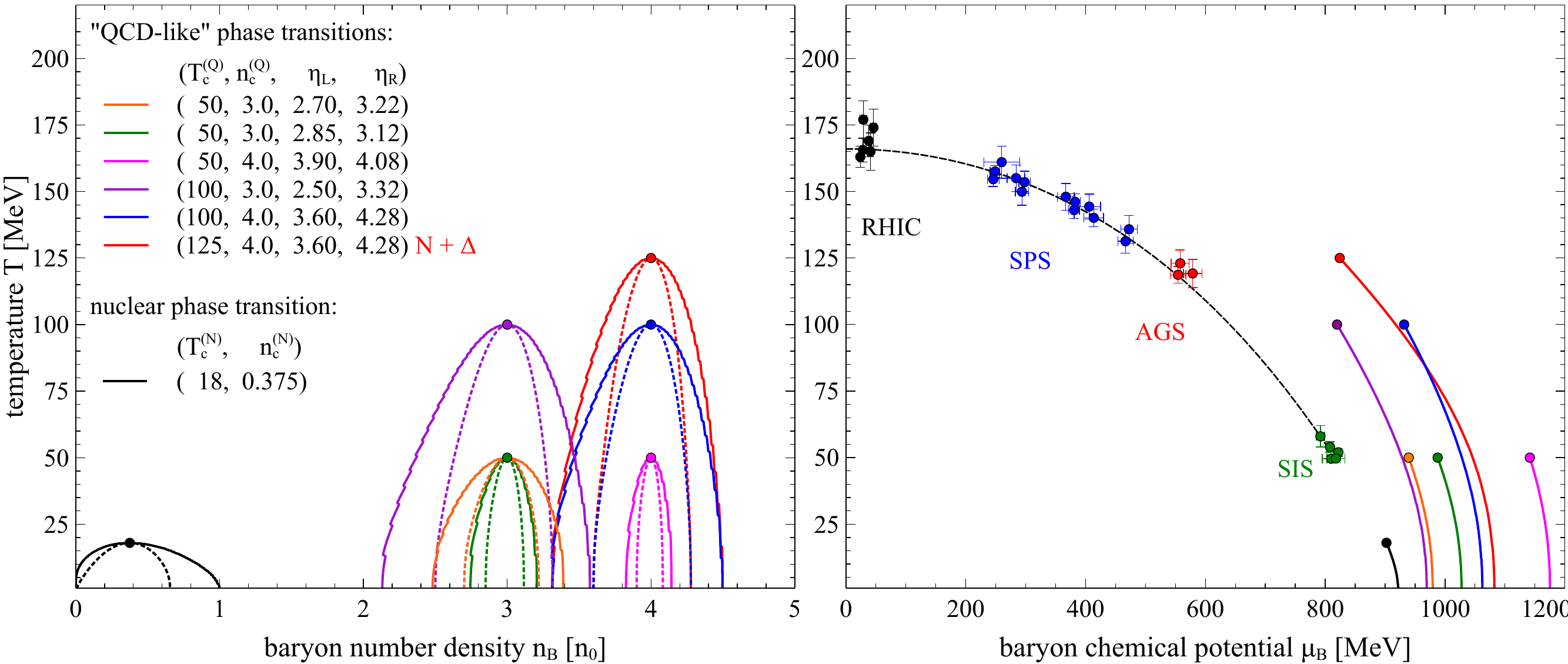}
\caption{Phase diagram in the $T$-$n_B$ (left panel) and $T$-$\mu_B$ (right panel) planes for
different nuclear matter equations of state, characterized by the properties given in the 
legend \cite{Sorensen:2020ygf}. Solid and dashed lines represent the boundaries of the coexistence and spinodal regions,
respectively. In the legend, the critical temperature of the "QGP-like" phase transition 
$T_c^{(Q)}$ is given in MeV, while the critical density $n_c^{(Q)}$ and the boundaries of the 
spinodal region, $\eta_{L}$ and $\eta_R$, are given in units of saturation density, $n_0 = 
0.160~\text{fm}^{-3}$. The coexistence and spinodal regions of the nuclear phase transition,
depicted with black solid and dashed black lines, respectively, are common to all presented EOSs. 
Also shown are chemical freeze-out points obtained in experiment and a parameterization of the 
freeze-out line, taken from~\cite{Cleymans:2005xv}. }
\label{phase_diagram} 
\end{figure}

To address this issue, we developed a vector density functional (VDF) model of the nuclear matter equation 
of state (EOS) \cite{Sorensen:2020ygf}. This functional can be easily parameterized to reproduce a given 
set of the properties of a nuclear matter EOS, and at the same time it leads to relativistic single-particle dynamics 
that allows for a numerically efficient implementation in a hadronic transport code.

For applications to heavy-ion collision simulations, we fit the VDF EOS to describe hadronic matter 
with a phase diagram that contains two first-order phase transitions. The first transition is the
experimentally observed low-temperature, low-density phase transition in nuclear matter, sometimes 
known as the nuclear liquid-gas transition. The second is a postulated high-temperature, high-density 
phase transition that is intended to correspond to the QCD phase transition. Because the degrees of 
freedom employed in the VDF model are baryons and not quarks and gluons, we will refer to the latter 
of the described phase transitions as a "QGP-like" phase transition.

In this variant of the VDF model the pressure takes a simple form
\begin{eqnarray}
P &=& g\int \frac{d^3 p}{(2\pi)^3}~ T ~ 
\ln \Big[ 1 + e^{-\beta (\varepsilon_{\mathbf{p}} - \mu)}  \Big]  +  
\sum_{i=1}^4 C_i \frac{b_i - 1}{b_i} n_B^{b_i} ~,
\end{eqnarray}
where $\varepsilon_{\mathbf{p}}$ is the quasiparticle energy and $n_B$ denotes the baryon number density. 
The corresponding single-particle equations of motion are
\begin{eqnarray}
&& \frac{dx^i}{dt} = \frac{p^i - \sum_{n=1}^N (A_n)^i}{\epsilon^{(N)}_{\mathrm{kin}}}~,
\label{equation_of_motion_x_generalized} \\
&& \frac{dp^i}{dt}  = \frac{\Big(p^k - \sum_{n=1}^N (A_n)^k\Big)}{ \epsilon^{(N)}_{\mathrm{kin}}} 
\bigg( \sum_{n=1}^N \frac{\partial(A_n)_k}{\partial x_i}\bigg) 
  + \sum_{n=1}^N \frac{\partial A_n^0}{\partial x_i}  ~,
\label{equation_of_motion_p_generalized}
\end{eqnarray}
where $A^{\lambda}$ is a vector field associated with the baryon current $j^{\lambda}$,
\begin{eqnarray}
A_{n}^{\lambda} (x; C_n, b_n) \equiv C_n 
\big( j_{\mu} j^{\mu}\big)^{\frac{b_n}{2} - 1} j^{\lambda}~.
\end{eqnarray}
In the above formulae, $\{ C_i, b_i\}$ are interaction parameters that are fixed by requiring that
the EOS reproduces the desired behavior of nuclear matter, including features of the phase transitions.

\begin{figure}
\includegraphics[width=\textwidth]{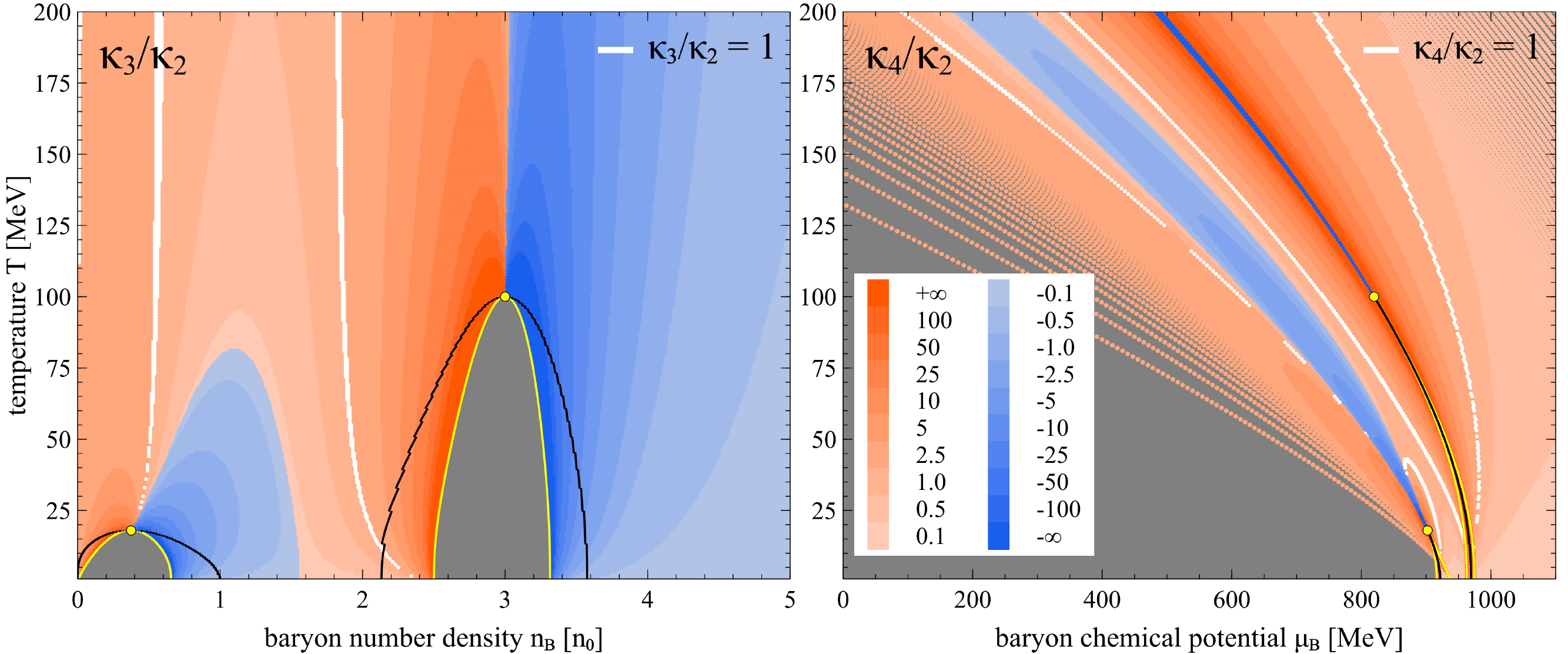}
\caption{Contour plots of cumulant ratios for a chosen example of a VDF EOS \cite{Sorensen:2020ygf}, represented by a purple line on 
Fig.~\ref{phase_diagram}. Left panel: $\kappa_3/\kappa_2$, shown in the $T$-$n_B$ plane. 
Right panel: $\kappa_4/\kappa_2$, shown in the $T$-$\mu_B$ plane. Black lines denote coexistence 
regions, while yellow lines denote spinodal regions; critical points are indicated with yellow 
dots. White regions correspond to values of cumulant ratios close to the Poissonian limit,
$\kappa_i/\kappa_j= 1 \pm 0.03$. Grey color signifies regions of the phase diagram in which either 
the cumulant calculation is invalid, or where data has not been produced. The legend entries denote 
upper (lower) boundaries of ranges of positive (negative) values of cumulant ratios.}
\label{Cumulants_diagrams}
\end{figure}

In Fig.\ \ref{phase_diagram} we show coexistence and spinodal region lines for several representative
a QGP-like phase transitions, specified by the position of the QGP-like critical point and the 
boundaries the corresponding spinodal region. It is evident that the VDF model is able to produce 
an array of phase diagrams corresponding to different proposed properties of the QCD phase transition. 
In Fig.~\ref{Cumulants_diagrams}, we plot cumulant ratios in the $T$-$n_B$ and $T$-$\mu_B$ planes, 
calculated using one of the representative EOSs. The behavior of the cumulant ratios, in particular 
their values exceeding or falling below the Poissonian limit of $\kappa_i/\kappa_j = 1$ in 
specific regions around the phase transition, agrees with well-known expectations 
\cite{Asakawa:2009aj,Stephanov:2011pb}.

\begin{figure}[b]
\includegraphics[width=\textwidth]{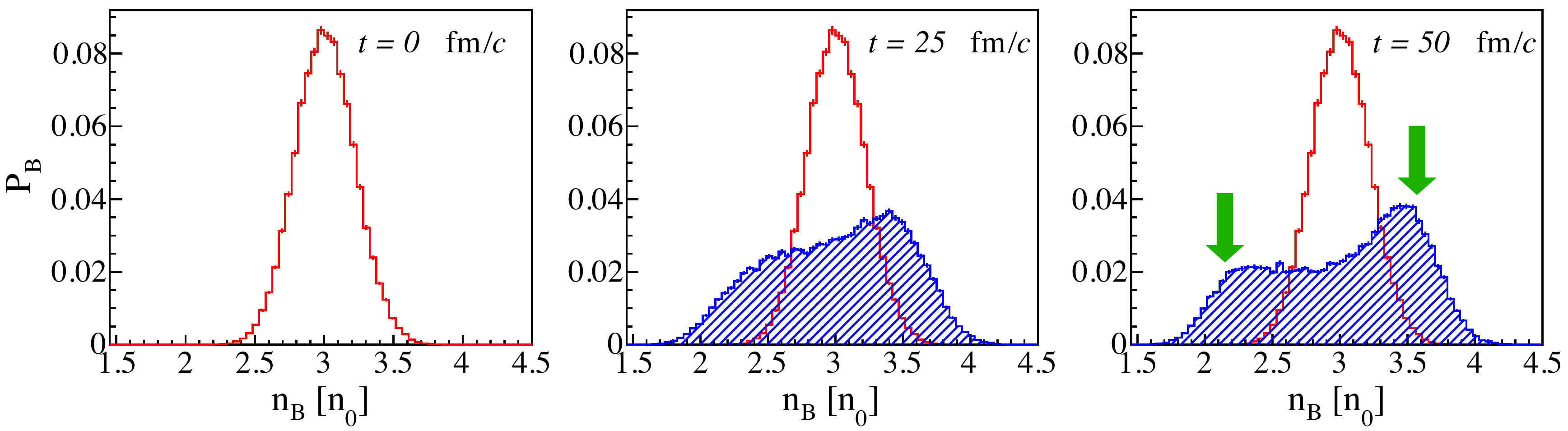}
\caption{Time evolution of the baryon number distribution for a system initialized inside the
QGP-like spinodal region (at baryon number density $n_B = 3 n_0$ and temperature $T = 1 
\ \textrm{MeV}$), averaged over $N_{\textrm{ev}} = 500$ events. Red histograms correspond to 
the baryon distribution at initialization ($t=0$), while blue shaded histograms correspond to 
baryon distributions at different times during the evolution ($t = \{25, 50 \} \ \textrm{fm}/c$). 
The system, initialized in a mechanically unstable region of the phase diagram, undergoes a 
spontaneous separation into a "less dense" and a "more dense" nuclear liquid, resulting in 
a double-peaked baryon number distribution. The green arrows point to values of baryon number 
densities corresponding to the boundaries of the coexistence region at $T=1~\textrm{MeV}$, 
$n_L = 2.13 n_0$ and $n_R = 3.57 n_0$.}
\label{spinodal_decomposition_hadron_histograms_full}
\end{figure}

The VDF equations of motion (\ref{equation_of_motion_x_generalized} and
\ref{equation_of_motion_p_generalized}) have been implemented in the hadronic transport code SMASH
\cite{Weil:2016zrk}. We verified that the mean-field hadronic transport reproduces the known properties 
of ordinary nuclear matter, such as the value of the binding energy at the saturation density, and 
the spinodal lines that characterize the unstable region of the nuclear phase transition. We 
show the evolution of the baryon number density for a system initialized inside the spinodal region 
of the proposed QGP-like phase transition in Fig.~\ref{spinodal_decomposition_hadron_histograms_full}. 
In this figure, the red curve corresponds to the distribution at time $t = 0$, while the blue curves
delineate the distribution at times $t>0$. At $t=0$, the distribution is peaked at the initialization
density $n_B = 3 n_0$, but in the course of the evolution the system separates into two coexisting 
phases, a "less dense" and a "more dense" nuclear liquid. As a result, the final distribution displays 
two peaks largely coinciding with the theoretical values of the coexistence region boundaries, 
$n_L = 2.13 n_0$ and $n_R = 3.57 n_0$; in the figure, these values are pointed to by green arrows. 

Our exploratory studies show that mean-field hadronic transport is sensitive to critical
behavior in nuclear matter, and that this behavior is exactly what is expected based on the
underlying theory. The correct description of both thermodynamics and non-equilibrium phenomena
implies that hadronic transport can be used as a tool with unique capabilities to investigate
the dynamic evolution of matter created in heavy-ion collisions.

The next step, currently in progress, is to employ these adjustable potentials in heavy-ion
collisions and determine to what extent the QGP-like phase transition affects observables. 
Preliminary findings show that at $\sqrt{s_{\mathrm{NN}}} = 7.7$ GeV these effects are small.
This is not surprising, given that in these collisions the system spends a very short time 
at densities above $2 n_0$, as one can see for example in Fig. \ref{fig:hydro:phase_diagram}.
However, at the energies explored in fixed target experiments at RHIC, and at the HADES 
experiment at GSI \cite{HADES:2019auv}, larger densities are explored and the QGP-like 
phase transition may have stronger effects on observables. Whether the effects due to 
a phase transition can be distinguished from the effect of uncertain parameters in the 
model has to be studied carefully, for example by using the Bayesian analysis method.

\section{Global modeling and analysis framework}
\label{sec:data}

The data from BES I and II runs are voluminous and heterogeneous. Measurements span a wide range 
of beam energies and centralities for a variety of beams. The experiments comprise numerous target 
and projectile combinations, and are analyzed by hundreds of collaborators within STAR and PHENIX.
Theoretical models of heavy-ion collisions are similarly complex, as the final observables depend 
on the three stages of the collision: pre-thermal evolution, hydrodynamic evolution and the final
decoupling stage. Each stage requires a different modeling paradigm. The decoupling stage is typically
described by a microscopic simulation using hadronic degrees of freedom, in contrast to the 
hydrodynamic degrees of freedom for the middle stage. The initial stage might be described by the
evolution of classical fields, a microscopic simulation involving partons, or simply a parametric 
form. Thus, one needs to carefully design and test interfaces between each stage to faithfully 
model the behavior of the degrees of freedom.

The fundamental questions addressed by the RHIC program do not easily map onto specific measurements
that can be isolated and addressed with only a single type of observable. Instead, all observables
(especially those related to soft physics) must be considered simultaneously. For example, changing 
the shear viscosity affects the anisotropic flow coefficients, the mean transverse momentum, the
multiplicity and femtoscopic correlations. The anisotropic flow coefficients are sensitive to the
viscosity, the equation of state, and details of saturation and stopping in the pre-thermal stage. 

Global analyses of higher-energy data have now been performed for data from $\sqrt{s}_{\rm NN}=200$ GeV
RHIC collisions and from LHC collisions. By simultaneously addressing several classes of observables,
these analyses are enabling rigorous scientific determination of fundamental quantities such as the
viscosity or equation of state.  At the highest energies, the hydrodynamic stage alone provides a major
impact with respect to affecting final-state observables. For BES data, the system spends a larger
fraction of its time in the final-state hadronic stage and in the initial pre-thermal stage, thus
increasing the importance of how these two stages are treated. Because of the larger non-uniformity 
in rapidity and the increased baryon density, the stopping and thermalization stage of low-energy
collisions is inherently more difficult to model and more rife with theoretical uncertainty. For these
reasons, along with the fact that a larger range of beam energies is considered, a global BES analysis
will be significantly more challenging, both numerically and theoretically, than the high-energy 
analyses.

The BEST Collaboration addresses the challenge laid out above in two ways. First, the BEST modeling
infrastructure is modular and the individual components are being thoroughly tested. Secondly, the 
design will accommodate a global Bayesian analysis aimed at rigorous expression of fundamental 
parameters describing the bulk properties and evolution of high-density QCD matter. In the following
section, we outline the structure and status of the BEST modeling framework, then describe how this 
will be applied to the interpretation of BES data.

\subsection{The BEST modeling framework}

\begin{figure}
\centerline{\includegraphics[width=1.0\textwidth]{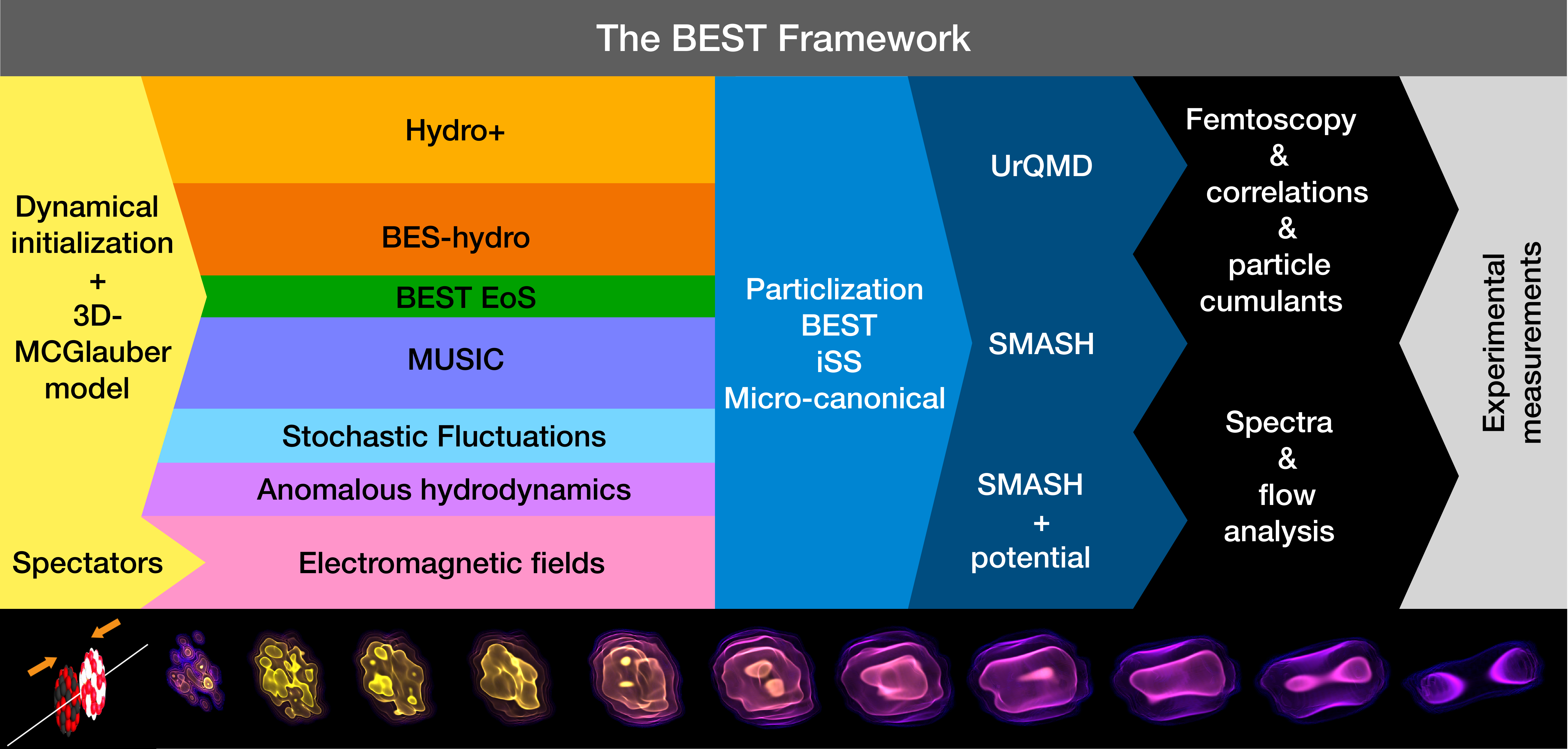}}
\caption{\label{fig:bestframework} The BEST modeling and analysis framework is highly modular. It permits the user to mix and match modules, and to insert improved or alternate versions seamlessly.}
\end{figure}

Modeling heavy ion collisions requires accurate descriptions of three phases, pre-thermal evolution,
hydrodynamics, and a hadronic simulation describing the final evolution and decoupling. Additionally,
interfaces between different phases must be developed. Rigorous extraction of fundamental parameters
requires a careful and thoughtful Bayesian comparison of experimental data to model output. The models
used in such analysis must faithfully express the entire breadth of reasonable possibilities. This
necessitates a modular design of the modeling framework, so that competing theoretical paradigms can 
be compared and distinguished. For that reason, the three principal modeling components are each 
designed as interchangeable modules with well defined and carefully tested interfaces between modules.
Figure \ref{fig:bestframework} illustrates the design, emphasizing both the modularity and workflow.

\begin{figure}
\centerline{\includegraphics[width=0.5\textwidth]{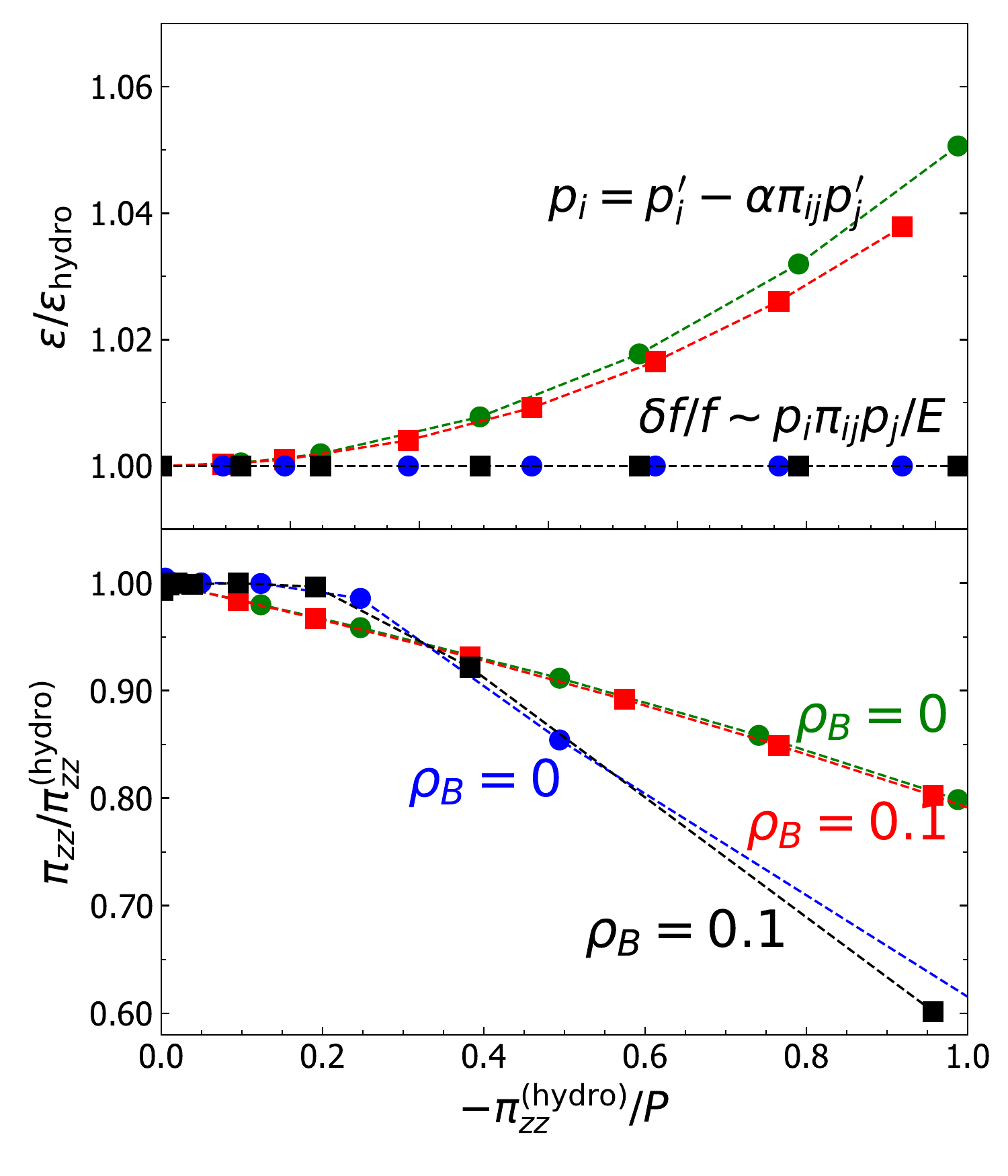}}
\caption{\label{fig:samplershear}
The particlization routine in the default BEST sampler offers different schemes for implementing
modifications in the phase space density due to viscous shear. One option is that the correction to 
the distribution function is $\delta f/f\sim p_i\pi_{ij}p_j/E$, where $\pi_{ij}$ is the shear correction
to the stress energy tensor (blue and black curves), see for example Ref.~\cite{Teaney:2003kp}.  A 
second option is to produce particles without shear correction, but to multiply their momenta by 
a matrix proportional to $\pi_{ij}$ (red and green curves), as described in Ref.~\cite{Pratt:2010jt}.
Both methods are accurate as long as the viscous correction to the stress energy tensor, in this 
case $\pi_{zz}$ where $\pi_{xx}=\pi_{yy}=-\pi_{zz}/2$, is less than half the pressure. The average 
energy density (upper panel) and the average stress-energy tensor (lower panel) of the sampled 
particles is compared to the hydrodynamic values for both zero baryon density $\rho_B=0$, and 
for $\rho_B=0.1$ fm$^{-3}$.}
\end{figure}

The utilization of standard formats enables the plug-and-play functionality. For example, the 
final-state particles resulting from the hadronic simulation are written in OSCAR format
\cite{Oscar}, which 
provides both the asymptotic momentum of each particle and its last point of interaction. From 
this information, the analysis code can construct single-particle observables like spectra and 
flow, two-particle femtoscopic correlations, and multiplicity fluctuations. Hydrodynamic codes 
produce a list of hyper-surface elements in a standard format from which the particlization codes 
produce a set of hadrons consistent with the stress-energy tensor at the interface between the
hydrodynamic- and hadronic-simulations.  The interfaces are flexible by design. For example, 
Fig.~\ref{fig:samplershear} illustrates how accurately the viscous correction to the stress-energy 
tensor is reproduced by the sampled particles at the boundary between the hadronic simulation 
and the hydrodynamics description. The codes permit the user to choose between different
representations of the viscous shear corrections to the phase space density. 

\begin{figure}
\centerline{\includegraphics[width=0.5\textwidth]{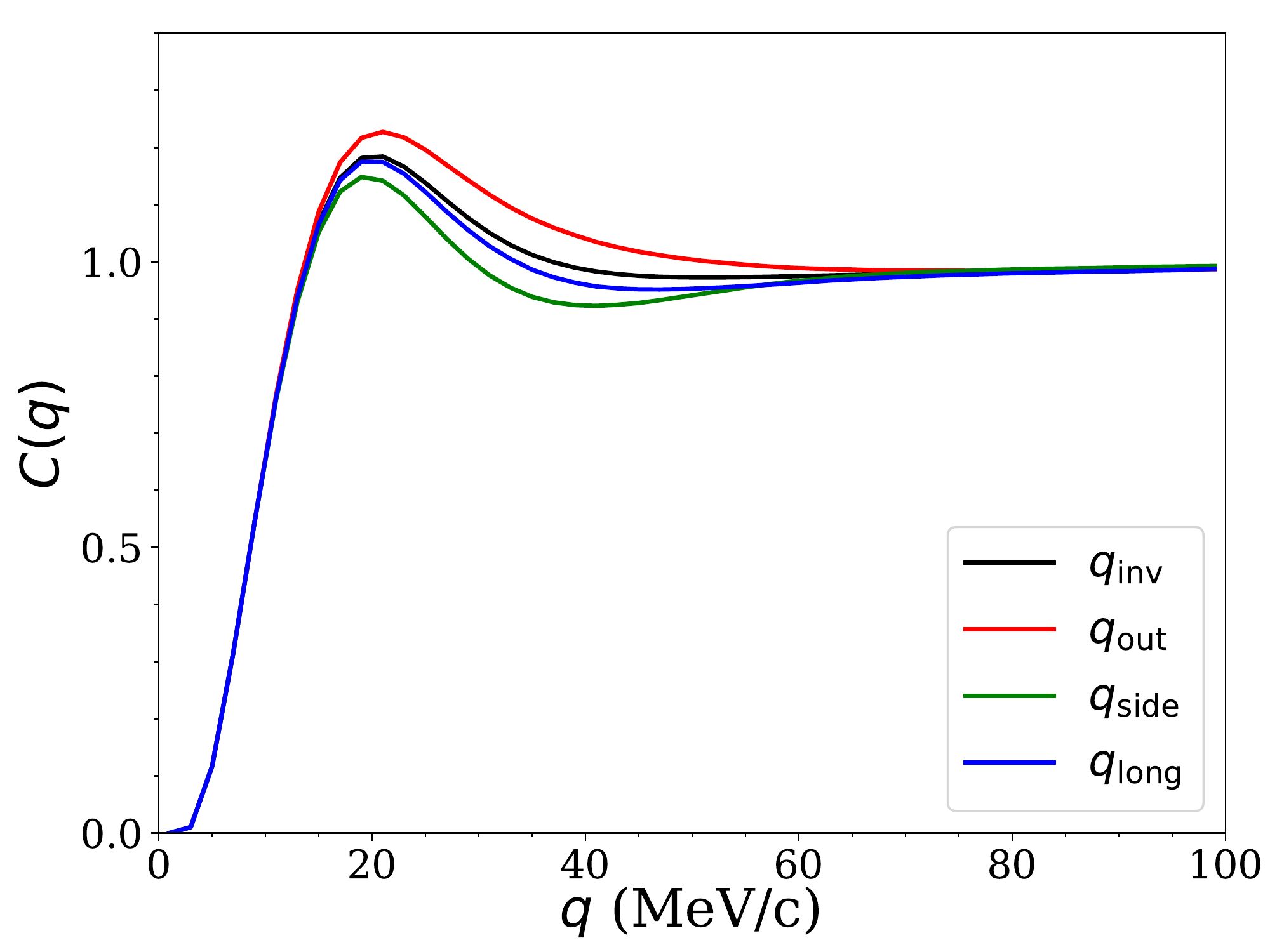}}
\caption{\label{fig:pphbt}BEST has developed analysis codes to generate spectra, flow coefficients 
and femtoscopic correlations from the OSCAR output of the hadronic simulations. Here, proton-proton
correlations are displayed for Au+Au collisions at a beam energy $\sqrt{s}_{NN}=19.6$ GeV. Femtoscopic
correlations can also be generated for $\pi^+\pi^-,\pi p,\pi K, K^+K^+, K^+K^-, Kp, pn, nn,
\Lambda\Lambda, p\Lambda$ and $p\bar{p}$. }
\end{figure}

The modeling framework produces lists of emitted hadrons. This includes the PID (particle identification
code), momentum and the space-time coordinates of its last interaction. BEST software allows the user 
to quickly produce spectra and flow coefficients. This will include composite particles, i.e. light
nuclei \cite{Oliinychenko:2020znl}. Femtoscopic correlations, as illustrated in Fig. \ref{fig:pphbt} are also readily generated.
Analysis codes can also be easily exchanged and are unaffected by choices of modules from earlier 
stages of the collision, as long as the observables consider final-state hadrons emerging from the
hadronic simulation. For particles emitted from earlier stages such as photons and dileptons, analysis
codes can use hydrodynamic histories, but these formats might vary, or not even exist, for some 
modules.

\subsection{Global Bayesian analysis}

Once the modeling infrastructure is complete, the BEST Collaboration will perform a global analysis 
of BES data, focusing on soft observables such as flow coefficients, spectra and femtoscopic source
sizes. The procedure will follow that described in Refs.~\cite{Bernhard:2019bmu,Xu:2018cvp,
Auvinen:2017nrm,Xu:2017nmc,Ke:2017xyi,Auvinen:2017fjw,Bass:2017zyn,Ke:2016jrd,Bernhard:2016tnd,
Pratt:2015zsa,Sangaline:2015isa,Novak:2013bqa,Pratt:2015jsa,Park:2020mkl}. Those analyses provide a
sampling of the model parameters weighted by the posterior Bayesian likelihood. In previous analyses
the number of model parameters, which we will collectively denote $\vec{\theta}$, have exceeded a 
dozen in some analyses. However, for a global BES analysis the number of model parameters will 
be much larger, on the order of several dozen. The majority of parameters are needed to describe 
the pre-thermal stage, which is poorly known. For example, in Ref.~\cite{Pratt:2015zsa} five 
parameters were used to describe the initial state of the hydrodynamic evolution at a single 
beam energy. These included a weight between two saturation models, the dependence on 
having asymmetric thickness functions, the energy density scale, the initial transverse flow, and 
the initial anisotropy of the stress-energy tensor. The Beam Energy Scan covers a wide variety of
energies, plus at each energy one must also model the rapidity dependence of both the energy and 
baryon densities. 

Bayesian analyses follow a fairly standard procedure:
\begin{enumerate}
\item Distill the data to a list of observables. This may involve reducing numerous graphs to a 
few numbers using principal component analysis as a guide. By applying a principal component 
analysis (PCA), one can identify linear combinations of observables which are insensitive to the 
change of parameters. The original observables, $y_a$, are first scaled by their variances, 
$\tilde{y}_a\equiv y_a/\sigma_a$. One then constructs a covariance matrix,
$S_{ab}=\langle\delta\tilde{y}_a\delta\tilde{y}_b\rangle$, where the averaging is performed over 
several hundred model runs throughout the model space. One then chooses linear combinations of
$\tilde{y}$ that are eigenstates of $S$. The observables are thus represented by these principal
components $z_a$, which are characterized by the corresponding eigenvalues of $S$, 
$\lambda_a$. The combinations $z_a$ with larger eigenvalues, $\lambda_a\gtrsim 1$, vary significantly
throughout the parameter space, and thus provide significant resolving power. Those with $\lambda_a
\ll 1$ are effectively useless, and can be ignored in the next two steps. Typically, for applications 
in heavy-ion physics, the number of significant principal components is smaller than the number of 
model parameters. Thus, Bayesian analyses in heavy-ion physics are typically under-constrained 
problems despite the immense size of the experimental data set.

\item A model emulator must be designed and constructed for use in the Markov-Chain Monte Carlo 
(MCMC) procedure described below. The purpose of an emulator is to provide the ability to estimate 
the observables coming from the model, or the principal components, as a function of the model 
parameters quickly, without having to always run the full model. The emulator is built by running 
the full model several hundred, or perhaps a few thousand, times semi-randomly throughout the parameter
space. These could be the same runs used in (1) to determine the principal components. The emulator
effectively provides an interpolation from the full-model evaluations, giving $z^{\rm(emu)}_a
(\vec{\theta}) \approx z^{\rm(mod)}_a(\vec{\theta})$. The emulator then takes the place of the full 
model in the MCMC procedure described below. Gaussian Process emulators are popular \cite{Novak:2013bqa},
but given the smooth, usually monotonic, response of observables to parameters in these applications, 
one can use various linear or quadratic fits just as well. 

\item The MCMC procedure provides a weighted walk through parameter space. At each step, the likelihood
is calculated,
\begin{equation}
\mathcal{L}(\vec{\theta}) \sim
\exp \left\{-\sum_a[z_a^{\rm(exp)}-z_a^{\rm(mod)}(\vec{\theta})]^2/2\right\}.
\end{equation}
Here, $z_a$ refer to linear combinations of observables, which after being scaled by their uncertainty,
are chosen according to PCA. In order to calculate the likelihood at a given point $\vec{\theta}$, one
must either run the full model to determine $z_a^{\rm(mod)}$, or use an emulator to estimate
$z_a^{\rm(mod)}$. Typically a metropolis algorithm is applied. This generates a set of points
$\vec{\theta}^{\rm(post)}$ that represents the posterior probability. The algorithm typically represents 
millions of such points. The mean value of the parameters are then calculated as
\begin{equation}
\langle\vec{\theta}\rangle = \frac{1}{N_s}\sum^{N_s}_n\vec{\theta}_n,
\end{equation}
where $N_s$ is the number of points sampled. 
Other moments of the posterior parameter distribution can be similarly extracted from the MCMC trace. 

\end{enumerate}
BEST's statistical analysis is based on that developed by the MADAI Collaboration (Modeling and Data
Analysis Initiative), see \cite{Novak:2013bqa} and \cite{madai}. The MADAI toolset assists with 
building and designing an emulator, and performing the MCMC trace. The software also includes 
code for analyzing the resolving power of specific observables in regards to constraining a 
given parameter \cite{Sangaline:2015isa}.

Performing a global Bayesian analysis on BES data is significantly more challenging than similar 
analyses at higher energies. The physics, especially the initial stage, is far more uncertain and 
will likely involve more than twice as many model parameters. In addition to that, the data 
cannot be modeled using a boost invariant, two-dimensional, approximations of the three-dimensional
hydrodynamic evolution. This means that the work is numerically 1-2 orders of magnitude more demanding
than any analysis previously performed in this field. As the full end-to-end model components were assembled in late 2020, the BEST Bayesian analysis will appear no sooner than late 2021.
\section{Present Status, outlook and conclusions}
\label{sec:conclusion}

Over the last five years the BEST Collaboration has made tremendous strides towards
developing a dynamical framework for a quantitative description of heavy ion collisions 
at energies relevant to the RHIC beam energy scan. Most of the essential elements the 
BEST Collaboration set out to address 
have been addressed: 

\begin{itemize}
    
\item A model for complex initial conditions considering baryon stopping and the finite 
time interval for the transition to hydrodynamics has been developed, implemented and 
tested by comparing with available experimental data.  
    
\item Viscous hydrodynamics has been extended to propagate the relevant conserved currents
and their respective dissipative corrections. Also, the time evolution of anomalous currents
has been included, and the inclusion of the corresponding dissipative terms is close to 
completion.
    
\item The time evolution of fluctuations has been addressed using both stochastic
hydrodynamics as well as a deterministic framework for propagating correlation functions. 
Within the deterministic approach we have studied the backreaction of fluctuations on the
hydrodynamic evolution using exploratory calculations in the Hydro+ approach.
    
\item We have constructed a flexible model equation of state which contains a critical 
point in the Ising universality class, and which reproduces available lattice QCD results 
at  vanishing chemical potential. We have implemented this equation of state in a 
hydrodynamic code.
    
\item The transition from hydrodynamic fields to particle degrees of freedom, often 
referred to as particlization, has been extended to allow for local conservation of all 
conserved quantities. This allows for a faithful mapping of fluctuations from stochastic 
hydrodynamics to kinetic theory. In addition, considerable progress has been made towards 
mapping the correlation functions in the deterministic approach to particles.
       
\item The kinetic evolution of the hadronic phase has been extended to allow for mean 
field interactions between the particles. This allows for a non-trivial EOS in the hadronic
phase, and for a proper mapping of the EOS used in hydrodynamics to the hadronic phase. 
To this end a flexible density functional model has been developed. 
    
\item A Bayesian analysis framework has been adjusted and modified to the needs for a 
comprehensive data comparison with the soon to be expected data from BESII. 
\end{itemize}

While the entire framework is not complete as of this writing, some parts of it have 
already been used extensively. For example, hydrodynamics including anomalous currents, the 
AVFD model, is being used by the STAR Collaboration to test the sensitivity of the various 
observables considered for the analysis of the isobar run. In addition parts of the current 
framework have been utilized to provide a baseline for the isobar run, which does not 
include any anomalous currents but accounts for background effects such as momentum and 
local charge conservation. The following points need to be elucidated in order to 
complete the dynamical framework

\begin{itemize}
\item Extend the hydrodynamic code to include the Hydro+ framework for the propagation of 
the two-point functions, necessary for the deterministic description of fluctuations. In 
addition, the transition from Hydro+ to particle degrees of freedom needs to be addressed. 
Both of these points are presently under development. 

\item The EOS with a critical point needs to be extended to allow for higher baryon 
densities. Also, the first order co-existence region including the unstable spinodal region 
needs to be modeled. This also requires the inclusion of finite range (or derivative) terms
in the EOS.

\item The mean field for the kinetic description needs to be chosen such that it matches 
the EOS used in hydrodynamics at particlization. This requires also an efficient algorithm 
to allow a flexible choice of the EOS in the Bayesian analysis. 

\item The propagation of the anomalous currents (AFVD) needs to be extended to be able to 
deal with systems at the lowest energies by properly including baryon currents and initial
conditions for the axial charges. 

\item In order to consider third  and fourth order cumulants of the baryon number, the 
Hydro+ formalism needs to be extended to include three- and four-point functions. 
\end{itemize}

Of course the ultimate goal is to carry out a Bayesian analysis of the experimental data 
to constrain the model parameters and thus the  possible existence and location of a QCD 
critical point as well as the presence of anomalous transport. The first step of such an
analysis is to constrain model parameters by comparing with a set of physical measurements 
which are not sensitive to either CP nor to anomalous transport, such as spectra and flow.
This will reduce the parameter space for the final comparison including fluctuations and 
correlation observables.

\section*{Note added in proof}

 After this work was submitted the first data from the RHIC isobar run 
 appeared \cite{STAR:2021mii}.
 
\section*{Acknowledgments}

\hspace*{\parindent}
This material is based upon work supported by the U.S. Department of Energy, 
Office of Science, Office of Nuclear Physics, within the framework of the 
Beam Energy Scan Theory (BEST) Topical Collaboration, and under contract numbers
DE-FG88ER40388 (D.K.),
DE-SC0012704 (D.K., S.M., B.S.),
DE-SC0018209 (H.U.Y.),
DE-SC0020633 (J.N.)
DE-FG0201ER41195 (M.S., H.U.Y., S.L., X.A., M.P.),
DE-SC0004286 (L.D., U.H., M.M.),
DE-AC02-05CH11231 (V.K., V.V.),
DE-FG02-03ER41259 (S.P.),
DE-SC0013460 (C.S.),
DE-SC0021969 (C.S.),
DE-SC0020081 (V.S.),
DE-FG02-03ER41260 (T.S., M.M.),
DE-SC0011090 (K.R., G.R., R.W., Y.Y.),
DE-FG02-87ER40328 (M.S.),
and Office of Science National Quantum Information Science Research Centers 
under the "Co-design Center for Quantum Advantage" award (D.K.),
and in part by the U.S. National Science Foundation Grant numbers
PHY-1913729 (J.L.),
PHY-2012922 (C.S.),
PHY-1654219 (C.R.),
and within the framework of the JETSCAPE Collaboration under Award 
No.~\rm{ACI-1550223} (U.H.),
and in part by the Natural Sciences and Engineering Research Council of 
Canada (C.G., S.S., M.S.),
and in part by the program ``Etoiles montantes en Pays de la Loire 2017'' 
(M.B., M.N.).
U.H. also acknowledges support by a Research Prize from the Alexander von 
Humboldt Foundation, and 
V.V. received support by the Alexander von Humboldt Foundation through the 
Feodor Lynen Program.
S.S. and M.S. also acknowledges the Fonds de recherche du Qu\'ebec - Nature et 
technologies (FRQNT) through the Programmede Bourses d'Excellencepour 
\'Etudiants \'Etrangers (PBEEE) scholarship.
Y.Y. is supported by the Strategic Priority Research Program of Chinese 
Academy of Sciences.
D.M. is supported by National Science Foundation Graduate Research Fellowship Program 
under Grant No. DGE–1746047.
P.P. acknowledges support by the DFG grant SFB/TR55 and by the Federal Ministry 
of Education and Research (through BMBF projects 05P18PXFCA and 05P21PXFCA)
This research used resources of the National Energy Research Scientific 
Computing Center, which is supported by the Office of Science of the U.S. 
Department of Energy under Contract No. DE-AC02-05CH11231, resources 
provided by the Open Science Grid~\cite{Pordes:2007zzb, Sfiligoi:2009cct}, 
which is supported by the National Science Foundation award 2030508, 
resources provided by the Ohio Supercomputer Center (Project PAS0254)
\cite{OhioSupercomputerCenter1987}, 
and resources of the high performance computing 
services at Wayne State University.

\bibliography{BEST.bib}

\end{document}